\documentclass[11pt, oneside]{article}   	
\usepackage{slashed}
\usepackage{jheppub}
\usepackage{amsmath}
\usepackage{amssymb}
\usepackage{amsfonts}
\usepackage{amsthm}
\usepackage{amsbsy}
\usepackage{array}
\usepackage{mathtools}
\usepackage[usenames,dvipsnames]{xcolor}
\usepackage{subcaption}
\usepackage{dsdshorthand}
\usepackage{graphicx}
\usepackage[vcentermath]{youngtab}
\usepackage{hyperref}
\usepackage{cleveref}
\usepackage{cancel}
\usepackage[normalem]{ulem}

\usepackage{tikz}
\usetikzlibrary{snakes}
\usetikzlibrary{decorations}
\usetikzlibrary{trees}
\usetikzlibrary{decorations.pathmorphing}
\usetikzlibrary{decorations.markings}
\usetikzlibrary{external}
\usetikzlibrary{intersections}
\usetikzlibrary{shapes,arrows}
\usetikzlibrary{arrows.meta}
\usetikzlibrary{calc}
\usetikzlibrary{shapes.misc}
\usetikzlibrary{decorations.text}
\usetikzlibrary{backgrounds}
\usetikzlibrary{fadings}
\usetikzlibrary{tikzmark,calc,arrows,shapes,decorations.pathreplacing}
\usetikzlibrary{patterns}
\tikzset{
        cross/.style={cross out, draw=black, minimum size=2*(#1-\pgflinewidth), inner sep=0pt, outer sep=0pt},
	branchCut/.style={postaction={decorate},
		snake=zigzag,
		decoration = {snake=zigzag,segment length = 2mm, amplitude = 2mm}	
    }}
\usepackage{booktabs}
\usepackage{array}

\newcommand{\bea}{\setlength\arraycolsep{2pt} \begin{eqnarray}}
\newcommand{\eea}{\end{eqnarray}}

\def\fft#1#2{{\frac{#1}{#2}}}

\newcommand{\baa}{\begin{align}}
\newcommand{\eaa}{\end{align}}

\makeatletter
\def\@fpheader{\ }
\makeatother

\newcommand{\mydiagram}[9]{%
  \begin{scope}[xshift=#1cm, yshift=#2cm]
    \coordinate (v1) at (0,0);
    \coordinate (v2) at (1.4,0);
    \draw[thick] (v1) to[bend left=20]  (v2);
    \draw[thick] (v1) to[bend right=20] (v2);
    \draw[dashed, thick] (v1) -- ++( 135:1.2);
    \draw[dashed, thick] (v1) -- ++(-135:1.2);
    \draw[dashed, thick] (v2) -- ++(  45:1.2);
    \draw[dashed, thick] (v2) -- ++( -45:1.2);
    \node at (-0.5,  0.25) {#3};
    \node at (-0.5, -0.25) {#4};
    \node at ( 1.9,  0.25) {#5};
    \node at ( 1.9, -0.25) {#6};
    \node at ( 0.2,  0.25) {#7};
    \node at ( 0.2, -0.25) {#8};
    \node at ( 1.2,  0.25) {#9};
  \end{scope}
}
\newcommand{\mydiagramX}[2]{%
  \begin{scope}[xshift=#1cm]
    \node at ( 1.2, -0.25) {#2};
  \end{scope}
}

\usetikzlibrary{calc}

\newcommand{\SKdiag}[8]{%
\begin{tikzpicture}[baseline=-0.5ex]
  \def\L{1.6}\def\A{45}
  \tikzset{lbl/.style={font=\footnotesize, inner sep=1pt}}
  \coordinate (V)  at (0,0);
  \coordinate (S)  at (-\L,0);
  \coordinate (eU) at ({\L*cos(\A)}, { \L*sin(\A)});
  \coordinate (eM) at ( \L,0);
  \coordinate (eD) at ({\L*cos(\A)}, {-\L*sin(\A)});
  \draw[thick] (S)--(V) (V)--(eU) (V)--(eM) (V)--(eD);
  \node[left=2pt] at (S) {$\phi^{c}_{\mathrm{EFT}}$};   
  \node[lbl,above=1pt] at ($(S)!0.20!(V)$)  {$#1$};
  \node[lbl,above=1pt] at ($(V)!0.85!(eU)$) {$#6$};
  \node[lbl,above=1pt] at ($(V)!0.85!(eM)$) {$#7$};
  \node[lbl,below=1pt] at ($(V)!0.85!(eD)$) {$#8$};
  \node[lbl,above=1pt] at ($(S)!0.80!(V)$)  {$#2$};
  \node[lbl,above=1pt] at ($(V)!0.20!(eU)$) {$#3$};
  \node[lbl,above=1pt] at ($(V)!0.22!(eM)$) {$#4$};
  \node[lbl,below=1pt] at ($(V)!0.20!(eD)$) {$#5$};
\end{tikzpicture}}

\title{Stochastic inflation as an open quantum system II: open effective field theory and stochastic matching
}
\author{Yue-Zhou Li$^\phi$}
\affiliation{
${}^\phi$Department of Physics, Princeton University, Princeton, NJ 08544, USA 
}
\emailAdd{liyuezhou@princeton.edu}
\date{}
\abstract{We further develop the proposal in Phys.\ Rev.\ Lett.\ \textbf{136} 071501 that interprets stochastic inflation as an open quantum system, by constructing the open effective field theory for the reduced density matrix of long wavelength modes. We clarify that this open effective field theory enjoys two renormalization group flows: the conventional Wilsonian channel, and a stochastic channel arising from the openness that has no counterpart in ordinary Wilsonian effective field theory. Focusing on the stochastic channel in the hard cutoff scheme, we identify both Gaussian and non-Gaussian diffusion as effective operators in the influence functional, and show that they are required by matching onto correlators and form factors of the perturbative full theory through a method-of-region in time. Beyond Gaussian order, the matching data are no longer local Wilson coefficients but nonlocal and non-Markovian Wilson kernels. We then obtain the bare Hamiltonian density of this open effective field theory and derive its nonlocal functional master equations, including the Fokker-Planck equation for the diagonal density matrix and the Klein-Kramers equation for the Wigner functional, with their zero-modes simplifications discussed. Finally, we take a first step toward a continuum version of this open effective field theory, replacing the hard cutoff by an analytic regulator in the stochastic channel, and demonstrate stochastic renormalization using a massive scalar as an example.}

\begin{document}

\maketitle
\pagenumbering{roman}
\setcounter{page}{2}
\newpage
\pagenumbering{arabic}
\setcounter{page}{1}

\section{Introduction}

Effective field theories (EFTs) provide a powerful framework for describing long distance phenomena by integrating out heavy degrees of freedom that cannot be resolved at the relevant scale. In standard Wilsonian reasoning, EFTs are conventionally expected to be local and unitary. However, exceptions arise for phenomena involving long distance but high energy, where the physics is intrinsically Lorentzian and nonlocal EFTs can emerge beyond the Wilsonian paradigm. A prominent example is soft-collinear effective theory in QCD \cite{Bauer:2000ew,Bauer:2000yr,becher2015introduction,Rothstein:2016bsq}. In such cases, however, unitarity remains preserved.

On the other hand, an open quantum system describes a system interacting with an environment, whose bath degrees of freedom are traced out. The resulting EFT is often referred to as an open EFT. Such an open EFT manifestly breaks unitarity by incorporating dissipation and diffusion effects, encoded in the Feynman–Vernon influence functional \cite{feynman1963theory} within the Schwinger–Keldysh (SK) formalism \cite{schwinger1961brownian,keldysh2024diagram}. Unitarity can only be recovered after resumming the effective interactions from the environment, thereby reconstructing the environmental degrees of freedom. Examples include hydrodynamic EFTs \cite{Crossley:2015evo,Liu:2018kfw}, worldline EFTs for gravitating compact objects \cite{Goldberger:2009qd,Saketh:2023bul,Ivanov:2024sds,Glazer:2024eyi,Caron-Huot:2025tlq,Ivanov:2026icp}, and cosmological EFTs with unknown matter sectors as bath \cite{LopezNacir:2011kk,Burgess:2014eoa,Salcedo:2024smn,Colas:2024lse,Salcedo:2025ezu,Colas:2025app,Colas:2025ind,Salcedo:2026cqb,Salcedo:2026sdn,Cespedes:2025ple}.

It is certainly true that heavy degrees of freedom and rapidly fluctuating modes can also be viewed as an environment for lighter and more slowly fluctuating modes. It is therefore natural to ask whether there exists an example of an open system that arises from integrating out ``ultraviolet'' degrees of freedom. 

One such example arises in slow-roll inflationary cosmology, as originally shown by Starobinsky \cite{Starobinsky:1986fx}. Inflation describes an early period of exponential expansion of the universe, during which the inflaton slowly rolls down under an approximately flat potential. The associated quantum fluctuations are stretched to cosmological scales and seed the cosmic microwave background. The correlations of these perturbations can usually be computed using cosmological perturbation theory \cite{Mukhanov:1990me,Maldacena:2002vr,Acquaviva:2002ud,Durrer:2004fx,Baumann:2009ds,wang2014inflation}. However, in the regime of eternal inflation \cite{Linde:1986fd,Goncharov:1987ir,Guth:2007ng,Creminelli:2008es}, or in scenarios involving primordial black holes \cite{Carr:1974nx,Carr:1975qj,Sasaki:2018dmp}, standard cosmological perturbation theory can break down due to the strong nonlinearity of quantum fluctuations.

Starobinsky proposed to address these regimes using a random-walk model for the inflaton \cite{Starobinsky:1986fx,Starobinsky:1994bd}, described by the corresponding Langevin equation and Fokker-Planck (FP) equation. This framework is known as stochastic inflation, and has been widely used to understand eternal inflation \cite{Goncharov:1987ir,mijic1990random,Linde:1993nz,Linde:1993xx,Tolley:2008na,Cohen:2021jbo}, the resummation of secularly divergent observables \cite{Starobinsky:1994bd,tsamis2005stochastic,Gorbenko:2019rza,Woodard:2025cez,rey1987dynamics,nambu1988stochastic,nambu1989stochastic,kandrup1989stochastic,habib1992stochastic,Enqvist:2008kt,Hu:2018nxy,Baumgart:2019clc}, and, more recently, primordial black holes with $\delta N$ formalism \cite{Pattison:2017mbe,Kuhnel:2019xes,Ezquiaga:2019ftu,Vennin:2020kng,Ballesteros:2020sre,Figueroa:2021zah,Saha:2026cay,Ye:2026saa}. Nevertheless, many discussions rely on Langevin equations and their corresponding FP equations \cite{Calzetta:1993qe,PerreaultLevasseur:2013kfq,Fujita:2013cna,Fujita:2014tja,PerreaultLevasseur:2014ziv,Vennin:2015hra,Grain:2017dqa,Firouzjahi:2018vet,Pinol:2018euk,Firouzjahi:2020jrj,Pattison:2021oen,Cruces:2021iwq,Cohen:2021fzf,Aldabergenov:2025ulq,Panda:2025tpu,Kawasaki:2026hnx}, although there has been progress in developing a deep understanding of its quantum field theory (QFT) foundation over the past several decades \cite{morikawa1990dissipation,hosoya1989stochastic,Hu:1992ig,Hu:1994iw,Garbrecht:2013coa,Garbrecht:2014dca,Moss:2016uix,Prokopec:2017vxx,Collins:2017haz,Andersen:2021lii,Tokuda:2017fdh,Tokuda:2018eqs,Cruces:2022imf,Miyachi:2023fss,Kamenshchik:2024ybm,Pinol:2020cdp,Kamenshchik:2025ses,Cruces:2026yvs,Palma:2023idj,Palma:2025oux}.

Most recently, progress has been made toward more rigorous and systematic understandings of stochastic inflation, including treating it as an open quantum system \cite{Li:2025azq}, promoting the random walk to a quantum walk \cite{Green:2025hmo}, understanding it using exact renormalization group (RG) techniques \cite{Prokopec:2017vxx,Cespedes:2023aal,Green:2025hmo,Cespedes:2026fdp}, and interpreting FP equations as RG equations for infrared (IR) operator mixing \cite{Cohen:2021fzf,Beneke:2026ksj}.

In this paper, we follow \cite{Li:2025azq} to interpret the stochastic inflation as an open quantum system, and then further construct the corresponding open EFT. In particular, we go beyond the zero-modes limit and Gaussian order discussed in \cite{Li:2025azq}, and explicitly construct and match the non-Gaussian diffusion. We closely follow the EFT paradigm, while working in the hard cutoff scheme traditionally used for stochastic inflation. By following this paradigm, we clarify the renormalization structure of the open EFT, and further develop a consistent and convenient prescription showing that this EFT indeed plays its role by matching onto perturbative full theory observables, including correlation functions and form factors.

We refer readers to the roadmap of this paper below.

\subsection{Roadmap}

We summarize here the key conceptual point and the main results of each section, together with a brief guide to the Appendices.

\begin{itemize}

\item In section \ref{sec: path integral}, we provide a self contained review of the SK formulation of density matrices and master equations in Liouville space. The conceptual point is that tracing out a subsystem is identified with the integration out step of an EFT, so the reduced density matrix is described by an EFT with its effective Liouville Hamiltonian, which encodes the dissipation and diffusion effects. This provides the unified language used throughout the paper. Readers familiar with the SK formalism can skip this section.

\item In section \ref{sec: setup}, we first introduce the slow roll inflation setup. We then revisit the open system description of inflation, treating the long wavelength modes as the system and the short wavelength modes as the environment, split at the time-dependent scale $\Lambda(t)=\varepsilon a(t)H_{0}$. We explain three key approximations used throughout our construction: the Born approximation, the secular approximation, and large spatial scale separation. We explain that the openness arises from the time-dependent scale separation between the environment and the system, generating a $\dot\phi_{s}\dot\phi_{e}$ coupling localized on the crossing scale $\varepsilon a H_0$. This generates a genuinely \emph{open} EFT for the reduced density matrix of the long wavelength modes.

\item In section \ref{sec: EFT Gaussian}, we first show that, in this open EFT, the free theory already produces a nontrivial diffusion term, \eqref{eq: diffusion computation}, which is phrased as the fluctuation-dissipation relation for the Bunch-Davies vacuum in dS, \eqref{eq: FDR dS}. We then discuss the operator power counting in this open EFT and clarify that open EFT for stochastic inflation carries two RG flows: a standard Wilsonian channel in $\log b/\varepsilon$ and a new stochastic channel in $\log\varepsilon$, as summarized in Table \ref{tab:three-logs}. We construct the Gaussian diffusion terms and match them onto the Keldysh correlator and a form factor diagram in the perturbative full theory, using a massive scalar and $\lambda \phi^4$ theory as examples. This stochastic matching identifies the stochastic channel with a \emph{method-of-region in time}, split at the crossing region $t^{\ast}(k)=H_{0}^{-1}\log(k/\varepsilon H_{0})$.

\item In section \ref{sec:non-gaussian}, we go beyond the Gaussian sector and construct non-Gaussian diffusion terms in the influence functional with more than two response fields for $\lambda\phi^{4}$ theory. We find that, instead of local Wilson coefficients, the non-Gaussian influence functional contains Wilson kernels, which depend on multiple crossing times and on the history of response fields, and are therefore intrinsically nonlocal in space and non-Markovian in time. We identify the operators $\dot{\phi}^q_s\dot{\phi}^q_s\phi^q_s\phi^c_s$ in \eqref{eq: IF O1 1},
$\dot{\phi}^q_s\dot{\phi}^q_s\phi^q_s\phi^q_s$ in \eqref{eq: IF O1 2}, $\dot{\phi}^q_s\dot{\phi}^q_s\dot{\phi}^q_s\phi^c_s$ and
$\dot{\phi}^q_s\dot{\phi}^q_s\dot{\phi}^q_s\phi^q_s$ in \eqref{eq: IF O3 and O4}, and $(\dot{\phi}^q_s)^{4}$ in \eqref{eq: O5}, and verify the perturbative matching onto the equal-time form factor of $(\phi^c)^2$, the squeezed limit of the equal-time four-point function, and the full equal-time four-point function in the full theory of $\lambda\phi^{4}$ theory.

\item In section \ref{sec:eft-structure}, we connect the matched Gaussian and non-Gaussian influence functionals to the effective Liouville action, and derive the corresponding functional master equations for the reduced cosmological density matrix, including a functional FP equation for the diagonal density matrix, \eqref{eq: FP functional}, and a Klein-Kramers equation for the Wigner functional, \eqref{eq: Wigner eq}. We also identify the nonlocal Lindblad form for the Gaussian sector in \eqref{eq: Lindblad functional}. The main result is that these equations are intrinsically \emph{functional and non-Markovian}. Nevertheless, we present the zero-modes simplification in \eqref{eq: zero-modes IF}, recovering the results of \cite{Gorbenko:2019rza,Li:2025azq} for Gaussian diffusion. We also identify the zero-modes simplification of the cubic non-Gaussian diffusion term, $\partial^3/\partial \phi^3$, in \eqref{eq: C cubic}, which is, however, not renormalized and thus is different from \cite{Cohen:2021fzf}.

\item In section \ref{sec: continuum EFT}, we take a first step toward a continuum version of this open EFT. In particular, we replace the hard cutoff by an analytic regulator adapted to mode crossing, via the distributional replacement \eqref{eq: distributional replacement}. We propose that the open EFT is reorganized in terms of nonlocal stochastic operators $\mathcal{O}_{n,i}(\nu,\mu;p_{k})$ with Wilson coefficients $C_{n,i}(\nu,\mu)$ that satisfy closed RG equations along both the stochastic scale $\nu$ and the conventional scale $\mu$. We point out that stochastic renormalization is well defined in this continuum scheme, with new stochastic anomalous dimensions. This is illustrated in the warm-up example of chaotic inflation with $v=1/2 \, m^{2}\phi^{2}$.

\item In section \ref{sec: summary}, we summarize the paper and provide an outlook.

\item In appendix \ref{app: Brownian}, we provide a worked-out open quantum system warm-up for Brownian motion, deriving the FP, and Klein-Kramers equations as projections of Caldeira-Leggett equation from the SK formalism.

\item In appendix \ref{app: FDR massive}, we provide the all-orders Gaussian influence functional and the fluctuation-dissipation relation for a massive scalar in dS.

\item In appendix \ref{app: relation to SdSEFT}, we comment on the relation between our influence functional and the stochastic initial conditions of soft dS EFT: after localizing the time integrals at the crossing time and projecting onto equal-time, the Gaussian and non-Gaussian influence functionals reduce to the corresponding (non-)Gaussian stochastic initial conditions.

\item In appendix \ref{app: one and two-pt probability}, we identify the standard one-point probability distribution of stochastic inflation with the diagonal density matrix functional projected onto homogeneous configurations, reproducing the zero-modes FP equation. We also clarify that the two-point probability distribution is the two-point function of the field-space projector operator, which can be derived from the open EFT.

\end{itemize}

\section{Wavefunction and density matrix from the path integral}
\label{sec: path integral}

In this section, we review the basic principles of constructing wavefunctions and the density matrix from path integrals. Readers who are familiar with this subject may skip this section. The formalism presented here applies equally to quantum mechanics and QFTs; nevertheless, we will make comments when necessary regarding its application to ``quantum gravity''.

\subsection{Wavefunction and the Schrodinger equation}

We now briefly review how to consider wavefunction from the path integral perspective. Our particular focus is on describing the time evolution of the wavefunction. This is a preliminary for the discussion of density matrices.

Path integrals compute transition amplitudes by summing over histories of the configurations of the system under consideration. This allows us to interpret the wavefunction in a particular representation at a final time $t_f$ as a transition amplitude from an initial state into a state labeled by that representation. For a point particle in the position representation, we have
\be
\psi(x_f,t_f)=\langle x_f,t_f|\psi,t_i\rangle=N \int dx_i\int_{x(t_i)=x_i}^{x(t_f)=x_f} [Dx Dp] e^{i S[x]} \psi(x_i,t_i)\,,
\ee
where $N$ is the normalization factor to ensure $\int dx_f |\psi(x_f,t_f)|^2=1$. In the position representation, we can simply integrate out $p$.
The weight is the action of the system, for quantum mechanics of a single particle, we have
\be
S[x]=\int dt\left(p\dot{x}-H\right)=\int dt \left(\fft{1}{2}\dot{x}^2-V(x)\right)\,,\label{eq: simple model}
\ee
where we integrate out $p$ in the second equality.
We can also consider other representations, which then may have to retain the measure $p$\footnote{For brevity, in the following, we will not distinguish between the first and second formalisms in the action. Nevertheless, we will be careful with the measure to indicate which representation we are working in.}.

In this paper, we are particularly interested in the ``ground state'' i.e., the state of minimal energy. We therefore choose to regard the initial state as this ground state. It is well known that such a state can be prepared via a Euclidean path integral, which suppresses higher-energy contributions as $t \rightarrow -i\infty$. The ground-state wavefunction at any time can then be represented by a path integral along a complex time contour stretching from $-i\infty$ to $t_f$ (we now integrate over $p$)
\be
\langle x_f,t_f|0\rangle=\psi_g(x_f,t_f)=N\int^{x(t_f)=x_f}_{x(-i\infty)=0}[D x]e^{i S[x]}\,,\label{eq: ground state path-integral}
\ee
In the context of gravitational theory, the condition $x(-i\infty) = 0$ includes the no-boundary condition for the geometry \cite{hartle1983wave}, which imposes the ``minimal energy state'' of quantum gravity as a geometry without boundary.
Excited states can be obtained by inserting nontrivial operators into the path integral. 

We now derive the Schrödinger equation from the path integral, such as Eq.~\eqref{eq: ground state path-integral}. This derivation, however, is valid for any initial state, as long as the Hamiltonian used to prepare the initial state is the same as the one controlling the subsequent time evolution
\be
\fft{\partial \psi(x_f,t_f)}{\partial t_f}=N\int_{x(-i\infty)=0}^{x(t_f)=x_f}[Dx] \left(i\fft{\partial S}{\partial t_f}\right) e^{i S}\,.
\ee
We now use the Hamilton-Jacobi equation but inside the path integral for the simple model \eqref{eq: simple model}
\be
\fft{\partial S}{\partial t_f}=\fft{dS}{dt_f}-\fft{\partial S}{\partial x_f}\dot{x}_f=-H\left(x_f,p_f,t_f\right)=-\left(\fft{1}{2}p_f^2+V(x_f)\right)\,,
\ee
leading to
\be
\fft{\partial \psi(x_f,t_f)}{\partial t_f}=-i\left(-\fft{1}{2}\fft{\partial^2}{\partial x_f^2}+V(x_f)\right)\psi(x_f,t_f)\,,
\ee
where we used
\be
\delta S=\int dt\, {\rm EOM}\, \delta x + \left(p\, \delta x_f\right)_{t_f}\,.
\ee

Although this derivation may seem trivial to most readers, we present it explicitly because we will later show that this simple picture can correctly reproduce the master equations in open quantum systems, as well as the evolution equations in stochastic inflation.

All of our discussions here and in the following sections apply equally to QFTs, where $x(t)$ is naturally promoted to field configurations $\phi(t, x)$. In this case, we can derive the functional Schrödinger equation for wavefunctionals in the field-profile representation by fixing $\phi(t_f, x) = \phi_f(x)$.

\subsection{Schwinger-Keldysh formalism and density matrix}

\subsubsection{Schwinger-Keldysh formalism for the density matrix}

We now generalize the previous discussion to the density matrix. This generalization is straightforward if we consider a pure state. The idea is to use the Schwinger–Keldysh (SK) formalism, which involves two timefolds and doubles the degrees of freedom to describe the bra and ket, respectively \cite{schwinger1961brownian,keldysh2024diagram}
\be
\rho(x_f^+,x_f^-,t_f)&=\langle x_f^+,t_f|\psi,t_i\rangle\langle \psi,t_i|x_f^-,t_f\rangle=\psi(x_f^+,t_f)\psi^{\ast}(x_f^-,t_f)\nn\\
&=N^2 \int dx_i^+ dx_i^-\int_{x^{\pm}(t_i)=x_i^{\pm}}^{x^{\pm}(t_f)=x_f^{\pm}}[Dx^+Dx^-] e^{i \left(S[x^+]-S[x^-]\right)}\rho(x_i^+,x_i^-,t_i)\,.
\ee
We again consider the initial state to be the ground pure state, and thus have in mind the Schwinger–Keldysh contour to both prepare and evolve the density matrix
\be
\rho_g(x_f^+,x_f^-,t_f)=N^2 \int_{x^{\pm}(\mp i \infty)=0}^{x^\pm(t_f)=x_f^\pm}[Dx^+Dx^-] e^{i\left(S[x^+]-S[x^-]\right)}\,.\label{eq: density matrix ground x}
\ee

It is beneficial to consider the Keldysh basis of the field variables \cite{kamenev2023field}
\be
x^q=x^+-x^-\,,\quad x^c=\fft{1}{2}(x^++x^-)\,,
\ee
where $x^q$ is called the response field and $x^c$ the classical field, whose physical interpretations will become clear later.

In this formalism, if we are only interested in the diagonal elements of the density matrix, we can impose the boundary condition $x_f^+ = x_f^-$, which corresponds to $x^q(t_f) = 0$ in \eqref{eq: density matrix ground x}. Similarly, we can directly project onto the Wigner function, which describes the phase space distribution in the semiclassical limit. This amounts to setting the boundary condition $x^c(t_f) = x_f$, $p^c(t_f) = p_f$. These are the two scenarios we encounter in this paper, and we summarize them below

\begin{itemize}
\item[I.]  Diagonal density matrix
\be
P_g(x_f,t_f)=\langle x_f,t_f|\rho_g|x_f,t_f\rangle=N^2 \int_{x^\pm(\mp i\infty)=0}^{x^q(t_f)=0,x^c(t_f)=x_f}[Dx^q Dx^c]e^{i\left(S[x^+]-S[x^-]\right)}\,.
\ee

\item[II.] Wigner function
\be
W_g(x_f,p_f,t_f)&=\int \fft{dy_f}{\pi}\langle x_f-y_f,t_f|\rho_g|x_f+y_f\rangle e^{2ip_f y_f}\nn\\
&=N^2 \int_{x^\pm(\mp i\infty)=0}^{x^c(t_f)=x_f,p^c(t_f)=p_f}[Dx^q Dx^c Dp^q Dp^c]e^{i\left(S[x^+]-S[x^-]\right)}\,.
\ee

\end{itemize}

Scenario I is the one typically considered in stochastic inflation \cite{Starobinsky:1986fx,Starobinsky:1994bd}. Physically, it describes the population probability of finding the system at position $x_f$ at time $t_f$. However, it inevitably loses some information about the state, for example, it is blind to the decoherence. In contrast, Scenario II, i.e., the Wigner function, retains a complete set of information, as it is simply a Fourier transform of \eqref{eq: density matrix ground x}. Generalizing them to arbitrary initial states is straightforward.

\subsubsection{The Liouville space and the von Neumann equations}

Pure states are simply outer products of wavefunctions; thus, their time evolution is governed by the von Neumann equation $\frac{\partial \rho}{\partial t_f} = -i[H, \rho]$. In the position representation, this becomes
\be
\fft{\partial\rho(x_f^+,x_f^-,t_f)}{\partial t_f}=-i N^2\int_{x^\pm(\mp i\infty)=0}^{x^\pm(t_f)=x_f^\pm}[Dx^+Dx^-]\left(H^+-H^-\right)e^{i\left(S[x^+]-S[x^-]\right)}\,,\label{eq: von Neumann position}
\ee
where we should replace the momentum by the differential form $p^\pm=-i\partial/\partial x^\pm$.
It is often convenient and conceptually insightful to represent the density matrix as a vector in the so-called Liouville space \cite{manzano2020short}, $\rho \rightarrow |\rho\rangle\rangle$, so that the von Neumann equation takes the form of a standard Schrödinger equation in Liouville space
\be
\fft{\partial |\rho\rangle\rangle}{\partial t_f}=-i H_L |\rho\rangle\rangle\,,
\ee
Here, $H_L$ is the Liouville superoperator, playing the role of the Hamiltonian in Liouville space. We emphasize that this perspective is manifest in the SK formalism, as shown in Eq.~\eqref{eq: von Neumann position}. The state $|\rho\rangle\rangle$ can still be projected onto any representation we choose. The punchline is to interpret the combination $S[x^+] - S[x^-]$ as the Liouville action $S_L$, and to identify its associated Hamiltonian as $H_L = H^+ - H^-$. For example, in a simple model \eqref{eq: simple model}, we indeed find
\be
S_L=\dot{x}^q \dot{x}^c-V(x^+)+V(x^-)\,,\quad H_L=p^c p^q+V(x^+)-V(x^-)=H^+-H^-\,,
\ee
where $p^{c/q} = \dot{x}^{c/q}$, and their differential forms in the position representation are $p^{c/q} = -i\partial/\partial x^{q/c}$. In Liouville space, we can directly follow the previous section to derive the von Neumann equation in any representation. Although this derivation is trivial for pure states, as simply a rewriting of the Schrödinger equation, we find it useful to present it explicitly here as a baseline for the master equation that we will discuss soon.

\begin{itemize}
\item[I.] Diagonal density matrix
\be
&\fft{\partial P_g(x_f,t_f)}{\partial t_f}=-i\left\langle H_L\left[x^c(t_f)=x_f,x^q(t_f);p^c(t_f),p^q(t_f)=-i\fft{\partial}{\partial x_f}\right]\right\rangle_{P_g}\,,\nn\\
&\langle \mathcal{O}\rangle_{P_g}=N^2 \int_{x^\pm(\mp i\infty)=0}^{x^q(t_f)=0,x^c(t_f)=x_f}[Dx^q Dx^c] \mathcal{O} e^{i S_L[x^+,x^-]}\,.\label{eq: FP general}
\ee
This equation is not closed, because we only impose the boundary conditions $x^q(t_f) = 0$ and $x^c(t_f) = x_f$, which means we must evaluate $H_L$ quantum mechanically at the final time $t_f$. Nevertheless, we can expand this expression in the semiclassical limit, where the equations of motion can be solved under these boundary conditions to determine $p^c(t_f)$. For the simple model in eq.~\eqref{eq: simple model}, we then have 
\be
\fft{\partial P_g(x_f,t_f)}{\partial t_f} \ = -\fft{\partial}{\partial x_f} \langle \dot{x}^c(t_f) \rangle_{P_g}\,.
\ee

\item[II.] Wigner function
\be
\fft{\partial W_g(x_f,p_f,t_f)}{\partial t_f}=-i H_L\left[x^c(t_f)=x_f,x^q(t_f)=i\fft{\partial}{\partial p_f}; p^c(t_f)=p_f, p^q(t_f)=-i\fft{\partial}{\partial x_f}\right] W_g(x_f,p_f,t_f)\,.\label{eq: Wigner general}
\ee
For the simple model \eqref{eq: simple model}, this equation indeed becomes the Moyal equation for the Wigner function
\be
\fft{\partial W_g(x_f,p_f,t_f)}{\partial t_f}=\{\{H(x_f,p_f),W_g(x_f,p_f,t_f)\}\}\,,\quad \{\{A,B\}\}=2A \sin\left(\fft{\overleftarrow{\partial_x} \overrightarrow{\partial_p}-\overleftarrow{\partial_p} \overrightarrow{\partial_x}}{2}\right)B\,.
\ee
\end{itemize}

\subsubsection{Open quantum systems and master equations}

Let’s now generalize our discussion to open quantum systems. Instead of considering only a single system $x$, we consider two systems, $(x, y)$, coupled together. It is important to note that in the context of QFTs, especially with gravity, $y$ need not be a distinct object. For example, $y$ could represent particular modes of the same system, such as the short-wavelength modes we will encounter in inflation, as discussed in \cite{Li:2025azq} and in section \ref{sec: setup}. We consider
\be
S[x,y]=S_1[x]+S_2[y]+S_{\rm int}[x,y]\,,
\ee
where $S_i$ refers to the actions for the individual systems $(x, y)$, and $S_{\rm int}$ describes the coupling between them. We now regard $y$ as the environment with which the system $x$ interacts, and we trace it out. In this way, we are describing the ``effective'' density matrix of the open system $x$, in the sense that we are ignorant of its possible interactions with $y$. We assume that the initial state of the joint system is factorizable, $\rho(t_i) = \rho_x(t_i)\rho_y(t_i)$, and in particular, we consider the ground state
\be
\rho(x_f^+,x_f^-,t_f)&=\int dy_f \int_{y^\pm(\mp i\infty)=0,x^\pm(\mp i\infty)=0}^{x^\pm(t_f)=x_f^\pm,y^{\pm}(t_f)=y_f} [Dx^+ Dx^-Dy^+ Dy^-] e^{i\left(S[x^+,y^+]-S[x^-,y^-]\right)}\nn\\
& = \int_{x^\pm(\mp i\infty)=0}^{x^\pm(t_f)=x_f^\pm}[Dx^c Dx^q]e^{iS^{\rm eff}_L[x^c,x^q]}\,,\label{eq: density matrix open}
\ee
Other representations follow similarly. It is reasonable to absorb $\int dy_f$ into the definition of the measure $Dy^+Dy^-$, and thus fully integrate over $y^{\pm}$. In our language, ``tracing out'' is completely equivalent to ``integrating out'', which yields an effective theory, as shown in the second line of \eqref{eq: density matrix open}, where
\be
S^{\rm eff}_L[x^c,x^q]=S_1[x^+]-S_1[x^-]+S_{\rm IF}[x^+,x^-]=\int dt\left(p^q \dot{x}^c+p^c \dot{x}^q-H_{L}^{\rm eff}\right)\,.
\ee
The last term, $S_{\rm IF}$, encodes the communication between the system $x$ and the environment $y$, including effects such as noise, diffusion, and dissipation. The quantity $e^{i S_{\rm IF}}$ is famously known as the Feynman–Vernon influence functional \cite{feynman1963theory}. This ``integrating-out'' procedure resembles the construction of Wilsonian effective field theories (EFTs). Indeed, in the case where $y$ represents high-energy modes and $x$ low-energy modes, this prescription precisely corresponds to Wilsonian coarse-graining, which preserves locality and unitarity and is therefore not open in many senses. Nevertheless, the prescription here represents a more general coarse-graining procedure, in which locality and unitarity may not be preserved in the effective theory $S_L^{\rm eff}$. In the case of stochastic inflation, as we will show, $y$ corresponds to short-wavelength modes and $x$ to the nearly zero-modes. The time-dependent scale separation pushes this setup beyond the Wilsonian paradigm, resulting in an open-system description of the long wavelength inflaton modes.

We can also consider variants of \eqref{eq: density matrix open} in other representations, such as by restricting to diagonal elements or projecting onto the Wigner function. In all cases, we interpret these objects in Liouville space, and the master equation then generally takes the form
\be
\fft{\partial \rho(r[x,p],t_f)}{\partial t_f}&=-i \int_{x^\pm(\mp i\infty)}^{r[x,p]}[Dx^c Dx^q Dp^c Dp^q] H_L^{\rm eff} e^{iS_L^{\rm eff}[x^c,x^q]}\nn\\
&=-i \int_{x^\pm(\mp i\infty)}^{r[x]}[Dx^c Dx^q Dp^c Dp^q] \left(H_{L1}+H_{L,{\rm IF}}\right) e^{iS_L^{\rm eff}[x^c,x^q]}\,,
\ee
where we use $r[x]$ to denote any boundary conditions that project the density matrix into a specific representation, and $H_{L1}$ is the original Liouville Hamiltonian for $x$ without interactions with $y$. This expression is essentially the master equation for a mixed density matrix describing an open quantum system, where $H_{L,{\rm IF}}$ captures all dissipation, diffusion, and memory effects. In principle, these effects can be nonlocal in both time (referred to as non-Markovian) and space (in the context of QFTs). Nevertheless, we can always attempt a local expansion in time and space, which resembles a multipole expansion in Wilsonian EFT.

A particular class of effective theories with $H_{L,{\rm IF}}$, which has been extensively studied in the context of open quantum systems (and is also our focus in this paper), is the Lindbladian master equation. This essentially comes with a structure of
\be
H_{L,{\rm IF}}=-i\sum_{i}\fft{\gamma_i}{2}\left(L_i^q\right)^2\,,\quad S_{\rm IF}=-\int dt H_{L,{\rm IF}}\,,\label{eq: Lindbladian}
\ee
where $L_i$ refers to certain Hermitian operators in the theory (more general operators can also be considered). This is because the Lindbladian is
\be
\fft{\partial\rho}{\partial t}=-i[H,\rho]+\sum_{i}\gamma_{i}\left(L_i \rho L_i-\fft{1}{2}\{L_i L_i,\rho\}\right)\,,\label{eq: Lindbladian eq}
\ee
which always leads to terms of the form $(L_i^q)^2$ when projecting onto the bra and ket $\langle +|\rho|-\rangle$ in SK formalism. We note that $L_i^q$ can be combinations of operators whose coefficients are functionals of some ``classical'' operator $\hat{\mathcal{O}}^c$, since we can imagine expanding around the limit $x^q \rightarrow 0$, which is a semi-classical limit that perturbatively controls the quantum fluctuations $x^q\sim \hbar$, such that, for example, $\mathcal{O}(x^+) - \mathcal{O}(x^-) \sim \hat{\mathcal{O}}(x^c) x^q + \cdots$. We take Brownian particle as an example to demonstrate this framework in appendix \ref{app: Brownian}.

\section{Starobinsky's stochastic inflation as an open EFT: setup}
\label{sec: setup}

We now adopt the formalism of the previous section to revisit Starobinsky's stochastic inflation. We will show that stochastic inflation is naturally an open quantum system, where the short-wavelength modes serve as the environment through a generalized Wilsonian coarse-graining procedure. As in ordinary quantum mechanics discussed earlier, this leads to an EFT for the long wavelength modes, encoding the diffusion. This perspective was also discussed in \cite{Li:2025azq}. Nevertheless, the conceptual improvement, together with the SK formalism presented in the previous section, clarifies the approximations entering Starobinsky's stochastic inflation.

\subsection{Set-up: single-field slow-roll inflation}

\subsubsection{Slow-roll in semi-classical and eternal regime}

We consider the single-field inflation model
\be
S=\fft{1}{16\pi G}\left(\int d^4x\sqrt{-g}R-2\int d^3x\sqrt{h}K\right)-\int d^4x\sqrt{-g}\left(\fft{1}{2}(\nabla\phi)^2+V(\phi)\right)\,.
\ee
We have in mind a potential that satisfies the slow-roll approximation
\be
\epsilon=\fft{1}{16\pi G}\fft{V'^2}{V^2}\ll 1\,.
\ee

To proceed, we consider the semi-classical regime of gravity, avoiding quantum gravity effects and the trans-Planckian problem. Accordingly, we take the starting point of inflation to be when a given mode has just crossed the horizon, denoted by $\varphi_0$, with Hubble constant $H_0$. Prior to this, the inflationary trajectory curves into the Euclidean contour. This evolution prepares the initial state, which is essentially the Hartle–Hawking state, and then the semi-classical limit $G H_0^2\sim H_0^2/M_{pl}^2\ll 1$ ensures a classical background geometry for inflation.

Let us further comment on the regime we consider. We emphasize that in standard slow-roll inflationary cosmology, including in some previous work on stochastic inflation, not only is the background treated classically, but the inflationary trajectory itself is also treated semi-classically, followed by a slow-roll expansion. This regime corresponds to a scale separation
\be
\fft{1}{H_0^2}\gg \fft{\epsilon}{H_0^2}\gg \fft{1}{M_{pl}^2}\,.
\ee
In this regime, the standard procedure is to first quantize the fluctuations of the inflaton around the classical trajectory, and then study cosmological perturbation theory in the slow-roll limit \cite{Mukhanov:1990me,Maldacena:2002vr,Acquaviva:2002ud,Durrer:2004fx,Baumann:2009ds,wang2014inflation}.

Nevertheless, we note that stochastic inflation plays a powerful role in describing the eternal inflation, where the quantum fluctuations of the inflaton are not suppressed \cite{Linde:1986fd,Goncharov:1987ir,Guth:2007ng,Creminelli:2008es}. That is,
\be
\frac{1}{H_0^2} \gg \frac{\epsilon}{H_0^2} \sim \frac{1}{M_{\rm pl}^2}\,,
\ee
even though gravity remains semi-classical. Since the background remains classical and adiabatically evolving, we can still perform a controlled expansion in $|V - V_0| / V_0 \sim \sqrt{\epsilon}\, \Delta\phi\ll 1$.

\subsubsection{Schwinger-Keldysh and an emergent time}

For simplicity, we consider a gauge where we fix the classical background to be homogeneous
\be
ds^2=-N^2 dt^2+a^2 dx^2\,.
\ee
In this background, we have the action
\be
S=-3 \int dtd^3x \frac{a \dot{a}^2}{N}-\int dtd^3x\left(\fft{1}{2N}a^3 \dot{\phi}^2-\fft{1}{2}N a (\partial\phi)^2-a^3 N V(\phi)\right)\,,
\ee
where we take $8\pi G=1$.
To discuss the cosmological density matrix, we have to extend the single inflation theory to SK geometry \cite{Jana:2020vyx,Ivo:2024ill}
\be
(ds^\pm_{\rm SK})^2=-(N^\pm)^2 dt^2+ (a^\pm)^2 dx^2\,.
\ee
where we have $S_L=S_{L,{\rm grav}}+S_{L,\phi}+S_{L,{\rm res}}$. The reason we have $a^q, N^q$ is that the no-boundary Hartle-Hawking geometry \cite{hartle1983wave} for the density matrix is a complex geometry \cite{Witten:2021nzp}, as it goes into Euclidean region differently for bra and ket \cite{Ivo:2024ill}. We consider the semi-classical gravity, therefore we only keep linear terms in $g_{\mu\nu}^q$ and take $N^c=1$:
\begin{align}
S_{L,{\rm grav}}&=-3 \int dt d^3x \dot{a^c}\left(a^q \dot{a}^c+a^c\left(2\dot{a}^q-N^q \dot{a}^c\right)\right)\,,\nn\\
S_{L,\phi}&=\int dt d^3x\left((a^c)^3 \dot{\phi}^c \dot{\phi}^q-a^c \partial_i \phi^c \partial^i \phi^q+(a^c)^3\left(v^+-v^-\right)\right)\,,\nonumber\\
 S_{L,{\rm res}}&=-\fft{1}{4}\int dt d^3x a^q\left(\left((\partial\phi^+)^2-3(a^c)^2 (\dot{\phi}^+)^2+6 (a^c)^2 V^+\right)+(\phi^+\rightarrow \phi^-)\right)\nonumber\\
& -\fft{1}{4}\int dt d^3x a^c N^q\left(\left((\partial\phi^+)^2+(a^c)^2 (\dot{\phi}^+)^2+2(a^c)^2 V^+\right)+\left(\phi^+\rightarrow \phi^-\right)\right)\,,\label{eq: SK S inflation}
\end{align}
where $v=V(\phi)-V(\phi_0)$ and $V^\pm=V(\phi^\pm)$.
The last term $S_{L,{\rm res}}$ refers to the linear response of the metric with dependence of $(a^q,N^q)$.

The Hamiltonian constraint for the Liouville superoperator is
\be
\left(H_{L,{\rm grav}}+H_{L,\phi}+H_{L,{\rm res}}\right)\rho(a^\pm,\phi^\pm)=0\,.
\ee
It is straightforward to see that the Liouville Hamiltonian for the semi-classical gravity is:
\be
H_{L,{\rm grav}}=-\frac{p^c_a p^q_a}{6a^c}\rightarrow H_{L,{\rm grav}}\rho=-i\dot{a}^c \partial_{a^c}\rho\,.
\ee
Therefore, the classical time flow for the inflaton can be understood as being generated by the operator $H a\partial_a=\dot{a}\partial_a:= \partial_t$.

In our approximation, we neglect the fluctuations in $S_{L,{\rm res}}$. This can be justified because integrating out $\phi$ in $S_{L,{\rm res}}$ yields spacetime diffusion terms $\mathcal{O}\left((a^q)^2, a^q N^q, a^q \phi^q, N^q \phi^q\right)$, which are essentially quantum gravity effects scaling as $\mathcal{O}(G_N)$ and are dropped in the semi-classical limit. Nevertheless, we emphasize that the classical scale factor $a^c$ should be determined by varying $(a^q, N^q)$ in $S_{L,{\rm grav}} + S_{L,{\rm res}}$, yielding the Friedmann equations. At leading order around $\phi=\varphi_0$, these equations give $a^c = e^{H_0 t}$. Therefore, in this paper, we will only work with $S_{L,\phi}$ in an exact dS background with $a^c = e^{H_0 t}$. Discussions of deviations from exact dS can be found in \cite{Li:2025azq}, and a covariant formalism including this deviation is developing in \cite{Ye:2026saa, Saha:2026cay}.

\subsection{EFT of long wavelength modes}

This section clarifies why there is an open EFT for the long wavelength sector. We first specify the environment, the coarse-graining prescription and the approximations underlying the construction in subsection \ref{sec:eft-setup}. We then explain why openness is generic in this setup in subsection \ref{sec:beyond-wilsonian}, even though tracing out short modes is the Wilsonian step. 

\subsubsection{Environment, coarse-graining and assumptions}\label{sec:eft-setup}

The original picture of stochastic inflation is to think of quantum effects as noise inducing a random walk of the inflationary trajectory \cite{Starobinsky:1986fx}. This picture can be improved to describe stochastic inflation as a quantum walk \cite{Green:2025hmo}. Our paradigm makes these pictures precise by treating the long wavelength modes as an open system that defines the trajectory, and the short-wavelength modes act as the environment bath. In other words, we split the inflaton field as $\phi = \phi_s + \phi_e$, where
\be
\phi_e=\phi\big|_{\text{large}\, k}\,,\quad  \phi_s=\phi\big|_{\text{small}\, k}\,.
\ee

It is worth noting that this setup is different from standard open quantum systems such as the Brownian particle discussed in appendix \ref{app: Brownian} or open EFTs of cosmology \cite{LopezNacir:2011kk,Burgess:2014eoa,Salcedo:2024smn,Colas:2024lse,Salcedo:2025ezu,Colas:2025app,Colas:2025ind,Salcedo:2026cqb,Salcedo:2026sdn,Cespedes:2025ple}, as the environment here is not invisible material or particles; instead, it is part of the original system with significant fluctuations. In other words, it is a Wilsonian separation of short and long modes, but in the case of the density matrix for dS. For simplicity, we consider the step function separation as in Wilsonian EFT. Since in inflation, the scale is defined relative to the horizon, therefore we consider
\be
\phi_e(k,t)=\phi(k,t)\theta(k-\varepsilon a H_0)\,,\label{eq: step window}
\ee
where $\varepsilon$ is a bookkeeping parameter to parameterize the scale boundary of the open system. To describe very long wavelength modes, we would take $\varepsilon\rightarrow 0$. We should then trace out $\phi_e$ using SK path integral to obtain a theory for the cosmological reduced density matrix. This corresponds to performing the coarse-graining and obtaining a long wavelength EFT that controls the behavior of the reduced density matrix.

In the QFT description of stochastic inflation, the function $\theta(k-\varepsilon aH_0)$ is often referred to as the window function. One may replace the sharp step function by a smooth window $W_\varepsilon(k)=1-K\left(k/(\varepsilon aH_0)\right)$ such that $\phi_e(k,t)=\phi(k,t)W_\varepsilon(k)$, where
\be
K\left(\fft{k}{\Lambda}\right)\Big|_{k\gg \Lambda}=0\,,\quad K\left(\fft{k}{\Lambda}\right)\Big|_{k\ll \Lambda}=1\,.
\ee
It has been emphasized that such smooth window functions can lead to colored noise and apparent non-Markovian effects in the effective dynamics \cite{Winitzki:1999ve}. From an EFT perspective, however, these differences should be understood as scheme dependence associated with the choice of coarse-graining. As long as long- and short-wavelength modes are parametrically separated, the resulting effective description should be insensitive to the details of the window function at the level of physical observables. Different choices of window functions may lead to different intermediate effective actions, but these descriptions are expected to be equivalent up to scheme-dependent terms that can be absorbed into non-observable parameters or higher-order operators in the EFT. In this sense, using a smooth window function can be viewed as moving from a sharp Wilsonian separation to an exact RG scheme of the type introduced by Polchinski \cite{polchinski1984renormalization}, which has recently been explored in the context of stochastic inflation \cite{Green:2025hmo, Cespedes:2026fdp}. We will also discuss a continuum EFT scheme in section \ref{sec: continuum EFT}, which makes the EFT nature of the open system more transparent. 

We should now follow the procedure outlined in section \ref{sec: path integral}, for example, replacing $(x,y)$ by $(\phi_s,\phi_e)$ in \eqref{eq: density matrix open}
\be
S_{L}^{\rm eff}[\phi^c_s,\phi^q_s]=S_{L,\phi_s}[\phi^c_s,\phi^q_s]+S_{\rm IF}[\phi^c_s,\phi^q_s]\,,
\ee
where $S_{L,\phi}$ can be found in \eqref{eq: SK S inflation} and
\be
& e^{i S_{\rm IF}}=\int_{\rm BD}D\phi_e^\pm e^{i\left(S_{Le0}+S_{\rm int}\right)}\,,\quad  S_{Le0}=\int dtd^3x\left(a^3\dot{\phi}_e^q \dot{\phi}_e^c-a\partial_i\phi_e^q \partial^i\phi_e^c\right)\,,\nn\\
& S_{\rm int}=\int 
dtd^3x\left(a^3\dot{\phi}^q_{(s}\dot{\phi}^c_{e)}-a\partial_i\phi^q_{(s}\partial^i\phi^c_{e)}+a^3 \left(v(\phi_s^+)-v(\phi_s^-)-v(\phi^+)+v(\phi^-)\right)\right)\,.\label{eq: get IF integral}
\ee
Our goal is then to explicitly determine $S_{\rm IF}$ by tracing out $\phi_e$, which corresponds to performing the path integral in \eqref{eq: get IF integral} with the Bunch–Davies vacuum \cite{bunch1978quantum}, where we have
\be
& G^{cc}(k, t_1,t_2)=\frac{H_0^2 \left(\left(\eta _1 \eta _2 k^2+1\right) \cos \left(\left(\eta _1-\eta _2\right) k\right)+\left(\eta _1-\eta _2\right) k \sin \left(\left(\eta _1-\eta _2\right) k\right)\right)}{2k^3}\,,\nn\\
& G^{cq}(k,t_1,t_2)=-i\frac{H_0^2\left(\left(\eta _1 \eta _2 k^2+1\right) \sin \left(\left(\eta _1-\eta _2\right) k\right)+\left(\eta _2-\eta _1\right) k \cos \left(\left(\eta _1-\eta _2\right) k\right)\right)}{k^3}\theta\left(t_1-t_2\right)\,,\nn\\
&  G^{qc}(k,t_1,t_2)=G^{cq}(k,t_2,t_1)\,.\label{eq: green function dS}
\ee
In this notation, $G^{cq}$ is the retarded propagator, while $G^{qc}$ is the advanced propagator.
For simplicity, we use the conformal time $\eta H_0=-e^{-H_0 t}$.
In the presence of a general potential, this is not a feasible task. Therefore, as in more conventional open quantum systems and EFTs, approximations are required in order to perform the path integral perturbatively. We now outline main approximations and properties that enter in the EFT of inflationary open system.

\paragraph{Born approximation (weak coupling)}

The Born approximation basically assumes the interactions between the system and environment are small, therefore we can perturbatively compute the path integral \eqref{eq: get IF integral} in terms of Feynman
diagrams.

\paragraph{Secular approximation (large time-scale separation)}

The secular approximation, also known as the rotating wave approximation, essentially suppresses fast oscillating terms compared to the time scale of the system. In Wilsonian EFT language, this corresponds to a large time-scale separation, allowing a multipole expansion for $\phi_s^c$ in time, for example
\be
\phi_s^\alpha(t_1)\phi_s^c(t_2)\mathcal{O}_e(t_2) = \left(\phi_s^c(t_1) +t_{21} \partial_{t_1} \phi_s^c(t_1) + \cdots\right)\phi_s^\alpha(t_1)\mathcal{O}_e(t_2)\,,
\ee 
where $\alpha$ takes $c$ or $q$. 
Therefore, the system effectively acts as a static source for the environment and does not enter the $t_2$ integral. This expansion is controlled systematically by $\dot{\phi}^c/H_0\ll 1$, which is justified in the slow-roll limit $\dot{\phi}^c\sim v'(\phi^c)/H_0\sim \sqrt{\epsilon}\ll 1$.\footnote{If a frequency representation is allowed (which is not the case in dS!), this corresponds to projecting the Keldysh correlator of the environment operators to the low-frequency Taylor expansion, because we generically have 
\be
\int dt_2 t_2^n \mathcal{O}_e(\omega)e^{i \omega t_2}\sim \lim_{\omega\rightarrow 0}\partial_\omega^n \mathcal{O}_e(\omega). 
\ee}

We emphasize that this leads to the Markov property of the Keldysh two-point correlator of the environment and generates higher time-derivative operators in the EFT for Gaussian diffusion. Indeed, this approximation is important for ensuring Markovian Langevin dynamics in stochastic inflation \cite{Figueroa:2021zah,Cruces:2022imf}. However, it is worth noting that the temporal multipole expansion for $\phi_s^q$ is not justified because the fluctuation of $\dot{\phi}_s^q$ is comparable to that of $\phi_s^q$. As we will see in section \ref{sec:non-gaussian}, the matching calculation requires knowledge of the history of $\phi_s^q$. In other words, non-Gaussian diffusion is genuinely non-Markovian.

\paragraph{Large spatial-scale separation}

This approximation suppresses rapidly vibrating terms relative to the characteristic length scale of the system. Such an approximation is not usually adopted in zero–spatial-dimensional open systems. Nevertheless, it is standard in EFT, where one assumes a large separation of spatial scales, allowing for a multipole expansion in space, for example
\be
\phi_s^\alpha(x_1)\phi_s^\beta(x_2)\mathcal{O}_e(x_2) = \left(\phi_s^\beta(x_1) + x_{21}\cdot \partial_{x_1} \phi_s^\beta(x_1) + \cdots\right)\phi_s^\alpha(x_1)\mathcal{O}_e(x_2)\,.
\ee 
As long as spatial translation invariance is preserved, which is the case in inflation, one can work in a momentum-space representation. The spatial multipole expansion then amounts to projecting the total spatial momentum of the environment operator to be small
\be
\int d^{d-1}x_2 x_2^{\mu_1}\cdots x_2^{\mu_n}\mathcal{O}_e(k)e^{-ik\cdot x_2}\sim \lim_{k\rightarrow 0} \partial_k^{\mu_1}\cdots \partial_{k}^{\mu_n}\mathcal{O}_e(k)\,.
\ee
Thus this expansion is controlled by $k/\Lambda_{\rm EFT}$.

In conventional EFT, this leads to spatial locality of the effective action. However, as we will see later, in stochastic inflation one must resum some of this expansion to generate nontrivial diffusion terms, because short wavelength kernel is strongly peaked at the horizon-crossing point $k=\varepsilon a H$, where long–short modes mixing happens. In this case, the higher order corrections are systematically expanded by $k/(a H)$.\footnote{A similar phenomenon appears in soft-collinear effective theory \cite{Bauer:2000ew,Bauer:2000yr,becher2015introduction,Rothstein:2016bsq}.} This is a genuinely novel feature that goes beyond conventional Wilsonian EFT and signals the intrinsically open nature of the system.

It is worth noting that in conventional Wilsonian EFT, where the Lorentz symmetry is preserved, approximations 2 and 3 are essentially merged and ensure the locality of the EFT. This is no longer the case in dS, and not even in flat space for highly boosted targets.

\paragraph{Comments on Markov approximation}

In open quantum system language, the Markov approximation for the evolution of the density matrix usually refers to the assumption that the response function of the environment to the system decays rapidly, $G_R(t,t')\sim e^{-|t-t'|/\tau_{\rm env}}$. In inflationary cosmology, as we see in \eqref{eq: green function dS}, the retarded Green function of $\phi_e$ in the BD vacuum vanishes identically at equal times, $G_R(t,t)=0$. As a result, Markovianity arises from the way the window function and the secular approximation project the environment’s response. For the step window function \eqref{eq: step window}, the Markov approximation is perturbatively ensured at Gaussian order with the secular approximation: both the environment’s response and its noise are local in time, proportional to $\delta(t-t')$, and therefore project onto $G_{R,e}=G_R(t,t)=0$. A smooth window function does not destroy this property. Instead, it makes the system more closely resemble a conventional open system, since it can provide the projection through exponential decay.\footnote{We emphasize that the Markovian approximation we adopt here follows the language of open quantum systems and refers to the time locality of the density matrix evolution \cite{kamenev2023field,breuer2002theory}. This is conceptually different from the use of “(non)Markovian” in earlier stochastic inflation literature \cite{Figueroa:2021zah,Cruces:2022imf}, where it is used in the context of Langevin dynamics and denotes that the amplitude of the white noise is history dependent.} Beyond the Gaussian order, the Markov approximation breaks down.

\subsubsection{Beyond the Wilsonian paradigm: open EFTs}\label{sec:beyond-wilsonian}

It is important to comment on why openness is generic in our setup. Tracing out the environment is understood as integrating out short-distance modes, which is essentially the coarse-graining procedure underlying Wilsonian EFT. It is therefore natural to ask why, unlike ordinary Wilsonian EFT where the resulting effective theory is local and unitary, the EFT obtained here is expected to describe a reduced density matrix that is non-unitary and, in general, not spatially local.

The ordinary local and unitary Wilsonian EFT describes in-out observables and thus usually fails to manifestly capture effects in in-in observables, such as dissipation and diffusion, unless one specifies particular boundary terms or encodes them in nontrivial branch cuts of in-out amplitudes \cite{Green:2024cmx}. The openness then arises from the SK formalism for in-in observables whenever the coarse-graining generates couplings between the bra and ket. This also happens in gravitational wave physics such as \cite{Goldberger:2009qd,Saketh:2023bul,Ivanov:2024sds,Glazer:2024eyi,Caron-Huot:2025tlq,Ivanov:2026icp} and in inflation with unknown material content or heavy particles \cite{LopezNacir:2011kk,Burgess:2014eoa,Salcedo:2024smn,Colas:2024lse,Salcedo:2025ezu,Colas:2025app,Colas:2025ind,Salcedo:2026cqb,Salcedo:2026sdn,Cespedes:2025ple}. Nevertheless, this heavily depends on the correlations and excitations of the environment. For example, when the environment consists of short-wavelength modes or heavy degrees of freedom in the Minkowski vacuum, we simply recover the conventional Wilsonian description, because any couplings between the bra and ket have to be generated by Wightman functions of the environment, which are on-shell and incompatible with soft external kinematics. For example, the coupling $\partial^n \phi_s \cdot \partial^n \phi_e$ is impossible in this case because their momentum supports are disjoint (in a sharp step-function window, this term is exactly zero).

In the case of inflation, the novel feature that goes beyond the Wilsonian paradigm is that the separation scale depends on time $\Lambda(t)\sim a(t) H_0$. Thus, the main source of openness simply comes from the fact that the effective long wavelength modes accumulate from the bath, as short-wavelength modes are forced to cross the horizon and become part of the long wavelength sector over time. This keeps the coupling $\dot{\phi}_s \dot{\phi}_e$ nontrivial because their supports overlap around $k=\varepsilon a H_0$. This coupling precisely leads to the key diffusion effect in stochastic inflation.

\section{EFT and matching at Gaussian order}
\label{sec: EFT Gaussian}

\subsection{Free diffusion and the fluctuation-dissipation relation in dS}\label{sec:fdr-free}

In this subsection we restrict to the \emph{free theory}: we set the interaction potential $v$ to zero and identify the leading diffusion contribution that openness already generates at Gaussian order. Interaction-dressed diffusion contributions are deferred to \ref{sec:diff-interactions}. 
Even at this Gaussian order, openness is already nontrivial: as we will see, tracing over the short-wavelength sector of the BD vacuum yields a nonzero influence functional whose structure is fixed by a fluctuation--dissipation relation.

The free-theory Feynman rules needed below are
\begin{align}
& \begin{tikzpicture}[baseline={(0,-0.5ex)}]
\draw[thick,dashed] (0,0) -- (0.7,0);
  \draw[thick] (0.7,0) -- (1.4,0);
  \node at (0.6,0.2) {$q$};
  \node at (0.8,0.2) {$c$};
  \node at (0.7,-0.3) {$t$};
\end{tikzpicture}=a^3 \dot{\phi}_s^q \partial_t-a k^2 \phi_s^q\,,\quad \begin{tikzpicture}[baseline={(0,-0.5ex)}]
\draw[thick,dashed] (0,0) -- (0.7,0);
  \draw[thick] (0.7,0) -- (1.4,0);
  \node at (0.6,0.2) {$c$};
  \node at (0.8,0.2) {$q$};
  \node at (0.7,-0.3) {$t$};
\end{tikzpicture}=a^3 \dot{\phi}_s^c \partial_t-a k^2 \phi_s^c\,,\nn\\
& \begin{tikzpicture}[baseline={(0,-0.5ex)}]
  \draw[thick] (0,0) -- (1.4,0);
  \node at (0.1,0.2) {$c$};
  \node at (1.3,0.2) {$c$};
  \node at (0.1,-0.2) {$t_1$};
  \node at (1.3,-0.2) {$t_2$};
  \node at (0.75,0.25) {$k$};
\end{tikzpicture}=G^{cc}(k,t_1,t_2)\theta\left(k+\frac{\varepsilon}{\eta_1}\right)\theta\left(k+\frac{\varepsilon}{\eta_2}\right)\,,\nn\\
& \begin{tikzpicture}[baseline={(0,-0.5ex)}]
  \draw[thick] (0,0) -- (1.4,0);
  \node at (0.1,0.2) {$c$};
  \node at (1.3,0.2) {$q$};
  \node at (0.1,-0.2) {$t_1$};
  \node at (1.3,-0.2) {$t_2$};
   \node at (0.75,0.25) {$k$};
\end{tikzpicture}=G^{cq}(k,t_1,t_2)\theta\left(k+\frac{\varepsilon}{\eta_1}\right)\theta\left(k+\frac{\varepsilon}{\eta_2}\right)\,.\label{eq:Feynman-rules-free}
\end{align}
Note that the second two-point vertex is actually irrelevant, since it gives $G^{cq}$, which vanishes once we eventually collapse to equal time at leading order in the temporal multipole expansion. On the other hand, the first two-point vertex serves as the key ingredient for deriving the long wavelength influence functional in stochastic inflation, as emphasized in \cite{Garbrecht:2013coa,Garbrecht:2014dca, Moss:2016uix, Tokuda:2017fdh, Tokuda:2018eqs}. Interaction vertices (e.g.\ background-potential insertions of $v''(\phi_s^c)$) will be introduced in \ref{sec:diff-interactions} when they are needed; they play no role in this subsection. The key identity that produces the overlap between system and environment at the crossing shell is
\begin{equation}
\partial_t\,\theta\!\left(k-\Lambda(t)\right) = -\dot\Lambda(t)\,\delta\!\left(k-\Lambda(t)\right).
\label{eq:shell-identity}
\end{equation}
For the sharp window function, it is important to specify the prescription for the distributional contact at the crossing point, since the value of $\theta(0)$ is in general ambiguous. We use the one-sided prescription $[1-\theta(k-\Lambda(t))]\delta(k-\Lambda(t))\equiv \delta(k-\Lambda(t))$, or equivalently $\theta(0)=0$. Since the long wavelength modes are the complement of the environmental modes, this assigns the crossing contribution entirely to the long wavelength sector. In this prescription, whenever a window-function propagator is acted on by a time derivative, it produces a delta function localized on the instantaneous separation scale.

In free theory, the unique diagram that generates a quadratic $(\phi^q)^2$ contribution to $S_{\rm IF}$ via the system--environment $\dot\phi_s^q\dot\phi_e^c$ coupling is
\begin{equation}
\begin{tikzpicture}[baseline={(0,-0.5ex)}]
\draw[thick,dashed] (-0.5,0.5)--(0,0);
\draw[thick] (0,0)--(1,0);
\draw[thick,->] (0,0)--(0.6,0);
\draw[thick,dashed] (1,0)--(1.5,0.5);
\node at (-0.05,0.25) {$q$};
\node at (1+0.05,0.25) {$q$};
\node at (0.15,0.15) {$c$};
\node at (0.85,0.15) {$c$};
\node at (0,-0.3) {$t_1$};
\node at (1,-0.3) {$t_2$};
\end{tikzpicture}\label{eq:free-diffusion-diagram}
\end{equation}
This single free-theory diagram already produces a nontrivial diffusion term, as we now show. Evaluating it gives
\begin{align}
\begin{tikzpicture}
[baseline={(0,-0.5ex)}]
\draw[thick,dashed] (-0.5,0.5)--(0,0);
\draw[thick] (0,0)--(1,0);
\draw[thick,dashed] (1,0)--(1.5,0.5);
\node at (-0.05,0.25) {$q$};
\node at (1+0.05,0.25) {$q$};
\node at (0.15,0.15) {$c$};
\node at (0.85,0.15) {$c$};
\node at (0,-0.3) {$t_1$};
\node at (1,-0.3) {$t_2$};
\end{tikzpicture}=\int dt_1 dt_2 a_1^3 a_2^3 \dot{\phi}_s^q(t_1,x_1)\dot{\phi}_s^q(t_2,x_2)\times \int \frac{d^3k}{(2\pi)^3}\partial_{t_1}\partial_{t_2} G_e^{cc}(t_1,t_2,k)\,,\label{eq: leading diff}
\end{align}
where $a_i = a(t_i)$ and $G^{cc}_{e}$ is the Keldysh propagator of the environment. Note that the term proportional to $k^2$ vanishes identically, because the system contracts directly with the environment with disjoint momentum support, $\theta(k+\varepsilon/\eta)\,\theta(\varepsilon/\eta-k) \equiv 0$. In contrast, the contribution in \eqref{eq: leading diff} is nonzero, because, via \eqref{eq:shell-identity}, the time derivative produces a Dirac delta $\delta(k+\varepsilon/\eta)$ that enforces short-wavelength modes to mix with long wavelength modes at $k=-\varepsilon/\eta$. However, we emphasize that this invalidates the naive spatial multipole expansion that sets $k=0$, since $\delta(k-\varepsilon a H)$ is peaked around $k\sim \varepsilon a H$. As promised above, this leads to nontrivial diffusion terms even in the free theory. This diagram can be further decomposed into three contributions:
\begin{align}
& \int dt_1 dt_2 d^3x_1 d^3x_2\, a_1^3 a_2^3 \dot{\phi}_s^q(t_1,x_1)\dot{\phi}_s^q(t_2,x_2) \times \int \frac{d^3k}{(2\pi)^3}\partial_{t_1}\partial_{t_2} G_e^{cc} (t_1,t_2,k)e^{ik\cdot x_{12}}\nn\\
& =\int dt_1 dt_2 d^3x_1 d^3x_2\,a_1^3 a_2^3 \dot{\phi}_s^q(t_1,x_1)\dot{\phi}_s^q(t_2,x_2)\times \int \frac{d^3k}{(2\pi)^3}\varepsilon^2 \dot{a}_1\dot{a}_2 H_0^2 \delta_{12}(k) G^{cc}(t_1,t_2,k)e^{ik\cdot x_{12}}\nn\\
&\quad -2\int dt_1 dt_2\,d^3x_1 d^3x_2\, a_1^3 a_2^3 \dot{\phi}^q_s(t_1,x_1)\phi^q_s(t_2,x_2)\times \int \frac{d^3k}{(2\pi)^3}\varepsilon^2 \dot{a}_1\dot{a}_2 H_0^2 \delta_{12}(k)\partial_{t_2}G^{cc}(t_1,t_2,k)e^{ik\cdot x_{12}}\nn\\
&\quad +\int dt_1 dt_2\, d^3x_1 d^3x_2\, a_1^3 a_2^3 \phi^q_s(t_1,x_1)\phi^q_s(t_2,x_2)\times \int \frac{d^3k}{(2\pi)^3}\varepsilon^2 \dot{a}_1\dot{a}_2 H_0^2 \delta_{12}(k)\partial_{t_1}\partial_{t_2}G^{cc}(t_1,t_2,k)e^{ik\cdot x_{12}}\,,\label{eq: diffusion computation}
\end{align}
where $\delta_{12}(k)=\delta(k-\varepsilon a_1 H_0)\delta(k-\varepsilon a_2 H_0)$. We thus find \cite{Cespedes:2026fdp}:
\begin{align}
& S_{\rm IF}\supset i\frac{H_0^3}{8\pi^2} \int dt\, d^3x_1 d^3x_2\, a^6 \mathcal{G}(\varepsilon a H_0 |x_{12}|)\times \nn\\
& \left((1+\varepsilon^2)\partial_t \phi^q(x_1)\partial_t \phi^q(x_2)+H_0 \varepsilon^2 \left(\partial_t \phi^q(x_1) \phi^q(x_2)+\partial_t \phi^q(x_2) \phi^q(x_1)\right)+H_0^2 \varepsilon^4 \phi^q(x_1)\phi^q(x_2)\right)\,,\label{eq: SIF free}
\end{align}
where $\mathcal{G}(x)=\sin x/x$ and $\varepsilon$ here is a bookkeeping parameter that is not necessarily small. Therefore, the influence functional is nonlocal in space. The reason that nontrivial effective diffusion terms arise even in the free theory is that we trace out the short-wavelength modes together with the initial Hartle--Hawking state, while the remaining long wavelength modes are insensitive to excitations of the Hartle--Hawking state and therefore would naively see a vanishing Keldysh correlator. The free-theory influence functional \eqref{eq: SIF free} is precisely what enforces the correct Keldysh correlator for the system. The reproduction of the long-wavelength propagators from the long-short two-point vertex in \eqref{eq:Feynman-rules-free} was shown in \cite{Tokuda:2017fdh,Tokuda:2018eqs}. Indeed, we find
\begin{align}
& G^{cc}(p,t_1,t_2)=-\fft{H_0^3}{2} \int dt' a'^6 \fft{\delta(p-\varepsilon a' H_0)}{p^2}\Bigg(-\left(1+\varepsilon^2\right)\partial_{t'}G^{cq}(p,t_1,t')\partial_{t'}G^{cq}(p,t_2,t')\nn\\
&\quad +\varepsilon^2 H_0\left(\partial_{t'}G^{cq}(p,t_1,t')G^{cq}(p,t_2,t')-\left(1\leftrightarrow 2\right)\right) +\varepsilon^4 H_0^2 G^{cq}(p,t_1,t')G^{cq}(p,t_2,t')\Bigg) \,,\label{eq: FDR dS}
\end{align}
which is valid for any $\varepsilon$. We interpret \eqref{eq: FDR dS} as the fluctuation--dissipation relation (FDR) for the BD vacuum in dS.\footnote{If we first integrate over $t'$ and then take $\varepsilon\rightarrow 0$, the FDR \eqref{eq: FDR dS} can be understood as the split representation.} See also appendix \ref{app: FDR massive} for FDR of a massive scalar in dS. This free influence functional can also be understood as the leading Gaussian stochastic initial condition in the language of soft dS EFT (SdSEFT) \cite{Cohen:2020php}, see appendix \ref{app: relation to SdSEFT}. 

Higher powers of $\varepsilon$  in \eqref{eq: SIF free} should be understood as arising from higher-dimensional operators, $\varepsilon^{2n} \sim 1/(a H_0)^{2n}\partial^{2n} \sim \left(p/(a H_0)\right)^{2n}$. These contributions can therefore be neglected at leading order in the EFT expansion in $p/(a H_0)$, reproducing the result of \cite{Starobinsky:1986fx}. However, the spatial nonlocality is in general preserved, since there exists a regime with fixed $\varepsilon |x_{12}|a H_0$. If instead we consider $\varepsilon |x_{12}|a H_0\ll 1$, we effectively collapse into the zero-modes sector, reproducing the standard stochastic inflation diffusion term $i H_0^3/(8\pi^2)\int dt\, a^6 (\dot{\phi}^q)^2$ \cite{Starobinsky:1986fx}. We will return to this simplification in subsection \ref{subsec: zero-modes}.

Let's now comment on the absence of the drift-like terms $(\dot{\phi}^q_s \dot{\phi}^c_s,\dot{\phi}^q_s \phi^c_s,\dot{\phi}^c_s \phi^c_s,\phi_s^q \phi^c_s)$. Naively these terms can arise from the following diagram
\begin{equation}
\begin{tikzpicture}[baseline={(0,-0.5ex)}]
\draw[thick,dashed] (-0.5,0.5)--(0,0);
\draw[thick] (0,0)--(1,0);
\draw[thick,->] (0,0)--(0.6,0);
\draw[thick,dashed] (1,0)--(1.5,0.5);
\node at (-0.05,0.25) {$c$};
\node at (1+0.05,0.25) {$q$};
\node at (0.15,0.15) {$q$};
\node at (0.85,0.15) {$c$};
\node at (0,-0.3) {$t_1$};
\node at (1,-0.3) {$t_2$};
\end{tikzpicture}\nn
\end{equation}
However, it is obvious to see that the terms with $\dot{\phi}_s^q(t_1,x_1)\dot{\phi}_s^c(t_1,x_1)$ and $\phi_s^q(t_1,x_1)\phi_s^c(t_1,x_1)$ are respectively proportional to $\rho(t_1,t_1)=\partial_{t_1}\partial_{t_2}\rho(t_1,t_2)\Big|_{t_2=t_1}\equiv 0$, where $G^{cq}(t_1,t_2)=\rho(t_1,t_2)\theta(t_1-t_2)$.  The remaining terms attaching to $\dot{\phi}^q_s(t_1,x_1)\phi^c_s(t_1,x_2)$ and its permutation  contain $\theta(0) \partial_{t_1}\rho(t_2,t_1)\Big|_{t_2=t_1}$ and $\theta(0)\partial_{t_2}\rho(t_2,t_1)\Big|_{t_2=t_1}$. As specified below \eqref{eq:shell-identity}, we use the prescription $\theta(0)=0$, which also projects all precise equal-time retarded correlators to zero, i.e., $\langle\pi(t)\phi(t)\rangle_R=0$.

\subsection{Two RG channels and operator power counting}\label{sec:wilsonian-comments}

The free-theory analysis of \ref{sec:fdr-free} has already produced a nonzero diffusion sector. Once interactions are turned on, the open EFT acquires further structure that is most usefully organized by the logarithm in which it runs. Before performing any explicit calculation, it is worth emphasizing the structural picture. At the operator level, the open EFT contains two families of Wilson coefficients: unitary-sector coefficients (mass, coupling and wavefunction normalization etc), and diffusion coefficients. These coefficients run under two  renormalization group flows. A standard Wilsonian RG, with logarithm $\log(b/\varepsilon)$, renormalizes the unitary-sector coefficients as in any conventional EFT and also dresses the diffusion coefficients through sub-diagrams embedded in the diffusion-generating diagrams, where the bookkeeping UV scale is $\Lambda_{\rm UV}=b a H_0$ with $b>\varepsilon$. A new stochastic RG, with logarithm $\log\varepsilon$, generates and runs the diffusion coefficients and has no analogue in conventional Wilsonian EFT. The Wilsonian flow varies $\log(b/\varepsilon)$ at fixed $\log\varepsilon$, whereas the stochastic flow varies $\log\varepsilon$ at fixed $\log(b/\varepsilon)$. On the physical slice $b=1$, the two variations are no longer independent. We summarize the RG structures in Table \ref{tab:three-logs}.

\begin{table}[htbp]
\centering
\renewcommand{\arraystretch}{1.4}
{\small
\begin{tabular}{p{0.20\textwidth} p{0.17\textwidth} p{0.14\textwidth} p{0.40\textwidth}}
\toprule
\textbf{Effect} & \textbf{Channel} & \textbf{Logarithm} & \textbf{Description} \\
\midrule
(a) Diffusion coefficients &
Stochastic RG &
$\log\varepsilon$ &
Diffusion Wilson coefficients generated and run by dressed versions of \eqref{eq:free-diffusion-diagram}; secular accumulation from modes crossing $k=\varepsilon aH_0$; UV-independent. \\[4pt]
(b) Wilsonian dressing of diffusion coefficients &
Wilsonian RG &
$\log(b/\varepsilon)$ &
Self-energy sub-diagrams embedded in the diffusion-generating topology dress the same diffusion Wilson coefficients as in~(a). \\[4pt]
(c) Unitary-sector renormalization &
Wilsonian RG &
$\log(b/\varepsilon)$ &
Wilson coefficients of the unitary sector: mass, coupling, wavefunction normalization. \\
\bottomrule
\end{tabular}
}
\caption{Three logarithmic effects in the open EFT matching at the horizon shell. Effect (a) is governed by stochastic RG and produces secular $\log\varepsilon$ from modes crossing the comoving scale $k=\varepsilon aH_0$. Effects (b) and (c) are governed by Wilsonian RG with $\log(b/\varepsilon)$, dressing the diffusion coefficients and the unitary sector respectively.}
\label{tab:three-logs}
\end{table}

\paragraph{Power counting and scheme dependence.}
Before turning to explicit examples, it is useful to organize the operators in each channel by power counting, building directly on the controlling expansions of \ref{sec:eft-setup}. In the hard cutoff implementation we are using, three parametrically independent expansion parameters appear. The power counting discussed below refers to the controlled expansions used in the explicit matching calculations of this paper, rather than to a complete bottom-up ordering of the most general nonlocal influence functional. A general power counting for arbitrary non-Markovian kernels is beyond the scope of the present construction and remains an important problem for future investigation.

\begin{itemize}

\item{\emph{Wilsonian channel: $(b a H_0)^n$ and $(\varepsilon a H_0)^n$ expansion.}}

Higher-dimensional unitary-sector operators (and Wilsonian dressings of diffusion coefficients) are organized by powers of the ``EFT scale'' $\Lambda_{\rm EFT}^n = (\varepsilon a H_0)^n$ and UV scale $\Lambda_{\rm UV}^n = (b a H_0)^n$. In the hard cutoff scheme, these expansions show up as power-law, the $b^2 - \varepsilon^2$ term in the mass tadpole below is one example. These pieces are scheme-dependent artifacts of the hard cutoff and can be removed by a better regulator. The physical content of the Wilsonian channel sits in the logarithmic ratio $\log(b/\varepsilon)$.

\item{\emph{Stochastic channel: $k/(a H_0)$ expansion.}}

Higher-dimensional diffusion operators are organized by the soft expansion in $k/(a H_0) \sim \varepsilon$. This is the resummed version of the spatial multipole expansion (approximation~3 of \ref{sec:eft-setup}) around the crossing point $k=\varepsilon a H_0$ rather than around $k=0$. The $\varepsilon^{2n}$ tower visible in \eqref{eq: SIF free} comes precisely from this expansion: the leading $\varepsilon^0$ piece reproduces the standard Starobinsky diffusion $iH_0^3/(8\pi^2)\int dt\, a^6 (\dot\phi^q)^2$, and higher powers correspond to higher-dimensional diffusion operators. This is a genuine EFT expansion of the open sector and its RG sits in $\log\varepsilon$, although scheme-dependent artifacts may also be present.

\item{\emph{Temporal multipole expansion: $\dot{\phi}_s^c/H_0$.}}

The temporal expansion in $\dot{\phi}_s^c/H_0$ (approximation~2 of \S\ref{sec:eft-setup}) is logically independent of the channel structure and can appear in \emph{either} channel: it organizes higher time-derivative corrections both to unitary-sector and to diffusion Wilson coefficients, and acts orthogonally to the spatial $k/(aH_0)$ expansion. The $\partial_t/H_0$ corrections in the unitary sector should be also scheme dependent. However, in the stochastic sector, $\dot{\phi}_s^c/H_0$ corrections are in general a mixture of scheme-dependent and scheme-independent pieces in the hard cutoff scheme.

\end{itemize}

\bigskip

We now illustrate the two channels in concrete terms by taking $\lambda \phi^4$ theory as a working example. As mentioned, renormalization is not computationally easy in the hard cutoff scheme, nevertheless, the structural insights should still be valid.

\paragraph{Wilsonian channel acting on unitary sectors}
The simplest effect is mass generation that modifies the drift. At order $\mathcal{O}(\lambda)$, we have a simple tadpole diagram,
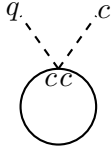
\begin{align}
\begin{tikzpicture}
\draw[thick] (0,0) circle (0.5cm);
  \draw[thick,dashed] (0,0.5) -- (-0.5, 1.2);
  \node at (-0.6, 1.25) {$q$};
  \node at (0.6, 1.25) {$c$};
  \draw[thick,dashed] (0,0.5) -- (0.5, 1.2);
  \node at (0.1, 0.35) {$c$};
  \node at (-0.1, 0.35) {$c$};
\end{tikzpicture}\nn
\end{align}
This effect was also included in stochastic inflation in \cite{Gorbenko:2019rza, Beneke:2026ksj}, nevertheless, it arises from conventional Wilsonian effects,
\begin{align}
i\fft{\lambda}{2} \int \fft{d^3k}{(2\pi)^3} G^{cc}(k,t,t)\theta(k-\varepsilon a H_0)\, a^3 \phi^q_s(t) \phi^c_s(t)=i\fft{a^3 H_0^2}{8\pi^2}\lambda\left(\log\fft{b}{\varepsilon}+\fft{b^2-\varepsilon^2}{2}\right)\phi^q_s \phi^c_s\,,\label{eq: mass generation}
\end{align}
where we regulate the UV divergence using a hard cutoff $\Lambda_{\rm UV}=b a H_0$ with $b>\varepsilon$. This UV divergence is the genuine ultraviolet divergence, which defines the cutoff scale of the scalar theory in dS. As usual in QFT with a hard cutoff scheme, the power-law divergent terms are scheme-dependent artifacts. We emphasize that they have to exist in the hard cutoff scheme in the long-distance EFT to ensure matching to the full theory, however, they can be removed by choosing a better scheme. On the other hand, the logarithmic term $\log(b/\varepsilon)$ signals the Wilsonian renormalization from the scale $b a H_0$ down to the scale $\varepsilon a H_0$ at which we define our EFT. Similarly, at $\mathcal{O}(\lambda^2)$, we have mass generation and wavefunction renormalization. At the same order, we also have the renormalization of $\lambda$ from the following Feynman diagrams and their permutations, up to symmetry factors,

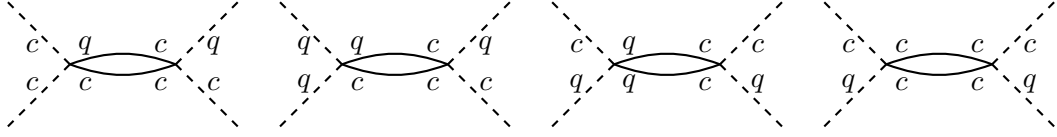
\begin{align}
\begin{tikzpicture}
\mydiagram{0}{0}   {$c$}{$c$}{$q$}{$c$}{$q$}{$c$}{$c$}
\mydiagramX{0}{$c$}
\mydiagram{3.6}{0} {$q$}{$q$}{$q$}{$c$}{$q$}{$c$}{$c$}
\mydiagramX{3.6}{$c$}
\mydiagram{7.2}{0} {$c$}{$q$}{$c$}{$q$}{$q$}{$q$}{$c$}
\mydiagramX{7.2}{$c$}
\mydiagram{10.8}{0}{$c$}{$q$}{$c$}{$q$}{$c$}{$c$}{$c$}
\mydiagramX{10.8}{$c$}
\end{tikzpicture}\nn
\end{align}
By simple power counting, without any explicit computation, we can see that the first two diagrams renormalize the coupling $\lambda$ by an amount of order $\lambda^2 \log(b/\varepsilon)$. The last two terms seem to generate new diffusive terms of the form $(\phi^q)^2 (\phi^c)^2$; however, they scale respectively as $\Lambda^3$ and $\Lambda^{-3}$, and are therefore merely scheme-dependent pieces.

\paragraph{Wilsonian channel acting on the diffusion.}
The same Wilsonian channel, with the same $\log(b/\varepsilon)$ running, also dresses diffusion Wilson coefficients. The relevant ingredient is the hard self-energy,
\begin{align}
\begin{tikzpicture}
\draw[thick] (0,0) circle (0.5cm);
  \draw[thick] (-1,-0.5) -- (1, -0.5);
  \node at (-0.15,-0.3) {$c$};
  \node at (0.15,-0.3) {$c$};
  \node at (-0.2,-0.65) {$c$};
  \node at (0.2,-0.65) {$q$};
\end{tikzpicture}
\end{align}
which gives the wavefunction renormalization of the unitary sector and is also embedded into diagrams that generate diffusion Wilson coefficients, such as
\begin{align}
\begin{tikzpicture}
[baseline={(0,-0.5ex)}]
\draw[thick,dashed] (-0.5,0.5)--(0,0);
\draw[thick] (0,0)--(2,0);
\draw[thick,dashed] (1+1,0)--(1.5+1,0.5);
\node at (-0.05,0.25) {$q$};
\node at (1+0.05+1,0.25) {$q$};
\node at (0.15,0.15) {$c$};
\node at (0.85+1,0.15) {$c$};
\node at (0,-0.3) {$t_1$};
\node at (1+1,-0.3) {$t_2$};
\draw[thick] (1,0.5) circle (0.5cm);
\node at (-0.15+1,-0.3+0.5) {$c$};
  \node at (0.15+1,-0.3+0.5) {$c$};
  \node at (-0.2+0.8,-0.65+0.5) {$c$};
  \node at (0.2+1.2,-0.65+0.5) {$q$};
  \node at (0.2+0.85,-0.65+0.45) {$t_3$};
\end{tikzpicture}\label{eq: diffusion Wilsonian RG}
\end{align}
which again contributes the logarithmic renormalization $\log(b/\varepsilon)$. In this embedding, the self-energy sub-diagram is the mechanism by which the Wilsonian channel acts on a \emph{diffusion} Wilson coefficient rather than on a unitary-sector coefficient.  

\paragraph{Stochastic channel on diffusion coefficients.}
Alongside the Wilsonian channel above, there is a second channel that acts only on the diffusion Wilson coefficients: the stochastic RG that runs in $\log\varepsilon$. It is generated by diagrams with the same topology as the free-theory diffusion diagram \eqref{eq:free-diffusion-diagram}, now dressed by insertions of $v''(\phi^c_s)$. These interactions do not create openness from nothing; the free-theory influence functional \eqref{eq: SIF free} already contains the open structure. What they do is renormalize the diffusion Wilson coefficients by generating new contributions that run in $\log\varepsilon$. This is the structural novelty: the diffusion coefficients are the operators of the open EFT on which both the Wilsonian and the stochastic RGs act simultaneously, and the stochastic channel has no counterpart in conventional Wilsonian EFT. The detailed computation of these $\log\varepsilon$ corrections is carried out in \ref{sec:diff-interactions}.

We interpret $\log\varepsilon$ as capturing secular effects associated with the continuous accumulation of long wavelength modes from the bath as modes cross $k=\varepsilon aH_0$. It is independent of the UV cutoff $b$ precisely because it has no ultraviolet origin; it is a property of the coarse-graining scale alone. The hard cutoff quantity $\log\varepsilon$ is a scheme-dependent proxy: in the continuum EFT it will be replaced by a scale $\nu$ characterizing the mode-crossing behavior of the long wavelength sector, to be developed in section \ref{sec: continuum EFT}.

\bigskip

We can already see the distinction between standard Wilsonian effects and the diffusive effects arising from openness. The Wilsonian channel is characterized by the logarithmic ratio $\log(b/\varepsilon)$ and acts on both ordinary Wilson coefficients and diffusion Wilson coefficients. By contrast, the stochastic channel is characterized by the shell logarithm $\log\varepsilon$, which is independent of the UV scale, and acts only on diffusion coefficients. Mixed double logarithms, such as $\log\varepsilon\,\log(b/\varepsilon)$, can therefore arise and evolve under both the Wilsonian and stochastic RGs.

In the hard cutoff scheme, the Wilsonian channel must be retained for consistency. More importantly, for $b=1$, which corresponds to the physical scale in inflation, the two types of renormalization become mixed. However, since the Wilsonian channel is in principle an independent RG flow for $b\neq 1$, in this paper we switch it off in order to cleanly extract the diffusion coefficients of the stochastic open EFT. We will keep track of it later when matching observables that are sensitive to both channels, such as Keldysh correlators. This treatment is justified by the fact that the Wilsonian channel admits a dimensional regularization scheme, in which it reduces to standard continuous RG flow procedure. In this sense, there is no actual ``EFT'' associated with this channel.

A useful way to understand this separation is through the method-of-region. The ordinary Wilsonian channel is associated with the usual hard and soft momentum regions and is therefore controlled by $\log(b/\varepsilon)$. The stochastic channel has a different origin. Since the system and environment are separated by the time-dependent scale
$\Lambda(t)=\varepsilon a H_0$, a physical mode crosses from the environment into the system when $k\simeq \varepsilon a(t)H_0$. The corresponding region is localized near this crossing shell rather than in a conventional hard momentum domain. This is why the stochastic channel runs in $\log\varepsilon$, is insensitive to the UV scale $b aH_0$, and acts directly on diffusion Wilson coefficients. We will make this region interpretation explicit in the matching discussion below.

\subsection{Gaussian diffusion from interactions}\label{sec:diff-interactions}

Having organized the renormalization structure abstractly in subsection \ref{sec:wilsonian-comments}, we now perform the explicit diagrammatic computation for diffusion coefficients in the stochastic channel. Throughout, we keep a generic interaction potential $v$ and work to $\mathcal{O}(v^2)$, and we again restrict to the leading order in the temporal multipole expansion. The new ingredient relative to subsection \ref{sec:fdr-free} is the background-potential vertex
\begin{align}
\begin{tikzpicture}[baseline={(0,-0.5ex)}]
  \draw[thick] (0,0) -- (0.6,0);
   \draw[thick] (0.8,0) -- (1.4,0);
  \draw[thick] (0.7,0) circle (0.1);
   \draw[thick] (0.7-0.07, 0-0.07) -- (0.7+0.07, 0+0.07);
  \draw[thick] (0.7-0.07, 0+0.07) -- (0.7+0.07, 0-0.07);
  \node at (0.5,0.2) {$c$};
  \node at (0.9,0.2) {$q$};
  \draw[thick,dashed] (0.7,0.1) -- (0.7,0.6);
  \node at (0.7,0.7) {$c$};
  \node at (0.7,-0.3) {$t$};
\end{tikzpicture}= v''(\phi_s^c(t))\,,\label{eq:vpp-vertex}
\end{align}
where $\begin{tikzpicture}[baseline={(0,-0.5ex)}]
  \draw[thick] (0.7,0) circle (0.1);
   \draw[thick] (0.7-0.07, 0-0.07) -- (0.7+0.07, 0+0.07);
  \draw[thick] (0.7-0.07, 0+0.07) -- (0.7+0.07, 0-0.07);
\end{tikzpicture}$ denotes a background potential insertion. Together with the propagators in \eqref{eq:Feynman-rules-free}, this is all we will need.

The low-lying relevant diagrams that generate diffusion up to $\mathcal{O}\left(v^2,(\phi^q)^2\right)$ are
\begin{equation}
\begin{tikzpicture}[baseline={(0,-0.5ex)}]
\draw[thick,dashed] (-0.5,0.5)--(0,0);
\draw[thick] (0,0)--(1,0);
\draw[thick] (0,0)--(0.6,0);
\draw[thick,dashed] (1,0)--(1.5,0.5);
\node at (-0.05,0.25) {$q$};
\node at (1+0.05,0.25) {$q$};
\node at (0.15,0.15) {$c$};
\node at (0.85,0.15) {$c$};
\node at (0,-0.3) {$t_1$};
\node at (1,-0.3) {$t_2$};
\end{tikzpicture}+2\times
\begin{tikzpicture}[baseline={(0,-0.5ex)}]
\draw[thick,dashed] (-0.5,0.5)--(0,0);
\draw[thick] (0,0)--(0.65,0);
\draw[thick] (0,0)--(0.45,0);
\draw[thick] (0.85,0)--(1.25,0);
\draw[thick] (0.85,0)--(1.5,0);
\draw[thick] (0.75,0) circle (0.1);
   \draw[thick] (0.75-0.07, 0-0.07) -- (0.75+0.07, 0+0.07);
  \draw[thick] (0.75-0.07, 0+0.07) -- (0.75+0.07, 0-0.07);
\draw[thick,dashed] (1.5,0)--(2,0.5);
\draw[thick,dashed] (0.75,0.1) -- (0.75,0.5);
\node at (-0.05,0.25) {$q$};
\node at (1+0.05+0.5,0.25) {$q$};
\node at (0.15,0.15) {$c$};
\node at (0.85+0.5,0.15) {$c$};
\node at (0,-0.3) {$t_1$};
\node at (1+0.5,-0.3) {$t_2$};
\node at (0.75,0.6) {$c$};
\node at (0.75,-0.3) {$t_3$};
\node at (0.55,0.15) {$c$};
\node at (0.95,0.15) {$q$};
\end{tikzpicture}+2\times
\begin{tikzpicture}[baseline={(0,-0.5ex)}]
\draw[thick,dashed] (-0.5,0.5)--(0,0);
\draw[thick] (0,0)--(0.65,0);
\draw[thick] (0.85,0)--(1.4,0);
\draw[thick] (1.6,0)--(2.25,0);
\draw[thick] (0.75,0) circle (0.1);
   \draw[thick] (0.75-0.07, 0-0.07) -- (0.75+0.07, 0+0.07);
  \draw[thick] (0.75-0.07, 0+0.07) -- (0.75+0.07, 0-0.07);
\draw[thick,dashed] (2.25,0)--(2.75,0.5);
\draw[thick,dashed] (0.75,0.1) -- (0.75,0.5);
\draw[thick] (1.5,0) circle (0.1);
   \draw[thick] (1.5-0.07, 0-0.07) -- (1.5+0.07, 0+0.07);
  \draw[thick] (1.5-0.07, 0+0.07) -- (1.5+0.07, 0-0.07);
\draw[thick,dashed] (2.25,0)--(2.75,0.5);
\draw[thick,dashed] (0.75,0.1) -- (0.75,0.5);
\draw[thick,dashed] (1.5,0.1) -- (1.5,0.5);
\node at (-0.05,0.25) {$q$};
\node at (1+0.05+0.5+0.75,0.25) {$q$};
\node at (0.15,0.15) {$c$};
\node at (0.85+0.5+0.75,0.15) {$c$};
\node at (0,-0.3) {$t_1$};
\node at (1+0.5+0.75,-0.3) {$t_2$};
\node at (0.75,0.6) {$c$};
\node at (1.5,0.6) {$c$};
\node at (0.75,-0.3) {$t_3$};
\node at (1.5,-0.3) {$t_4$};
\node at (0.55,0.15) {$c$};
\node at (0.95,0.15) {$q$};
\node at (1.3,0.15) {$c$};
\node at (1.7,0.15) {$q$};
\end{tikzpicture}+
\begin{tikzpicture}[baseline={(0,-0.5ex)}]
\draw[thick,dashed] (-0.5,0.5)--(0,0);
\draw[thick] (0,0)--(0.65,0);
\draw[thick] (0.85,0)--(1.4,0);
\draw[thick] (1.6,0)--(2.25,0);
\draw[thick] (0.75,0) circle (0.1);
   \draw[thick] (0.75-0.07, 0-0.07) -- (0.75+0.07, 0+0.07);
  \draw[thick] (0.75-0.07, 0+0.07) -- (0.75+0.07, 0-0.07);
\draw[thick,dashed] (2.25,0)--(2.75,0.5);
\draw[thick,dashed] (0.75,0.1) -- (0.75,0.5);
\draw[thick] (1.5,0) circle (0.1);
   \draw[thick] (1.5-0.07, 0-0.07) -- (1.5+0.07, 0+0.07);
  \draw[thick] (1.5-0.07, 0+0.07) -- (1.5+0.07, 0-0.07);
\draw[thick,dashed] (2.25,0)--(2.75,0.5);
\draw[thick,dashed] (0.75,0.1) -- (0.75,0.5);
\draw[thick,dashed] (1.5,0.1) -- (1.5,0.5);
\node at (-0.05,0.25) {$q$};
\node at (1+0.05+0.5+0.75,0.25) {$q$};
\node at (0.15,0.15) {$c$};
\node at (0.85+0.5+0.75,0.15) {$c$};
\node at (0,-0.3) {$t_1$};
\node at (1+0.5+0.75,-0.3) {$t_2$};
\node at (0.75,0.6) {$c$};
\node at (1.5,0.6) {$c$};
\node at (0.75,-0.3) {$t_3$};
\node at (1.5,-0.3) {$t_4$};
\node at (0.55,0.15) {$q$};
\node at (0.95,0.15) {$c$};
\node at (1.3,0.15) {$c$};
\node at (1.7,0.15) {$q$};
\end{tikzpicture}\label{eq: Feynman diagrams}
\end{equation}
The first diagram is the free-theory contribution already computed in \eqref{eq: SIF free}. Here we include the corrections from inserting the interaction potential $v$ by computing the last three diagrams in \eqref{eq: Feynman diagrams}. We emphasize that the intermediate times $(t_3, t_4)$ are integrated using a multipole expansion in time, for example $v(\phi^c(t_3)) \sim v(\phi^c(t_1)) + \cdots$. In addition, at leading order, the scale separation in wavelength sets all momenta equal, $k_1 = k_2 = \cdots$, so the background potential insertions effectively behave as static sources. In other words, we neglect higher-dimensional operators in the EFT expanded in $\varepsilon^2 \sim p^2/(aH_0)^2$.

We first compute the second diagram in \eqref{eq: Feynman diagrams}, which was computed in \cite{Tokuda:2017fdh} for $\lambda \phi^4$ theory. Expanding the vertices as in \eqref{eq: diffusion computation} we find
\begin{align}
&\begin{tikzpicture}[baseline={(0,-0.5ex)}]
\draw[thick,dashed] (-0.5,0.5)--(0,0);
\draw[thick] (0,0)--(0.65,0);
\draw[thick] (0.85,0)--(1.5,0);
\draw[thick] (0.75,0) circle (0.1);
   \draw[thick] (0.75-0.07, 0-0.07) -- (0.75+0.07, 0+0.07);
  \draw[thick] (0.75-0.07, 0+0.07) -- (0.75+0.07, 0-0.07);
\draw[thick,dashed] (1.5,0)--(2,0.5);
\draw[thick,dashed] (0.75,0.1) -- (0.75,0.5);
\node at (-0.05,0.25) {$q$};
\node at (1+0.05+0.5,0.25) {$q$};
\node at (0.15,0.15) {$c$};
\node at (0.85+0.5,0.15) {$c$};
\node at (0,-0.3) {$t_1$};
\node at (1+0.5,-0.3) {$t_2$};
\node at (0.75,0.6) {$c$};
\node at (0.75,-0.3) {$t_3$};
\node at (0.55,0.15) {$c$};
\node at (0.95,0.15) {$q$};
\end{tikzpicture}
= \int dt_1 dt_2 dt_3 d^3x_1 d^3x_2\, a_1^3 a_2^3 a_3^3\, v''(\phi_s^c(t_1,x_1))\nn\\
& \times \Bigg[ \dot{\phi}^q_s(t_1,x_1)\dot{\phi}^q_s(t_2,x_2)\int \fft{d^3 k}{(2\pi)^3}\varepsilon^2 \dot{a}_1\dot{a}_2 H_0^2 \delta_{12}(k)\, G^{cc}(k,t_1,t_3) G^{cq}(k,t_2,t_3) e^{ik\cdot x_{12}}\nn\\
& +\, 2\dot{\phi}^q_s(t_1,x_1)\phi^q_s(t_2,x_2)\int \fft{d^3 k}{(2\pi)^3}\varepsilon^2 \dot{a}_1\dot{a}_2 H_0^2 \delta_{12}(k)\, G^{cc}(k,t_1,t_3)\partial_{t_2}G^{cq}(k,t_2,t_3) e^{ik\cdot x_{12}}\Bigg]\,,
\label{eq: second-diagram}
\end{align}
where we used the multipole expansions in $(t,x)$ to set $k_2 = k_1 = k$ and to move $v''(\phi_s^c(t_3,x_3)) \simeq v''(\phi_s^c(t_1,x_1))$. Similarly, we should consider the other channel where the multipole expansion happens around $x_3 \sim x_2$. The resulting contribution to the influence functional is
\begin{align}
S_{\rm IF} \supset {} & i\fft{H_0}{48\pi^2} \int dt\, d^3x_1\, d^3x_2\, a^6\, \mathcal{G}(\varepsilon a H_0 |x_{12}|)\, \left[v''(\phi_s^c(t,x_1))+v''(\phi_s^c(t,x_2))\right]\nn\\
& \quad \times \Big[ \dot{\phi}^q_s(t,x_1)\dot{\phi}^q_s(t,x_2)\,\log_{\varepsilon} + H_0\,\dot{\phi}^q_s(t,x_1)\phi^q_s(t,x_2) + (x_1\leftrightarrow x_2)\Big]\,,
\label{eq: second-diagram-IF}
\end{align}
where we have introduced the shorthand
\begin{align}
\log_{\varepsilon} \equiv \log \fft{\varepsilon}{2} - \psi\!\left(\fft{3}{2}\right)\,,
\end{align}
and $\psi$ is the digamma function. The structural feature of \eqref{eq: second-diagram-IF} is that, on top of the spatial kernel $\mathcal{G}(\varepsilon a H_0 |x_{12}|)$ inherited from the free-theory result \eqref{eq: SIF free}, the interaction-dressed contribution carries the shell logarithm $\log_{\varepsilon}\sim \log\varepsilon$ characteristic of the stochastic RG channel anticipated in \ref{sec:wilsonian-comments}. This reproduces the structure of the explicit $\lambda\phi^4$ computation in  \cite{Tokuda:2017fdh} in the notation of a general slow-roll potential, and will be extended below to higher orders.

We now move to the third and fourth diagrams in \eqref{eq: Feynman diagrams}. It is useful to define
\begin{align}
& F(t_1,t_2, x_{12}) \equiv  2\int dt_3 dt_4\, a_1^3 a_2^3 a_3^3 a_4^3 \int \fft{d^3 k}{(2\pi)^3}\varepsilon^2 \dot{a}_1\dot{a}_2 H_0^2\, \delta_{12}(k)\, G^{cc}(k,t_1,t_3) G^{cq}(k,t_4,t_3) G^{cq}(k,t_2,t_4)\, e^{ik\cdot x_{12}}\nn\\
& + \int dt_3 dt_4\, a_1^3 a_2^3 a_3^3 a_4^3 \int \fft{d^3 k}{(2\pi)^3}\varepsilon^2 \dot{a}_1\dot{a}_2 H_0^2\, \delta_{12}(k)\, G^{cq}(k,t_1,t_3) G^{cc}(k,t_3,t_4) G^{cq}(k,t_2,t_4)\, e^{ik\cdot x_{12}}\,.
\label{eq: F-def}
\end{align}
The two diagrams together give
\begin{align}
2\times \text{(third)} + \text{(fourth)}
&= \fft{1}{4}\int dt_1 dt_2\, d^3x_1 d^3x_2\, \big[v''(\phi_s^c(t_1,x_1)) + v''(\phi_s^c(t_1,x_2))\big]^2\, \mathcal{D}_{t_1,t_2}(x_1,x_2)\, F(t_1,t_2,x_{12})\,,
\label{eq: third-fourth}
\end{align}
where the differential operator $\mathcal{D}$ is
\begin{align}
\mathcal{D}_{t_1,t_2}(x_1,x_2) = {} & \dot{\phi}^q_s(t_1,x_1)\dot{\phi}^q_s(t_2,x_2) + \phi^q_s(t_1,x_1)\phi^q_s(t_2,x_2)\partial_{t_1}\partial_{t_2}\nn\\
& - \dot{\phi}^q_s(t_1,x_1)\phi^q_s(t_2,x_2)\partial_{t_2} - \phi^q_s(t_1,x_1)\dot{\phi}^q_s(t_2,x_2)\partial_{t_1}\,.
\label{eq: D-def}
\end{align}
This leads to
\begin{align}
S_{\rm IF} \supset {} & \fft{i}{144 H_0\, \pi^2}\int dt\, d^3x_1\, d^3x_2\, a^6\, \mathcal{G}(\varepsilon a H_0 |x_{12}|)\,\big[v''(\phi_s^c(t,x_1)) + v''(\phi_s^c(t,x_2))\big]^2\nn\\
& \times \Bigg[\left(\log_{\varepsilon}^2 + \fft{\log_{\varepsilon}}{3} + \fft{\pi^2 - 8}{4}\right)\dot{\phi}^q_s(t,x_1)\dot{\phi}^q_s(t,x_2)
+ H_0\!\left(\log_{\varepsilon} + \fft{1}{6}\right)\dot{\phi}^q_s(t,x_1)\phi^q_s(t,x_2)\nn\\
& \quad + H_0\!\left(\log_{\varepsilon} + \fft{1}{6}\right)\dot{\phi}^q_s(t,x_2)\phi^q_s(t,x_1) + \fft{H_0^2}{2}\,\phi^q_s(t,x_1)\phi^q_s(t,x_2) \Bigg]\,.
\label{eq: third-fourth-IF}
\end{align}
The structural lesson is the same as that of \eqref{eq: second-diagram-IF}: the interaction-dressed contribution carries the same spatial kernel $\mathcal{G}(\varepsilon a H_0|x_{12}|)$ as the free-theory result \eqref{eq: SIF free}, while the dependence on the coarse-graining scale appears through powers of $\log_{\varepsilon}\sim \log\varepsilon$. In particular, \eqref{eq: third-fourth-IF} contains $\log_{\varepsilon}^2$ at $\mathcal{O}(v^2)$, exactly as expected from a two-loop stochastic-RG insertion.

It is important to note that there is also a vertex
\begin{align}
\begin{tikzpicture}[baseline={(0,-0.5ex)}]
\draw[thick,dashed] (0,0) -- (0.7,0);
  \draw[thick] (0.7,0) -- (1.4,0);
  \node at (0.6,0.2) {$c$};
  \node at (0.8,0.2) {$q$};
  \node at (0.7,-0.3) {$t$};
\end{tikzpicture}=a^3 \dot{\phi}_s^c \partial_t-a k^2 \phi_s^c\,,\nn
\end{align}
which requires us to compute relevant diagrams that provide corrections to the diffusive-drift and causality-violating terms, schematically as $F(\phi^q_s,\phi^c_s)\phi^q_s \partial_t^n \phi^c_s$ and $\partial_t^m(\phi_s^c)^n$. However, we can indeed show that those contributions are zero using our approximations. More precisely, what we essentially have are:
\begin{equation}
\begin{tikzpicture}[baseline={(0,-0.5ex)}]
\draw[thick,dashed] (-0.5,0.5)--(0,0);
\draw[thick] (0,0)--(0.5,0);
\node at (-0.05,0.25) {$c$};
\node at (0,-0.3) {$t_1$};
\draw[thick,fill] (0.7,0) circle (0.2);
\draw[thick,dashed] (0.9-0.2,0.5)--(0+0.7,0);
\draw[thick,dashed] (0.9,0.5)--(0+0.7,0);
\draw[thick,dashed] (0.9-0.4,0.5)--(0+0.7,0);
\node at (1,0.4) {$\cdot$};
\node at (1.1,0.3) {$\cdot$};
\node at (1.2,0.2) {$\cdot$};
\draw[thick] (0.9,0)--(1.4,0);
\draw[thick,dashed] (1.4,0)--(1.5+0.4,0.5);
\node at (0.7,-0.4) {$t_3$};
\node at (0.4,0.15) {$c$};
\node at (1.4,-0.3) {$t_2$};
\node at (0.15,0.15) {$q$};
\end{tikzpicture}\equiv 0\,,
\end{equation}
which is valid regardless of whether the remaining lines are $\phi^q$ or $\phi^c$. The reason is simply that these diagrams are constrained to have $t_3>t_1$ by the causality, while the black bubble contains $\theta(p-\varepsilon a_3 H_0)\sim \theta(t_1-t_3)$, giving $\theta(t_{13})\theta(t_{31})=0$.

Together, \eqref{eq: SIF free}, \eqref{eq: second-diagram-IF}, and \eqref{eq: third-fourth-IF} exhibit the full Gaussian diffusion sector at $\mathcal{O}(v^2)$. The $\log\varepsilon$ dependence visible in \eqref{eq: second-diagram-IF}--\eqref{eq: third-fourth-IF} is the concrete realization of the stochastic RG channel introduced in \ref{sec:wilsonian-comments}, and the absence of any $\log b$ in these expressions confirms that this channel is genuinely UV-independent.

\subsection{Matching and region interpretation for Gaussian diffusion}

Importantly, as an EFT, stochastic inflation has to match the long distance limit of the full theory. As in the usual method-of-region \cite{Beneke:2023wmt}, a full theory integral is decomposed into regions, and the EFT reorganizes these regions into two types of contributions. The late time and long wavelength region is reproduced dynamically by EFT evolution, including insertions of the influence functional. The complementary early-time and hard region does not appear explicitly in the EFT, but is instead encoded in Wilson coefficients, or more generally in Wilson kernels.

As we emphasized, there are two independent decompositions. The ordinary Wilsonian channel comes from splitting loop momenta into hard and soft regions. By contrast, the stochastic channel splits time integrals around the crossing time 
\[
t^\ast(k)=\frac{1}{H_0}\log\frac{k}{\varepsilon H_0}\,,
\]
which is the same crossing-time scale that appears in the perturbative long-short decomposition of \cite{Tokuda:2017fdh,Tokuda:2018eqs}.
For $t<t^\ast(k)$, the mode belongs to the environment and contributes to matching coefficients. For $t>t^\ast(k)$, the same mode belongs to the EFT. The Gaussian influence functionals \eqref{eq: second-diagram-IF} and \eqref{eq: third-fourth-IF} are precisely the matching pieces that reproduce the contribution of this crossing region to the Keldysh correlators and other simple observables.

\subsubsection{Simple matching in $m^2 \phi^2$}

Let us start with a warm up in the massive theory $v=m^2/2\, \phi^2$, where we treat $m$ as perturbatively small. 
Since the mass insertions are contact diagrams, there is no nontrivial loop momentum region to separate. The only nontrivial issue is how the full theory time integral is split at the crossing time into an early-time contribution, which fixes Wilson coefficients, and a late-time contribution, which is reproduced by EFT. This serves as a clean illustration of the matching procedure before turning to more complicated cases. 

Let's first present the known massive Green's functions
\be
&G^{cq}_{m}(p,t_1,t_2)=-\frac{1}{2} i H_0^2 \pi\, \eta_1^{3/2}\, \eta_2^{3/2} \left[ J_\nu(p\eta_2)\, Y_\nu(p\eta_1) - J_\nu(p\eta_1)\, Y_\nu(p\eta_2) \right]\,,\nn\\
& G^{cc}_m(p,t_1,t_2)=\frac{1}{4} H_0^2 \pi\, \eta_1^{3/2}\, \eta_2^{3/2} \left[ J_\nu(p\eta_1)\, J_\nu(p\eta_2) + Y_\nu(p\eta_1)\, Y_\nu(p\eta_2) \right]\,.
\ee
We now specify what is meant by the ``long wavelength limit'' that the EFT is designed to reproduce. The long distance physical scale is $p\ll m$, so we should first perform a small $p$ expansion, and then at each order in polynomials of $p$, we further expand in the $m\rightarrow 0$ limit. In the leading order in $p$, we have
\be
& G_m^{cq}(p,t,0)\simeq  \frac{i (1-  e^{-3 H_0 t})}{3 H_0}  - \frac{i e^{-3 H_0 t} m^2 \left(2 + 3 H_0 t + e^{3 H_0 t}(-2 + 3 H_0 t)\right)}{27 H_0^3} \nn\\
 & + \frac{i e^{-3 H_0 t} m^4 \left(-4 - 3 H_0 t(2 + H_0 t) + e^{3 H_0 t}(4 + 3 H_0 t(-2 + H_0 t))\right)}{162 H_0^5}+\mathcal{O}(p^2)\,,\nn\\
 & G_m^{cc}(p,t,t)\simeq \fft{1}{p^3}\left(\fft{H_0^2}{2}+\fft{m^2}{3}\log_p +\fft{m^4}{9 H_0^2}\left(\log_p^2+\fft{\log_p}{3}+\fft{\pi^2-8}{4}\right)\right)+\mathcal{O}(1/p)\,,\label{eq: expand massive correlator}
\ee
where $\log_p=\log \fft{p}{2H_0 a}-\psi\left(\fft{3}{2}\right)$. For simplicity, we consider the equal-time Keldysh correlator, the generic time ordering follows similarly.


The starting point is to perturbatively match the retarded Green's function. This is straightforward as we are simply inserting the retarded self-energy 
\be
\begin{tikzpicture}[baseline={(0,-0.5ex)}]
\draw[thick] (0,0) -- (0.7,0);
  \draw[thick] (0.7,0) -- (1.4,0);
  \node at (0.6,0.2) {$c$};
  \node at (0.8,0.2) {$q$};
  \draw[thick] (0.62,-0.08) -- (0.78,0.08);
  \draw[thick] (0.62,0.08) -- (0.78,-0.08);
\end{tikzpicture}
+ 
\begin{tikzpicture}[baseline={(0,-0.5ex)}]
\draw[thick] (0,0) -- (0.7,0);
  \draw[thick] (0.7,0) -- (1.4+0.7,0);
  \node at (0.6,0.2) {$c$};
  \node at (0.8,0.2) {$q$};
   \node at (0.6+0.7,0.2) {$c$};
  \node at (0.8+0.7,0.2) {$q$};
  \draw[thick] (0.62,-0.08) -- (0.78,0.08);
  \draw[thick] (0.62,0.08) -- (0.78,-0.08);
  \draw[thick] (0.62+0.7,-0.08) -- (0.78+0.7,0.08);
  \draw[thick] (0.62+0.7,0.08) -- (0.78+0.7,-0.08);
\end{tikzpicture}
+\cdots\,,\nn
\ee
where the cross denotes the mass insertion. We have
\be
& G^{cq}_{m}(p,t,0)\simeq G^{cq}(p,t,0)+i m^2 \int_0^t dt_2 G^{cq}(p,t,t_2)G^{cq}(p,t_2,0) a_2^3 \nn\\
&- m^4 \int_0^tdt_2 \int_0^{t_2}dt_3 G^{cq}(p,t,t_3)G^{cq}(p,t_2,t_3)G^{cq}(p,t_2,0) a_2^3 a_3^3+\mathcal{O}(m^6)\,.\label{eq: Gcq m expand}
\ee
We can easily verify that this equals to the small mass limit of the massive retarded Green function  \eqref{eq: expand massive correlator} to $m^2$ with any power of $p$.

The nontrivial matching is for the Keldysh correlator. We already showed in \eqref{eq: FDR dS} how the Keldysh correlator for the free theory can be matched using the leading diffusion operator in EFT. We now focus on the mass corrections to the order $m^4$ but at only leading order in $p/(a H_0)$. 

We now insert the effective action
\eqref{eq: second-diagram-IF}
and \eqref{eq: third-fourth-IF} to compute $\langle \phi^c \phi^c\rangle$ 
\be
\begin{tikzpicture}[baseline={(0,-0.5ex)}]
\draw[thick] (0,0) -- (1.4,0);
   \node at (0.6-0.6,0.15) {$c$};
  \node at (1.4,0.15) {$c$};
\end{tikzpicture}
=
\begin{tikzpicture}[baseline={(0,-0.5ex)}]
\draw[thick] (0,0) -- (0.7,0);
  \draw[thick] (0.7,0) -- (1.4,0);
  \node at (0.6,0.2+0.05) {$q$};
  \node at (0.8,0.2+0.05) {$q$};
   \node at (0.6-0.6,0.15) {$c$};
  \node at (1.4,0.15) {$c$};
  \draw[thick,fill] (0.7,0) circle [radius=0.1];
\end{tikzpicture}\nn
\ee
We thus find
\be
&G^{cc}_{m}(p,t,t)=\nn\\
&\fft{m^2 H_0}{3}\int dt' a'^6\fft{\delta(p-\varepsilon a' H_0)}{p^2} \left(-\partial_{t'} G^{cq}_m(p,t,t')\partial_{t'}G^{cq}_m(p,t,t')\log_\varepsilon+ H_0 \partial_{t'} G^{cq}_m(p,t,t')G^{cq}_m(p,t,t')\right)\nn\\
&+\fft{m^4}{9 H_0}\int dt' a'^6 \fft{\delta(p-\varepsilon a' H_0)}{p^2}\Bigg[-\left(\log_\varepsilon^2+\fft{\log_\varepsilon}{3}+\fft{\pi^2-8}{4}\right) \partial_{t'} G^{cq}_m(p,t,t')\partial_{t'}G^{cq}_m(p,t,t')\nn\\
& +2H_0\left(\log_\varepsilon+\fft{1}{6}\right) \partial_{t'} G^{cq}_m(p,t,t')G^{cq}_m(p,t,t')+\fft{H_0^2}{2} G^{cq}_m(p,t,t')G^{cq}_m(p,t,t')\Bigg]+\cdots\,.\label{eq: Gcc m expand}
\ee
The prescription for matching the Keldysh correlator is to first integrate over $t$, setting $t=t^\ast=\log(p/(\varepsilon H_0))/H_0$, and then take $p\rightarrow 0$ before taking $\varepsilon\rightarrow 0$, since the EFT is valid in the regime $p\ll \varepsilon aH_0$. Using \eqref{eq: Gcq m expand} and following this prescription, we verified that \eqref{eq: Gcc m expand} agrees with the second line of \eqref{eq: expand massive correlator}. We also verified that the unequal time Keldysh correlator can be matched. 

An important comment is that this prescription can be implemented directly at the level of the influence functional, for example in \eqref{eq: third-fourth-IF} and \eqref{eq: second-diagram-IF}. Under this prescription, the time integral localizes at $t^\ast$, so that the resulting effective action takes the form of an early-time ``boundary term'' involving only momentum integrals. This term is equivalent to the stochastic initial condition in SdSEFT \cite{Cohen:2020php,Cohen:2021fzf,Beneke:2026rtf}. See appendix \ref{app: relation to SdSEFT} for a brief comment on this connection.

Let us now comment on the region interpretation. In this case, it simply amounts to the factorization of logarithms
\be
\log \fft{p}{H_0 a}=\log \fft{p}{H_0a\varepsilon}+\log \varepsilon\,.\label{eq: log factorize}
\ee
In other words, as in standard QFTs, physical observables in the full theory can only contain logarithms of ratios of physical scales, such as $\log \Lambda_{\rm UV}/\Lambda_{\rm IR}$. In the EFT, however, this logarithm is split into two pieces: the EFT contribution contains the logarithm of the ratio between the physical IR scale and the EFT cutoff, while the matching contribution, encoded in the Wilson coefficients, contains the logarithm of the ratio between the UV scale and the EFT cutoff
\be 
\log \Lambda_{\rm UV}/\Lambda_{\rm IR}=\log \fft{\Lambda_{\rm EFT}}{\Lambda_{\rm IR}}+ \log \Lambda_{\rm UV}/\Lambda_{\rm EFT}\,.
\ee
In this, the ``UV physical scale'' of the two-point function is the late-time scale $aH_0$ (which is however not necessarily the UV scale of the scalar theory in dS), while the ``IR physical scale'' is $p$. The key physics is that we introduce an intermediate scale, $\varepsilon aH_0$, at which the long distance and short distance physics factorize. Indeed, we can see, for example 
\be
-\int dt' a'^6\fft{\delta(p-\varepsilon a' H_0)}{p^2}\partial_{t'}G^{cq}_m(p,t,t')\partial_{t'}G^{cq}_m(p,t,t')=\fft{1}{H_0 p^3}+ \fft{m^2}{9 H_0^3 p^3}\left(6\log\left(\fft{p}{\varepsilon a H_0 }\right)+2\right)\,.
\ee
This combines with the $\log\varepsilon$ term in the Wilson coefficients to produce the $\log p$ term in \eqref{eq: expand massive correlator}. This explicit factorization, $\log(p/(aH_0)) = \log(p/(\varepsilon a H_0)) + \log\varepsilon$, between the EFT result and the Wilson coefficient log, is the simplest example of the region structure that we will encounter in more nontrivial form in the $\lambda\phi^4$ case below.

We generalize this matching to all orders in $p/(a H_0)\sim\varepsilon$ expansion in appendix \ref{app: FDR massive}, which is precisely the FDR for a massive scalar in dS.

\subsubsection{Matching and region interpretation in $\lambda \phi^4$ theory}
\label{subsec: Gaussian matching}

We now show how the matching works for the influence functional \eqref{eq: third-fourth-IF} in $\lambda\phi^4$ theory $v=\lambda/4!\, \phi^4$. 
This example is more instructive because the two region decompositions are both active. The stochastic channel again splits the time integral at the crossing time and generates the diffusion insertion in the open EFT. However, the self-energy sub-diagram also contains the ordinary Wilsonian hard momentum region. Therefore, the long wavelength Keldysh correlator cannot be matched by the stochastic channel alone. The full matching requires the sum of the stochastic contribution and the Wilsonian contribution. This is the first explicit example in which the two RG channels introduced in section \ref{sec: EFT Gaussian} are simultaneously required by matching.  For simplicity, we only perform the matching at order $\lambda$; higher orders in $\lambda$ follow similarly, although the computations become more involved.

Let's first comment on the matching of the retarded correlator, which is rather straightforward with only Wilsonian channel. In particular,  the self-energy is
\be
\Sigma^{cc}_{\rm full}(t)=i\fft{\lambda}{2}\int \fft{d^3\ell}{(2\pi)^3} G^{cc}_{\rm full}(\ell,t,t)a^3=i\fft{H_0^2 a^3}{8\pi^2}\lambda \left(\log \fft{b}{b_{\rm IR}}+\fft{b^2}{2}\right)\,,\label{eq: self-energy}
\ee
where we regulate the IR divergence using the IR cutoff $b_{\rm IR}a H_0$. In EFT, $G^{cc}$ is provided by the leading diffusion operator (the first term in \eqref{eq: SIF free} at $\varepsilon\rightarrow 0$)
\be
G^{cc}_{\rm EFT}(p,t,t)\sim \fft{H_0^2}{2p^3}+\mathcal{O}(p^{-1})\,.
\ee
Thus we have
\be
\Sigma^{cc}_{\rm EFT}(t)=i\fft{\lambda}{2}\int \fft{d^3\ell}{(2\pi)^3} G^{cc}_{\rm EFT}(\ell,t,t)a^3=i\fft{H_0^2 a^3}{8\pi^2}\lambda \left(\log \fft{\varepsilon}{b_{\rm IR}}+\fft{\varepsilon^2}{2}\right)\,,
\ee
which can be combined with the ``Wilson coefficient'' of the mass generation \eqref{eq: mass generation} to match \eqref{eq: self-energy}. This achieves the retarded correlator matching because 
\be
\begin{tikzpicture}[baseline={(0,-0.5ex)}]
\draw[thick] (0,0) -- (2,0);
   \node at (0.6-0.6,0.15) {$c$};
  \node at (1.4+0.6,0.15) {$q$};
  \node at (1-0.15,0.15) {$c$};
  \node at (1+0.15,0.15) {$c$};
  \node at (1-0.15,-0.15) {$q$};
  \node at (1+0.15,-0.15) {$c$};
  \draw[thick] (0.7+0.3,0.5) circle [radius=0.5];
\end{tikzpicture}
=\delta G^{cq}(p,t,0)\sim \int_0^t dt' G^{cq}(p,t,t') \Sigma^{cc}(t') G^{cq}(p,t',0)\,.
\ee
In this case, the region of the loop integral from $(b_{\rm IR}aH_0,b a H_0)$ is simply decomposed into $(b_{\rm IR}aH_0,\varepsilon a H_0)$ as EFT and $(\varepsilon a H_0,b a H_0)$ as matching coefficients.

\paragraph{Match Keldysh correlator}

Let us first perform the matching for the Keldysh correlator, where, as for retarded correlator, the Wilsonian channel must also be included. The purpose is to show that the matching framework remains consistent even in the presence of this channel. Although we follow our momentum-space matching prescription, the same matching was already verified in the time domain in \cite{Tokuda:2017fdh}. Later, we will turn to observables that are insensitive to the Wilsonian channel. 

In the full theory, the corrected Keldysh correlator is 
\be
\begin{tikzpicture}[baseline={(0,-0.5ex)}]
\draw[thick] (0,0) -- (2,0);
   \node at (0.6-0.6,0.15) {$c$};
  \node at (1.4+0.6,0.15) {$c$};
  \node at (1-0.15,0.15) {$c$};
  \node at (1+0.15,0.15) {$c$};
  \node at (1-0.15,-0.15) {$q$};
  \node at (1+0.15,-0.15) {$c$};
  \draw[thick] (0.7+0.3,0.5) circle [radius=0.5];
\end{tikzpicture}\nn
\ee
which gives
\be
&\delta G^{cc}_{\rm full}(p,t,t)=2\int^t_{-\infty} dt' G^{cc}_{\rm full}(p,t,t')\Sigma^{cc}_{\rm full}(t') G^{cq}(p,t,t')= \fft{ \Sigma^{cc}_{\rm full}(t)}{3a^3 p^3}\left(\log\fft{p}{2aH_0}-\psi\left(\fft{3}{2}\right)\right)\,.\label{eq: delta Gcc full}
\ee

In EFT, we have a similar diagram by inserting the four-point vertex 
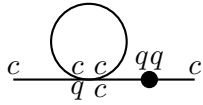
\be
\begin{tikzpicture}[baseline={(0,-0.5ex)}]
\draw[thick] (0,0) -- (2+0.4,0);
   \node at (0.6-0.6,0.15) {$c$};
  \node at (1.4+0.6+0.4,0.15) {$c$};
  \node at (1-0.15,0.15) {$c$};
  \node at (1+0.15,0.15) {$c$};
  \node at (1-0.15,-0.15) {$q$};
  \node at (1+0.15,-0.15) {$c$};
  \draw[thick] (0.7+0.3,0.5) circle [radius=0.5];
  \node at (1.7,0.2+0.05) {$q$};
  \node at (1.9,0.2+0.05) {$q$};
 \draw[thick,fill] (1.8,0) circle [radius=0.1];
\end{tikzpicture}\nn
\ee
This gives
\be
\delta G^{cc}_{{\rm EFT}1}=2\int_{H_0^{-1}\log\fft{p}{\varepsilon H_0}}^t dt' G^{cc}_{\rm EFT}(p,t,t')\Sigma^{cc}_{\rm EFT}(t')G^{cq}(p,t,t')=\fft{\Sigma^{cc}_{\rm EFT}(t)}{3a^3 p^3}\left(\log\fft{p}{\varepsilon a H_0}+\fft{1}{3}\right)\,,\label{eq: delta Gcc EFT 1}
\ee
where the lower limit of the $t'$ integral follows from the fact that, in the EFT, the mode must satisfy $p<\varepsilon a' H_0$. Furthermore, we should also insert further stochastic operator at the order of $\lambda$, i.e., \eqref{eq: second-diagram-IF}. This gives 
\be
\delta G^{cc}_{{\rm EFT}2}&= \fft{\Sigma^{cc}_{\rm EFT}(t)}{3}\int dt' a'^6\fft{\delta(p-\varepsilon a' H_0)}{p^2} \Big(\partial_{t'} G^{cq}(p,t,t')\partial_{t'}G^{cq}(p,t,t')\log_\varepsilon\nn\\
&+ H_0 \partial_{t'} G^{cq}(p,t,t')G^{cq}(p,t,t')\Big)= \fft{\Sigma^{cc}_{\rm EFT}(t)}{3 a^3 p^3}\left(\log_{\varepsilon}-\fft{1}{3}\right)\,.\label{eq: delta Gcc EFT 2}
\ee
This indeed has a direct region interpretation. The EFT contribution with the lower limit $t'>t^\ast$ reproduces the late-time region of the full theory diagram, where the external momentum mode has already entered the long wavelength EFT. The missing part of the full theory integral, namely the early-time region $t'<t^\ast$, is instead encoded in the diffusion terms generated by matching. Thus the stochastic diffusion coefficients are not an arbitrary added noise parameters, they are Wilson coefficients that represent the early-time integral region of the full theory diagram.

Combining two, we find
\be
\delta G^{cc}_{{\rm EFT}1}+\delta G^{cc}_{{\rm EFT}2}=\fft{\Sigma^{cc}_{\rm EFT}(t)}{3a^3 p^3}\left(\log\fft{p}{2a H_0}-\psi\left(\fft{3}{2}\right)\right)\,.
\ee
Nevertheless, this still does not match \eqref{eq: delta Gcc full}, because it is proportional to $\Sigma^{cc}_{\rm EFT}$ rather than $\Sigma^{cc}_{\rm full}$. To achieve the full matching, we need to include the Wilsonian channel. The mass insertion from $\delta m^2 \phi^c_s \phi^q_s$ in \eqref{eq: mass generation} gives
\be
\begin{tikzpicture}[baseline={(0,-0.5ex)}]
\draw[thick] (0,0) -- (0.7,0);
  \draw[thick] (0.7,0) -- (2,0);
  \node at (0.6,0.2+0.05) {$q$};
  \node at (0.8,0.2+0.05) {$q$};
   \node at (0.6-0.6,0.15) {$c$};
  \node at (2,0.15) {$c$};
  \draw[thick,fill] (0.7,0) circle [radius=0.1];
    \draw[thick] (0.62+0.7,-0.08) -- (0.78+0.7,0.08);
  \draw[thick] (0.62+0.7,0.08) -- (0.78+0.7,-0.08);
  \node at (0.62+0.7,0.2+0.05) {$c$};
  \node at (0.62+0.9,0.2+0.05) {$q$};
\end{tikzpicture}
=\delta G^{cc}_{{\rm EFT}1,\delta m}=\fft{\delta m^2}{3p^3}\left(\log \fft{p}{\varepsilon a H_0}+\fft{1}{3}\right)\,.
\ee
In addition, as emphasized in section \ref{sec: EFT Gaussian}, the diffusion coefficient also receives a Wilsonian renormalization, generated by the diagram in \eqref{eq: diffusion Wilsonian RG}. This gives
\be
\delta G^{cc}_{{\rm EFT}2,\delta m}=\fft{\delta m^2}{3p^3}\left(\log_\varepsilon-\fft{1}{3}\right)\,.
\ee
We thus arrived at the full matching
\be
\delta G^{cc}_{\rm full}=\delta G^{cc}_{{\rm EFT}1}+\delta G^{cc}_{{\rm EFT}2}+\delta G^{cc}_{{\rm EFT}1,\delta m}+\delta G^{cc}_{{\rm EFT}2,\delta m}\,.
\ee

This is the precise matching for Keldysh correlator in $\lambda \phi^4$ theory at Gaussian order. The first two terms reproduce the stochastic region decomposition of the Keldysh correlator: one part comes from long wavelength EFT computation of the same diagram, and the other from the diffusion operator generated at the ``horizon crossing'' region with early-time Wilson coefficients. The last two terms are required because the self-energy also contains the standard Wilsonian hard momentum region. Therefore, the full theory result is recovered only after both decompositions are included. This makes explicit that the open EFT reorganizes the full theory diagram by regions: late-time long wavelength modes are dynamically described by EFT, while early-time and hard contributions are encoded in Wilson coefficients.

\paragraph{Match a form factor diagram}

We have shown that Gaussian diffusion can generate the Keldysh correlator in $\lambda \phi^4$ theory, although this requires the Wilsonian channel. However, as we emphasized, the stochastic channel should be regarded as independent, and it can be matched even when the Wilsonian channel is turned off.

The observables that can be matched without Wilsonian renormalization are equal-time form factors. In particular, we aim to match $\langle \mathcal{O}_{\rm EFT}(q=0,t)\phi^c(p_1,t)\phi^c(p_2,t)\rangle$, where in this case $\mathcal{O}=(\phi^c)^2$. Importantly, we take the operator to lie within the EFT regime, so that all matching contributions come from the effective action. Otherwise, we would also have to match the operator from the full theory onto the EFT as $\mathcal{O}=\sum_i C_i \mathcal{O}_{{\rm EFT}}^i$\footnote{From the method-of-region perspective, the full form factor receives contributions from several distinct regions, involving both the hard/soft decomposition of loop momenta and the early/late decomposition of time integrals, see \cite{Beneke:2023wmt}. Here, this operator matching coefficients might be generally nonlocal as in soft-collinear effective theory \cite{becher2015introduction}.}. This does not contradict the previous matching of the Keldysh correlator. The equal-time two-point function is sensitive to both the stochastic channel and the Wilsonian channel. By contrast, the form factor considered here focuses on designed EFT operators and thus isolates the hard function associated with only the stochastic channel. It is therefore the appropriate observable for matching the diffusion Wilson coefficient with the Wilsonian channel turned off.

We note that the matching of form factors in the hard cutoff scheme works diagram by diagram. For the Gaussian diffusion in \eqref{eq: second-diagram-IF}, we only need to match the following diagram
\be
\begin{tikzpicture}[baseline={(0,-0.5ex)}]
  \coordinate (O) at (0,0);
  \coordinate (L) at (-0.8,0);
  \coordinate (R) at (0.8,0);
  \draw[thick] (195-5:0.8) arc[start angle=195-5,end angle=530,radius=0.8];
  \draw[thick] (L) circle [radius=0.14];
  \draw[thick] ($(L)+(-0.09,-0.09)$) -- ($(L)+(0.09,0.09)$);
  \draw[thick] ($(L)+(-0.09,0.09)$) -- ($(L)+(0.09,-0.09)$);
  \draw[thick] (R) -- ++(0.9,0.55);
  \draw[thick] (R) -- ++(0.9,-0.55);
  \node at (-0.05-0.5,0.15) {$c$};
  \node at (-0.05-0.5,-0.15) {$c$};
  \node at (0.67+0.2,0.18+0.1) {$c$};
  \node at (0.67+0.2,-0.18-0.1) {$q$};
   \node at (0.67-0.1,0.18) {$c$};
  \node at (0.67-0.1,-0.18) {$c$};
  \node at (1.35+0.2,0.45+0.2) {$c$};
  \node at (1.35+0.2,-0.45-0.2) {$c$};
  \node at (-1.7,0) {$(\phi^c)^2_{\rm EFT}$};
\end{tikzpicture}\label{eq: form factor diagram 1}
\ee
Other diagrams of the same form factor contributing to the form factor $\langle (\phi^c)^2_{\rm EFT}\phi^c \phi^c\rangle$ are captured by non-Gaussian diffusion, as discussed in the next section.

It is worth noting that, because of the insertion $(\phi^c)^2_{\rm EFT}$, this diagram factorizes as
\be
\begin{tikzpicture}[baseline={(0,-0.5ex)}]
  \coordinate (O) at (0,0);
  \coordinate (L) at (-0.8,0);
  \coordinate (R) at (0.8,0);
  \draw[thick] (195-5:0.8) arc[start angle=195-5,end angle=530,radius=0.8];
  \draw[thick] (L) circle [radius=0.14];
  \draw[thick] ($(L)+(-0.09,-0.09)$) -- ($(L)+(0.09,0.09)$);
  \draw[thick] ($(L)+(-0.09,0.09)$) -- ($(L)+(0.09,-0.09)$);
  \draw[thick] (R) -- ++(0.9,0.55);
  \draw[thick] (R) -- ++(0.9,-0.55);
  \node at (-0.05-0.5,0.15) {$c$};
  \node at (-0.05-0.5,-0.15) {$c$};
  \node at (0.67+0.2,0.18+0.1) {$c$};
  \node at (0.67+0.2,-0.18-0.1) {$q$};
   \node at (0.67-0.1,0.18) {$c$};
  \node at (0.67-0.1,-0.18) {$c$};
  \node at (1.35+0.2,0.45+0.2) {$c$};
  \node at (1.35+0.2,-0.45-0.2) {$c$};
\end{tikzpicture}
= \left(\int \fft{d^3\ell}{(2\pi)^3} \fft{H_0^4}{4\ell^6}\right)\times 2i\lambda \int dt' a'^3 G^{cc}_{\rm full}(p,t,t')G^{cq}_{\rm full}(p,t,t')\,.
\ee
The prefactor is a distribution that contains both IR and UV divergences and must be further renormalized to be well defined. Nevertheless, the same factor appears in the corresponding EFT computation. Therefore, we can factor out this prefactor and focus on the ``hard'' piece\footnote{This is analogous to deep inelastic scattering in particle physics, where the cross section at high transverse momentum factorizes into a UV and IR divergent distribution, namely the parton distribution function in partonic states, multiplied by a hard scattering cross section.}
\be
H(p,t)=2i\lambda \int dt' a'^3 G^{cc}(p,t,t')G^{cq}(p,t,t')\,.
\ee
Therefore, in both the full theory and the EFT, this is essentially the corrected Keldysh correlator with the self-energy factor stripped off, such as \eqref{eq: delta Gcc full} in the full theory and \eqref{eq: delta Gcc EFT 1}, \eqref{eq: delta Gcc EFT 2} in the EFT
\be
H(p,t)= \fft{a^3 \delta G^{cc}(p,t,t)}{\Sigma^{cc}(t)}\,.
\ee
We then see that the matching of \eqref{eq: delta Gcc EFT 1} and \eqref{eq: delta Gcc EFT 2} onto \eqref{eq: delta Gcc full} is achieved
\be
H_{\rm full}(p,t)=H_{{\rm EFT}1}(p,t)+H_{{\rm EFT}2}(p,t)\,.\label{eq: match form factor 1}
\ee
This form factor is also the one studied in SdSEFT in the context of generating Gaussian diffusion \cite{Cohen:2021fzf}.

The lesson of this subsection is that Gaussian stochastic inflation admits a standard EFT matching interpretation once the relevant regions are correctly identified. The ordinary Wilsonian channel is associated with hard momentum regions and renormalizes both unitary and diffusion Wilson coefficients. The stochastic channel is associated with the time-dependent crossing region $k\sim \varepsilon aH_0$ and directly generates diffusion Wilson coefficients in the influence functional. The Gaussian diffusion sector therefore has the same status as any EFT Wilson coefficient: it is fixed by matching to the long distance limit of full theory correlators. What is new is that the relevant region is not only a hard momentum region, but also an early-time region induced \cite{Beneke:2023wmt} by the time-dependent coarse graining.

\section{The non-Gaussian diffusion and matching }
\label{sec:non-gaussian}

In this section, we develop the non-Gaussian sector of the open EFT. 
Concretely, we consider influence-functional operators containing more than two response fields $\phi_s^q$. These operators encode non-Gaussian stochastic effects: the quadratic $\phi_s^q\phi_s^q$ terms discussed in section \ref{sec: EFT Gaussian} encode Gaussian diffusion, while cubic and quartic powers of $\phi_s^q$ encode higher-point diffusion.

The main new feature is that, beyond Gaussian order, the stochastic matching data are no longer ordinary local coefficients. 
They become Wilson kernels in the influence functional. 
These kernels remember the time at which each physical mode crosses the time-dependent scale $k\sim \varepsilon aH_0$, and they also remember the history of $\phi^q$ which has no valid temporal multipole expansion. 
Thus the non-Gaussian sector is the first place where the open EFT structure is genuinely more than a local Markovian theory.

We will organize the discussion by operators containing $\dot{\phi}^q_s$ and $\phi^q$. 
First we study operators $\dot{\phi}^q_s \dot{\phi}^q_s \phi^q_s \phi^c_s$ and $\dot{\phi}^q_s \dot{\phi}^q_s \phi^q_s \phi^q_s$. 
We then turn to operators with three $\dot\phi_s^q$, including $\dot{\phi}^q_s \dot{\phi}^q_s \dot{\phi}_s^q \phi_s^c$ and $\dot{\phi}^q_s \dot{\phi}^q_s \dot{\phi}_s^q \phi_s^q$, and finally to the quartic diffusion operator $(\dot\phi_s^q)^4$. 
In each case, the method-of-region identifies the early-time part of the full theory diagram that is missing from the long wavelength EFT. 
The open EFT promotes these missing early-time regions to definite Wilson kernels in the reduced density matrix.

\subsection{$\dot{\phi}^q_s \dot{\phi}^q_s \phi_s^q \phi_s^c$ and $\dot{\phi}^q_s \dot{\phi}^q_s \phi_s^q \phi_s^q$ operators}

We begin with the first corrections to Gaussian diffusion. They contain the same two $\dot\phi_s^q\dot\phi_s^q$ that generated the Gaussian diffusion, but now with additional $\phi^q_s$ insertions that lead to non-Gaussian diffusion in the phase space. 
The operators $\dot\phi_s^q\dot\phi_s^q\phi_s^q\phi_s^c$ and $\dot\phi_s^q\dot\phi_s^q\phi_s^q\phi_s^q$
therefore describe the first non-Gaussian deformation of the quadratic diffusion.

\subsubsection{Non-Markovian Wilson kernels}

We note that the first non-Gaussian diffusion already appear with the diagram in \eqref{eq: Feynman diagrams}. For $\dot{\phi}^q_s \dot{\phi}^q_s \phi_s^q \phi_s^c$, we have two diagrams and their permutations 
\be
\mathcal{O}_1=\begin{tikzpicture}[baseline={(0,-0.5ex)}]
\draw[thick,dashed] (-0.5,0.5)--(0,0);
\draw[thick] (0,0)--(1.5,0);
\draw[thick,dashed] (1.5,0)--(2,0.5);
\draw[thick,dashed] (0.8-0.2,0.5) -- (0.75,0);
\draw[thick,dashed] (0.8+0.1,0.5) -- (0.75,0);
\node at (-0.05,0.25) {$q$};
\node at (1+0.05+0.5,0.25) {$q$};
\node at (0.15,0.15) {$c$};
\node at (0.85+0.5,0.15) {$c$};
\node at (0.55,0.2) {$c$};
\node at (0.95,0.2) {$q$};
\node at (0.55,-0.15) {$c$};
\node at (0.95,-0.15) {$c$};
\end{tikzpicture}\qquad + \qquad 
\begin{tikzpicture}[baseline={(0,-0.5ex)}]
\draw[thick,dashed] (-0.5,0.5)--(0,0);
\draw[thick] (0,0)--(1.5,0);
\draw[thick,dashed] (1.5,0)--(2,0.5);
\draw[thick,dashed] (0.8-0.2,0.5) -- (0.75,0);
\draw[thick,dashed] (0.8+0.1,0.5) -- (0.75,0);
\node at (-0.05,0.25) {$q$};
\node at (1+0.05+0.5,0.25) {$q$};
\node at (0.15,0.15) {$c$};
\node at (0.85+0.5,0.15) {$c$};
\node at (0.55,0.2) {$c$};
\node at (0.95,0.2) {$q$};
\node at (0.55,-0.15) {$q$};
\node at (0.95,-0.15) {$q$};
\end{tikzpicture}
\ee
For $\dot{\phi}^q_s \dot{\phi}^q_s \phi_s^q \phi_s^q$, we have the following diagram with all permutations
\be
\mathcal{O}_2=\begin{tikzpicture}[baseline={(0,-0.5ex)}]
\draw[thick,dashed] (-0.5,0.5)--(0,0);
\draw[thick] (0,0)--(1.5,0);
\draw[thick,dashed] (1.5,0)--(2,0.5);
\draw[thick,dashed] (0.8-0.2,0.5) -- (0.75,0);
\draw[thick,dashed] (0.8+0.1,0.5) -- (0.75,0);
\node at (-0.05,0.25) {$q$};
\node at (1+0.05+0.5,0.25) {$q$};
\node at (0.15,0.15) {$c$};
\node at (0.85+0.5,0.15) {$c$};
\node at (0.55,0.2) {$q$};
\node at (0.95,0.2) {$q$};
\node at (0.55,-0.15) {$c$};
\node at (0.95,-0.15) {$q$};
\end{tikzpicture}
\ee

We keep only the leading order in $p/(a H_0)$. 
However, we note that the important difference from the Gaussian case is that the additional $\phi_s^q$ cannot be treated as a slowly varying classical source. 
For $\phi_s^c$, the temporal multipole expansion is justified by slow roll, and the system can be evaluated near the crossing time. 
For $\phi_s^q$, however, no such expansion is justified: the response field probes the history of the density matrix and its time derivative is not parametrically suppressed. Therefore the object generated by the diagram is not a local coefficient multiplying a nonlocal operator. 
It is a non-Markovian Wilson kernel.

We find it better to work directly in the momentum space for non-Gaussian diffusion. Straightforwardly, for $\mathcal{O}_1$ we find:
\be
S_{\rm IF}^{\mathcal{O}_1}=i \int dt \int \fft{d^3 p}{(2\pi)^3}\fft{d^3\ell}{(2\pi)^3} \delta(p-\varepsilon a H_0)\varepsilon \dot{a}H_0 a^6 C_{s,2}^{q}(p,\ell) \dot{\phi}^q_s(t,p)\dot{\phi}^q_s(t,-p)\phi^c_s(t,-\ell)\,,\label{eq: IF O1 1}
\ee
where the non-Markovian Wilson kernel $C^q$ is
\be
C_{s,2}^{q}(p,\ell)=\fft{i \lambda}{2} \int_{t^\ast(\ell)}^{t^\ast(p)} dt_3 a_3^3 \left(G^{cc}(p,t,t_3)G^{cc}(p,t,t_3)+\fft{1}{4}G^{cq}(p,t,t_3)G^{cq}(p,t,t_3)\right)\phi^q_s(t_3,\ell)\,.\label{eq: IF O1 2}
\ee
This should not be read as an ordinary Wilson coefficient. 
It is a functional of the long wavelength $\phi^q_s$ with all history. Therefore, in general, there is no direct way to determine $C^q_s$, except through inserting it into a long-distance observable, for example
\be
\langle C_{s,2}^{q}(p,\ell)\phi_s^c(t,-\ell)\rangle=\fft{H_0^2 \lambda}{96 p^6}\left(4\log(2\varepsilon)+4\gamma_E-5\right)\,,
\ee
where the scale hierarchy is $\ell\ll p\sim \varepsilon a H_0$. As expected, the ``Wilson coefficient" encode early-time physics $t<t^\ast(p)$.

Similarly, for $\mathcal{O}_2$, we have 
\be
&S_{\rm IF}^{\mathcal{O}_2}=i \int dt \int \fft{d^3 p}{(2\pi)^3} \delta(p-\varepsilon a H_0)\varepsilon \dot{a}H_0 a^6 C_{s,2}^{qq}(p;t) \dot{\phi}^q_s(t,p)\dot{\phi}^q_s(t,-p)\,,\nn\\
& C_{s,2}^{qq}=\fft{i\lambda}{8}\int \fft{d^3\ell}{(2\pi)^3}\int_{t^\ast(\ell)}^{t^\ast(p)} dt_3 a_3^3 G^{cq}(p,t,t_3)G^{cc}(p,t,t_3)\phi^q_s(t_3,\ell)\phi^q_s(t_3,-\ell)\,.\label{eq: IF O2}
\ee

Thus $S^{\mathcal{O}_1}_{\rm IF}$ and $S^{\mathcal{O}_2}_{\rm IF}$ represent the early-time region of the corresponding full theory diagrams in terms of an open EFT. The region analysis tells us where the contribution comes from, while the influence functional tells us how it acts on the reduced density matrix. For example, $S_{\rm IF}^{\mathcal{O}_1}$ and $S_{\rm IF}^{\mathcal{O}_2}$ describe non-Gaussian diffusion in the phase space, where $\phi^q\sim \delta/\delta \Pi$. See the section \ref{sec:eft-structure} for more discussions.

\subsubsection{Completing the form factor matching}

We now verify that these non-Markovian kernels \eqref{eq: IF O1 1}, \eqref{eq: IF O1 2}  and \eqref{eq: IF O2} are precisely the missing matching data for the equal-time form factor considered in subsection \ref{subsec: Gaussian matching}. 
This observable is useful because it isolates the stochastic channel. 
The operator $(\phi_s^c)^2_{\rm EFT}$ is already an operator in the long wavelength EFT, so the remaining mismatch between the full theory and the EFT is supplied by the influence functional rather than by an additional matching of the composite operator.

More concretely, the observable that we match onto is
\be
\langle (\phi^c)^2_{\rm EFT}(q\sim 0,t) \phi^c(p,t) \phi^c(-p,t)\rangle\,,\nn
\ee
but now for other diagrams in addition to \eqref{eq: form factor diagram 1}. In full theory, we find:
\be
&\begin{tikzpicture}[baseline={(0,-0.5ex)}]
  \coordinate (O) at (0,0);
  \coordinate (L) at (-0.8,0);
  \coordinate (R) at (0.8,0);
  \draw[thick] (195-5:0.8) arc[start angle=195-5,end angle=530,radius=0.8];
  \draw[thick] (L) circle [radius=0.14];
  \draw[thick] ($(L)+(-0.09,-0.09)$) -- ($(L)+(0.09,0.09)$);
  \draw[thick] ($(L)+(-0.09,0.09)$) -- ($(L)+(0.09,-0.09)$);
  \draw[thick] (R) -- ++(0.9,0.55);
  \draw[thick] (R) -- ++(0.9,-0.55);
  \node at (-0.05-0.5,0.15) {$c$};
  \node at (-0.05-0.5,-0.15) {$c$};
  \node at (0.67+0.2,0.18+0.1) {$c$};
  \node at (0.67+0.2,-0.18-0.1) {$c$};
   \node at (0.67-0.1,0.18) {$c$};
  \node at (0.67-0.1,-0.18) {$q$};
  \node at (1.35+0.2,0.45+0.2) {$c$};
  \node at (1.35+0.2,-0.45-0.2) {$c$};
  \node at (-1.7,0) {$(\phi^c)^2_{\rm EFT}$};
\end{tikzpicture}
+ \begin{tikzpicture}[baseline={(0,-0.5ex)}]
  \coordinate (O) at (0,0);
  \coordinate (L) at (-0.8,0);
  \coordinate (R) at (0.8,0);
  \draw[thick] (195-5:0.8) arc[start angle=195-5,end angle=530,radius=0.8];
  \draw[thick] (L) circle [radius=0.14];
  \draw[thick] ($(L)+(-0.09,-0.09)$) -- ($(L)+(0.09,0.09)$);
  \draw[thick] ($(L)+(-0.09,0.09)$) -- ($(L)+(0.09,-0.09)$);
  \draw[thick] (R) -- ++(0.9,0.55);
  \draw[thick] (R) -- ++(0.9,-0.55);
  \node at (-0.05-0.5,0.15) {$c$};
  \node at (-0.05-0.5,-0.15) {$c$};
  \node at (0.67+0.2,0.18+0.1) {$q$};
  \node at (0.67+0.2,-0.18-0.1) {$q$};
   \node at (0.67-0.1,0.18) {$c$};
  \node at (0.67-0.1,-0.18) {$q$};
  \node at (1.35+0.2,0.45+0.2) {$c$};
  \node at (1.35+0.2,-0.45-0.2) {$c$};
  \node at (-1.7,0) {$(\phi^c)^2_{\rm EFT}$};
\end{tikzpicture}
+ \begin{tikzpicture}[baseline={(0,-0.5ex)}]
  \coordinate (O) at (0,0);
  \coordinate (L) at (-0.8,0);
  \coordinate (R) at (0.8,0);
  \draw[thick] (195-5:0.8) arc[start angle=195-5,end angle=530,radius=0.8];
  \draw[thick] (L) circle [radius=0.14];
  \draw[thick] ($(L)+(-0.09,-0.09)$) -- ($(L)+(0.09,0.09)$);
  \draw[thick] ($(L)+(-0.09,0.09)$) -- ($(L)+(0.09,-0.09)$);
  \draw[thick] (R) -- ++(0.9,0.55);
  \draw[thick] (R) -- ++(0.9,-0.55);
  \node at (-0.05-0.5,0.15) {$c$};
  \node at (-0.05-0.5,-0.15) {$c$};
  \node at (0.67+0.2,0.18+0.1) {$c$};
  \node at (0.67+0.2,-0.18-0.1) {$q$};
   \node at (0.67-0.1,0.18) {$q$};
  \node at (0.67-0.1,-0.18) {$q$};
  \node at (1.35+0.2,0.45+0.2) {$c$};
  \node at (1.35+0.2,-0.45-0.2) {$c$};
  \node at (-1.7,0) {$(\phi^c)^2_{\rm EFT}$};
\end{tikzpicture}
\nn\\
&=\left(\int \fft{d^3\ell}{(2\pi)^3} \fft{1}{\ell^3}\right)\times \fft{H_0^4 \lambda}{12p^6}\left(\log \fft{2p}{a H_0}+\gamma_E-\fft{11}{12}\right)- \fft{H_0 \lambda}{576\pi^2 a^3 p^6}\,,
\label{eq: form factor diagram 2}
\ee
where we again factorized out the ``soft'' distribution.

Let's show how the EFT can match onto \eqref{eq: form factor diagram 2}. We first note that the second and the third diagrams in EFT are identically zero. The first part can be reproduced  by the first diagram in EFT and the matching piece contributed by $S_{\rm IF}^{\mathcal{O}_1}$, and the second part in \eqref{eq: form factor diagram 2} is fully matched by the influence functional. We find
\be
 \left[\begin{tikzpicture}[baseline={(0,-0.5ex)}]
  \coordinate (O) at (0,0);
  \coordinate (L) at (-0.8,0);
  \coordinate (R) at (0.8,0);
  \draw[thick] (195-5:0.8) arc[start angle=195-5,end angle=530,radius=0.8];
  \draw[thick] (L) circle [radius=0.14];
  \draw[thick] ($(L)+(-0.09,-0.09)$) -- ($(L)+(0.09,0.09)$);
  \draw[thick] ($(L)+(-0.09,0.09)$) -- ($(L)+(0.09,-0.09)$);
  \draw[thick] (R) -- ++(0.9,0.55);
  \draw[thick] (R) -- ++(0.9,-0.55);
  \node at (-0.05-0.5,0.15) {$c$};
  \node at (-0.05-0.5,-0.15) {$c$};
  \node at (0.67+0.2,0.18+0.1) {$c$};
  \node at (0.67+0.2,-0.18-0.1) {$c$};
   \node at (0.67-0.1,0.18) {$c$};
  \node at (0.67-0.1,-0.18) {$q$};
  \node at (1.35+0.2,0.45+0.2) {$c$};
  \node at (1.35+0.2,-0.45-0.2) {$c$};
\end{tikzpicture}\right]_{\rm EFT}&=\left(\int \fft{d^3\ell}{(2\pi)^3}\fft{1}{\ell^3}\right)\times \fft{H_0^4 \lambda}{12p^6}\left(\log\fft{p}{\varepsilon a H_0}+\fft{1}{3}\right)\,,\nn\\
 \begin{tikzpicture}[baseline={(0,-0.5ex)}]
  \coordinate (O) at (0,0);
  \coordinate (L) at (-0.8,0);
  \coordinate (R) at (0.8,0);
  \draw[thick] (195-5:0.8) arc[start angle=195-5,end angle=530,radius=0.8];
  \draw[thick] (L) circle [radius=0.14];
  \draw[thick] ($(L)+(-0.09,-0.09)$) -- ($(L)+(0.09,0.09)$);
  \draw[thick] ($(L)+(-0.09,0.09)$) -- ($(L)+(0.09,-0.09)$);
  \draw[thick] (R) -- ++(0.9,0.55);
  \draw[thick] (R) -- ++(0.9,-0.55);
  \node at (-0.05-0.5,0.15) {$c$};
  \node at (-0.05-0.5,-0.15) {$c$};
  \node at (0.67+0.2,0.18+0.1) {$c$};
  \node at (0.67+0.2,-0.18-0.1) {$q$};
   \node at (0.67-0.1,0.18) {$q$};
  \node at (0.67-0.1,-0.18) {$q$};
  \node at (1.35+0.2,0.45+0.2) {$c$};
  \node at (1.35+0.2,-0.45-0.2) {$c$};
  \node at (1.35+0.1,0) {$S_{\rm IF}^{\mathcal{O}_1}$};
  \draw[thick,fill] (0.8,0) circle [radius=0.1];
\end{tikzpicture}
& =\left(\int \fft{d^3\ell}{(2\pi)^3}\fft{1}{\ell^3}\right)\times \fft{H_0^4 \lambda}{12p^6}\left(\log(2\varepsilon)+\gamma_E-\fft{5}{4}\right)\,,\nn\\
 \begin{tikzpicture}[baseline={(0,-0.5ex)}]
  \coordinate (O) at (0,0);
  \coordinate (L) at (-0.8,0);
  \coordinate (R) at (0.8,0);
  \draw[thick] (195-5:0.8) arc[start angle=195-5,end angle=530,radius=0.8];
  \draw[thick] (L) circle [radius=0.14];
  \draw[thick] ($(L)+(-0.09,-0.09)$) -- ($(L)+(0.09,0.09)$);
  \draw[thick] ($(L)+(-0.09,0.09)$) -- ($(L)+(0.09,-0.09)$);
  \draw[thick] (R) -- ++(0.9,0.55);
  \draw[thick] (R) -- ++(0.9,-0.55);
  \node at (-0.05-0.5,0.15) {$c$};
  \node at (-0.05-0.5,-0.15) {$c$};
  \node at (0.67+0.2,0.18+0.1) {$q$};
  \node at (0.67+0.2,-0.18-0.1) {$q$};
   \node at (0.67-0.1,0.18) {$q$};
  \node at (0.67-0.1,-0.18) {$q$};
  \node at (1.35+0.2,0.45+0.2) {$c$};
  \node at (1.35+0.2,-0.45-0.2) {$c$};
  \node at (1.35+0.1,0) {$S_{\rm IF}^{\mathcal{O}_2}$};
  \draw[thick,fill] (0.8,0) circle [radius=0.1];
\end{tikzpicture}&=- \fft{H_0 \lambda}{576\pi^2 a^3 p^6}\,.
\ee

We emphasize that the matching works diagram by diagram. 
The reason is that the Wilson kernels in \eqref{eq: IF O1 2} and \eqref{eq: IF O2} are built from the same propagators and time integrals that appear in the full theory contribution \eqref{eq: form factor diagram 2}. 
The difference is how these integrals are organized. 
In the full theory, the crossing region and the early-time region are simply part of a single time integral. 
In the open EFT, the late-time part is reproduced by EFT evolution, while the early-time part is encoded in the non-Gaussian influence-functional kernels. 
Together with \eqref{eq: match form factor 1}, this completes the matching of the full form factor 
$\langle (\phi^c)^2_{\rm EFT}\phi^c\phi^c\rangle$.

More importantly, this fixes the interpretation of the Wilson kernels. 
The method-of-region identifies the early-time contribution that is not reproduced by the late-time EFT, while the open EFT organizes this contribution as Wilson kernels of the reduced density matrix.

\subsection{$\dot{\phi}^q_s \dot{\phi}^q_s \dot{\phi}_s^q \phi_s^c$ and $\dot{\phi}^q_s \dot{\phi}^q_s \dot{\phi}_s^q \phi_s^q$ operators}

We next consider operators with three $\dot{\phi}^q_s$. 
 Each factor of $\dot\phi_s^q$ is attached to a crossing region $p_i\sim \varepsilon a_iH_0$, so the corresponding Wilson kernels depend on several crossing times. 
The resulting influence functional is therefore nonlocal even at leading order in the long wavelength expansion.

\subsubsection{Nonlocal and non-Markovian Wilson kernels}

We have the following diagrams and their permutations
\be
& \mathcal{O}_3= \begin{tikzpicture}[baseline={(0,-0.5ex)}]
\draw[thick,dashed] (-0.5,0.5)--(0,0);
\draw[thick] (0,0)--(1.5,0);
\draw[thick,dashed] (1.5,0)--(2,0.5);
\draw[thick,dashed] (0.75,0) -- (0.75,1);
\node at (-0.05,0.25) {$q$};
\node at (1+0.05+0.5,0.25) {$q$};
\node at (0.15,0.15) {$c$};
\node at (0.85+0.5,-0.15) {$c$};
\node at (0.9,0.3) {$c$};
\node at (0.75-0.2,-0.15) {$c$};
\node at (0.55,0.15) {$c$};
\node at (0.95,-0.15) {$q$};
\node at (0.75-0.2,-0.5) {$c$};
\node at (0.75-0.2,-0.75) {$q$};
\draw[thick] (0.75,0) -- (0.75,-0.5);
\draw[thick,dashed] (0.75,-0.5) -- (0.75,-1);
\end{tikzpicture}
+
\begin{tikzpicture}[baseline={(0,-0.5ex)}]
\draw[thick,dashed] (-0.5,0.5)--(0,0);
\draw[thick] (0,0)--(1.5,0);
\draw[thick,dashed] (1.5,0)--(2,0.5);
\draw[thick,dashed] (0.75,0) -- (0.75,1);
\node at (-0.05,0.25) {$q$};
\node at (1+0.05+0.5,0.25) {$q$};
\node at (0.15,0.15) {$c$};
\node at (0.85+0.5,-0.15) {$c$};
\node at (0.9,0.3) {$c$};
\node at (0.75-0.2,-0.15) {$q$};
\node at (0.55,0.15) {$q$};
\node at (0.95,-0.15) {$q$};
\node at (0.75-0.2,-0.5) {$c$};
\node at (0.75-0.2,-0.75) {$q$};
\draw[thick] (0.75,0) -- (0.75,-0.5);
\draw[thick,dashed] (0.75,-0.5) -- (0.75,-1);
\end{tikzpicture}\,,\nn\\
& \mathcal{O}_4=
\begin{tikzpicture}[baseline={(0,-0.5ex)}]
\draw[thick,dashed] (-0.5,0.5)--(0,0);
\draw[thick] (0,0)--(1.5,0);
\draw[thick,dashed] (1.5,0)--(2,0.5);
\draw[thick,dashed] (0.75,0) -- (0.75,1);
\node at (-0.05,0.25) {$q$};
\node at (1+0.05+0.5,0.25) {$q$};
\node at (0.15,0.15) {$c$};
\node at (0.85+0.5,-0.15) {$c$};
\node at (0.9,0.3) {$q$};
\node at (0.75-0.2,-0.15) {$c$};
\node at (0.55,0.15) {$c$};
\node at (0.95,-0.15) {$c$};
\node at (0.75-0.2,-0.5) {$c$};
\node at (0.75-0.2,-0.75) {$q$};
\draw[thick] (0.75,0) -- (0.75,-0.5);
\draw[thick,dashed] (0.75,-0.5) -- (0.75,-1);
\end{tikzpicture}
+ \begin{tikzpicture}[baseline={(0,-0.5ex)}]
\draw[thick,dashed] (-0.5,0.5)--(0,0);
\draw[thick] (0,0)--(1.5,0);
\draw[thick,dashed] (1.5,0)--(2,0.5);
\draw[thick,dashed] (0.75,0) -- (0.75,1);
\node at (-0.05,0.25) {$q$};
\node at (1+0.05+0.5,0.25) {$q$};
\node at (0.15,0.15) {$q$};
\node at (0.85+0.5,-0.15) {$q$};
\node at (0.9,0.3) {$q$};
\node at (0.75-0.2,-0.15) {$c$};
\node at (0.55,0.15) {$c$};
\node at (0.95,-0.15) {$c$};
\node at (0.75-0.2,-0.5) {$c$};
\node at (0.75-0.2,-0.75) {$q$};
\draw[thick] (0.75,0) -- (0.75,-0.5);
\draw[thick,dashed] (0.75,-0.5) -- (0.75,-1);
\end{tikzpicture}\label{eq: non-Gaussian diagram qqq}
\ee
where $\mathcal{O}_3$ generates $\dot{\phi}^q_s \dot{\phi}^q_s \dot{\phi}_s^q \phi_s^c$, and $\mathcal{O}_4$ generates $\dot{\phi}^q_s \dot{\phi}^q_s \dot{\phi}_s^q \phi_s^q$. These non-Gaussian diagrams were also identified and estimated in \cite{Tokuda:2017fdh}, although the corresponding matching was not addressed there.

The essential new feature is that the interaction vertex and the crossing vertices are inserted at different times, where there is no reasonable assumption that can make it Markovian. The three $\dot\phi_s^q$ insertions localize the physical mode of the system near their crossing times, while the interaction vertex can still be inserted deeply early. 
 This is why the kernels are simultaneously nonlocal and non-Markovian. We have 
\be
 S_{\rm IF}^{\mathcal{O}_3}& =\fft{1}{3} \int dt_1 dt_2 dt_3 \int \fft{d^3p_1}{(2\pi)^3}\fft{d^3p_2}{(2\pi)^3} \fft{d^3p}{(2\pi)^3}\left(\prod_{i=1}^3\delta(p_i-\varepsilon a_i H_0)\varepsilon \dot{a}_i H_0 a_i^3\right)\times\nn\\
 & C_{s,3}(p_1,p_2,p_3) \dot{\phi}^q_s(t_1,p_1)\dot{\phi}^q_s(t_2,p_2)\dot{\phi}^q_s(t_3,p_3)\sum_{i=1}^3\phi^c_{s}(t_i,p)\,,\nn\\
 S_{\rm IF}^{\mathcal{O}_4}& = \int dt_1 dt_2 dt_3 \int \fft{d^3p_1}{(2\pi)^3}\fft{d^3p_2}{(2\pi)^3} \left(\prod_{i=1}^3\delta(p_i-\varepsilon a_i H_0)\varepsilon \dot{a}_i H_0 a_i^3\right)\times\nn\\
 &  C_{s,3}^{q}(p_1,p_2,p_3) \dot{\phi}^q_s(t_1,p_1)\dot{\phi}^q_s(t_2,p_2)\dot{\phi}^q_s(t_3,p_3)\,,\label{eq: IF O3 and O4}
\ee
where $\vec{p}=-\vec{p}_1-\vec{p}_2-\vec{p}_3\rightarrow 0$, and we thus have $\vec{p}_3\rightarrow-\vec{p}_1-\vec{p}_2$. More importantly, $C_s^q$ is the time integral of $\phi_{s}^q(t,p)$.

The nonlocal Wilson kernels $C_s^i$ can be computed from \eqref{eq: non-Gaussian diagram qqq}, with a transparent interpretation in terms of the method-of-region.
The momenta $p_i$ are connected to the system near their own crossing times, $p_i\simeq \varepsilon a_iH_0$. 
Therefore, at leading order in $\varepsilon\rightarrow 0$, the propagators attached to the system must be evaluated with $p_i\eta_i\to 0$. 
By contrast, the interaction time $t_4$ can fall into the early-time region, thus $p_i\eta_4$ must be kept finite. 
This is precisely the region of the full theory integral that is not reproduced by late-time EFT \cite{Beneke:2023wmt}. 
The Wilson kernels $C_s^i$ thus encode this missing early-time contribution in open EFT. We find
\be
C_{s,3}(p_i)=\frac{H_0^{2}\lambda}{36(p_1 p_2 p_3)^{3}}
\left[p_t^{\!3} 
\;-9p_1 p_2 p_3+\; 3\sum_{i=1}^{3} p_i^{3}\!\left(1 - \gamma_{E}
+ \log\!\frac{p_i}{\varepsilon p_t}\right)\right]\,,\label{eq: O3 C}
\ee
where $p_t=p_1+p_2+p_3$. The logarithms in this expression are associated with the stochastic channel: they arise from splitting the time integral around the crossing scale, rather than from an ordinary hard/soft split of loop momenta (it is simply tree-level).

The Wilson kernel $C_{s,3}^{q}$ is then defined by following the Feynman rules in \eqref{eq: non-Gaussian diagram qqq}. Since this kernel contains $\phi_s^q(t_4,p)$ inside the time integral, it keeps the history of $\phi^q_s$ and is genuinely non-Markovian
\be
C_{s,3}^{q}(p_i)&=-\fft{i\lambda}{6}\lim_{p_i\eta_i\rightarrow 0}\int_{-\infty}^{t^\ast(p_3)} dt_4 a_4^3 G^{cc}(p_1,t_1,t_4)G^{cc}(p_2,t_2,t_4)G^{cc}(p_3,t_3,t_4) \int \fft{d^3p}{(2\pi)^3}\phi^q_s(t_4,p)\nn\\
&-\fft{i\lambda}{8}\lim_{p_i\eta_i\rightarrow 0}\int_{-\infty}^{t^\ast(p_3)} dt_4 a_4^3 G^{cq}(p_1,t_1,t_4)G^{cq}(p_2,t_2,t_4)G^{cc}(p_3,t_3,t_4) \int \fft{d^3p}{(2\pi)^3} \phi^q_s(t_4,p)\nn\\
&+{\rm perm}\,.\label{eq: O4 C}
\ee

Therefore, the two kernels in \eqref{eq: O3 C} and \eqref{eq: O4 C} have slightly different structures. 
The kernel $C_{s,3}$ becomes a nonlocal function of the momenta after the slow-roll expansion of $\phi_s^c$. 
By contrast, $C_{s,3}^{q}$ still contains the history of $\phi_s^q$ and is therefore non-Markovian. 
Both kernels, however, have the same EFT interpretation: they encode the early-time region of the full theory diagram that is absent from the late-time long wavelength EFT computation.

\subsubsection{Matching onto four-point function in the squeezed limit}

Let us now show that the non-Gaussian influence functional \eqref{eq: IF O3 and O4}, with the Wilson kernels defined in \eqref{eq: O3 C} and \eqref{eq: O4 C}, provides the early-time matching contribution to a full theory observable. 
A useful observable is the equal-time four-point function in the squeezed limit, because the soft leg is already in the long wavelength system, while the other momenta enter the EFT through the stochastic channel
\be
\lim_{p\rightarrow 0}\langle \phi^c(t,p) \phi^c(t,p_1)\phi^c(t,p_2)\phi^c(t,p_3)\rangle\sim \langle \phi^c_{s}(t,p) \phi^c(t,p_1)\phi^c(t,p_2)\phi^c(t,p_3)\rangle\,.\nn
\ee
We should understand this as the density for the form factor of $(\phi^c)^3$ \cite{Cohen:2021fzf}
\be
\langle (\phi^c)^3(-p,t) \phi^c_{\rm EFT}(p,t)\rangle=\int \fft{d^3p_1}{(2\pi)^3}\fft{d^3p_2}{(2\pi)^3}\langle \phi^c_{s}(t,p) \phi^c(t,p_1)\phi^c(t,p_2)\phi^c(t,p_3)\rangle\,.
\ee
We have diagrams with the same structure as \eqref{eq: non-Gaussian diagram qqq}

\be
&\SKdiag{c}{c}{c}{c}{q}{c}{c}{c}\;+\;\SKdiag{c}{c}{q}{q}{q}{c}{c}{c}\nn\\
&+\;\SKdiag{c}{q}{c}{c}{c}{c}{c}{c}\;+\;\SKdiag{c}{q}{q}{q}{c}{c}{c}{c}\label{eq: squeezed diagram}
\ee

Let's first compute the first line of \eqref{eq: squeezed diagram} with $p_i\eta\rightarrow 0$
\be
\text{First line of \eqref{eq: squeezed diagram}}=\frac{H_0^{4}\lambda}{72\,(pp_1 p_2 p_3)^{3}}\!\left[ -p_t^{\!3}
  +9 p_1 p_2 p_3 -2\sum_i p_i^{3}
   + 3\sum_i p_i^{3}\,
    \log\!\frac{e^{\gamma_E}p_t}{a H_0}\right]\,.
\ee
This is indeed the leading order at the limit $p_4=p\rightarrow 0$ for the late equal-time four-point correlator \cite{Cohen:2021fzf,Beneke:2023wmt}, see \eqref{eq: 4pt full} below. In EFT, the second diagram in \eqref{eq: squeezed diagram} is vanishing in the leading order $p_i\eta\rightarrow 0$, and the first diagram gives
\be
\text{First line of \eqref{eq: squeezed diagram}}\Big|_{\rm EFT}=\fft{H_0^4\lambda}{24(p p_1 p_2 p_3)^3}\sum_i\left(\log \fft{p_i}{\varepsilon a H_0}+\fft{1}{3}\right)p_i^3\,.
\ee
Comparing with (5.11), the logarithmic factorization is manifest. The difference between the full theory result and the late-time EFT contribution is exactly the early-time contribution encoded in $S_{\rm IF}^{\mathcal{O}_3}$. 
We therefore find
\be
\text{First line of \eqref{eq: squeezed diagram}}\Big|_{\rm EFT}+\langle \phi^c_s(t,p) \phi^c(t,p_1)\phi^c(t,p_2)\phi^c(t,p_3) i S_{\rm IF}^{\mathcal{O}_3}\rangle=\text{First line of \eqref{eq: squeezed diagram}}\,.
\ee

Now let's match the second line of \eqref{eq: squeezed diagram}\,,
\be
& \text{Second line of \eqref{eq: squeezed diagram}}=\frac{\lambda H_0^{4}}{24(p_1 p_2 p_3)^3}\!\left[\,
\log\frac{e^{\gamma_E }p_t}{a H_0}-\fft{p_1 p_2+p_1 p_3+p_2 p_3}{p_t^2}-\fft{2 p_1 p_2 p_3}{p_t^3}
-\fft{2}{3}\right]\,,\nn\\
& \text{Second line of \eqref{eq: squeezed diagram}}\Big|_{\rm EFT}=\fft{\lambda H_0^4 }{72 (p_1 p_2 p_3)^3}\left(1+\sum_i \log\fft{p_i}{\varepsilon a H_0}\right)\,.
\ee
This agrees with the sub-sub-leading order at the limit $p_4=p\rightarrow 0$ for the late equal-time four-point correlator \cite{Cohen:2021fzf,Beneke:2023wmt}, see \eqref{eq: 4pt full} below.\footnote{The subleading order is encoded rather in the gradient expansion of $\mathcal{O}_3$.} The remaining difference is compensated by the influence-functional term $S_{\rm IF}^{\mathcal{O}_4}$ in \eqref{eq: IF O3 and O4}, yielding
\be
\text{Second line of \eqref{eq: squeezed diagram}}\Big|_{\rm EFT}+\langle \phi^c_s(t,p) \phi^c(t,p_1)\phi^c(t,p_2)\phi^c(t,p_3) i S_{\rm IF}^{\mathcal{O}_4}\rangle=\text{Second line of \eqref{eq: squeezed diagram}}\,.
\ee

We have therefore completed the matching for the equal-time four-point function in the squeezed limit. 
The result illustrates the same logic as in the Gaussian matching, but now for non-Gaussian diffusion. 
The late-time part of the full theory time integral is reproduced by long wavelength dynamics, while the early-time part is encoded in the non-Gaussian Wilson kernels in the influence functional. 
Thus the non-Gaussian stochastic terms are not additional assumptions, they are required and fixed by matching to the full theory.

\subsection{$\dot{\phi}^q_s\dot{\phi}^q_s\dot{\phi}^q_s\dot{\phi}^q_s$ operator and four-point matching}

Finally, let us consider the quartic diffusion operator $\dot\phi_s^q\dot\phi_s^q\dot\phi_s^q\dot\phi_s^q$. The corresponding diagrams are:
\be
\mathcal{O}_5=\begin{tikzpicture}[baseline={(0,-0.5ex)}]
\draw[thick,dashed] (-0.5,0.5)--(0,0);
\draw[thick] (0,0)--(1.5,0);
\draw[thick,dashed] (1.5,0)--(2,0.5);
\draw[thick] (0.75,0) -- (0.75,0.5);
\draw[thick,dashed] (0.75,0.5) -- (0.75,1);
\node at (-0.05,0.25) {$q$};
\node at (1+0.05+0.5,0.25) {$q$};
\node at (0.15,0.15) {$c$};
\node at (0.85+0.5,-0.15) {$c$};
\node at (0.9,0.15) {$c$};
\node at (0.75-0.2,-0.15) {$c$};
\node at (0.55,0.15) {$c$};
\node at (0.95,-0.15) {$q$};
\node at (0.75-0.2,-0.5) {$c$};
\node at (0.75-0.2,-0.75) {$q$};
\draw[thick] (0.75,0) -- (0.75,-0.5);
\draw[thick,dashed] (0.75,-0.5) -- (0.75,-1);
\node at (0.9,0.45) {$c$};
\node at (0.75-0.2,0.6) {$q$};
\end{tikzpicture}
+
\begin{tikzpicture}[baseline={(0,-0.5ex)}]
\draw[thick,dashed] (-0.5,0.5)--(0,0);
\draw[thick] (0,0)--(1.5,0);
\draw[thick,dashed] (1.5,0)--(2,0.5);
\draw[thick] (0.75,0) -- (0.75,0.5);
\draw[thick,dashed] (0.75,0.5) -- (0.75,1);
\node at (-0.05,0.25) {$q$};
\node at (1+0.05+0.5,0.25) {$q$};
\node at (0.15,0.15) {$c$};
\node at (0.85+0.5,-0.15) {$c$};
\node at (0.9,0.15) {$q$};
\node at (0.75-0.2,-0.15) {$c$};
\node at (0.55,0.15) {$q$};
\node at (0.95,-0.15) {$q$};
\node at (0.75-0.2,-0.5) {$c$};
\node at (0.75-0.2,-0.75) {$q$};
\draw[thick] (0.75,0) -- (0.75,-0.5);
\draw[thick,dashed] (0.75,-0.5) -- (0.75,-1);
\node at (0.9,0.45) {$c$};
\node at (0.75-0.2,0.6) {$q$};
\end{tikzpicture}\label{eq: O5}
\ee

We have
\be
& S_{\rm IF}^{\mathcal{O}_5}=i\int \prod_{i=1}^4 dt_i \int \prod_{i=1}^3 \fft{d^3p_i}{(2\pi)^3}\left(\prod_{i=1}^4\delta(p_i-\varepsilon a_i H_0)\varepsilon \dot{a}_i H_0 a_i^3\right)C_{s,4}(p_1,p_2,p_3,p_4)\prod_{i=1}^4 \dot{\phi}^q(t_i,p_i)\,.\label{eq: IF O5}
\ee
We can already see from \eqref{eq: O5} that these diagrams contain the four-point correlator of the environmental modes.
Therefore, the Wilson kernel $C_{s,4}$ is determined by the early-time part of Keldysh four-point function, with each external momentum being soft $p_i \eta_i\rightarrow 0$
\be
 C_{s,4}(p_i)=\frac{\lambda H_0^{4}}{8\,(p_1 p_2 p_3 p_4)^{3}}\!\left[-\frac{p_1 p_2 p_3 p_4}{p_t}
+p_t\,\Bigl(\!\sum_{i<j} p_i p_j - \fft{4}{9}p_t^{\!2}\Bigr)
+\fft{1}{3}\sum_{i=1}^{4} p_i^{3} \log\frac{e^{\gamma_E}\varepsilon p_t}{p_i}\right]\,,\label{eq: Cs2 O5}
\ee
where $p_t=p_1+p_2+p_3+p_4$.

Let's now show it indeed matches with equal-time four-point function in the full theory. The standard result \cite{Cohen:2021fzf,Beneke:2023wmt} in the regime of $p_i \eta\rightarrow 0$ is
\be
\langle 
& \phi^c(t,p_1)\phi^c(t,p_2)\phi^c(t,p_3)\phi^c(t,p_4)\rangle=
\begin{tikzpicture}[baseline={(0,-0.5ex)}]
\draw[thick] (0,0)--(1.5,0);
\draw[thick] (0.75,0) -- (0.75,0.75);
\node at (0.15,0.15) {$c$};
\node at (0.85+0.5,-0.15) {$c$};
\node at (0.9,0.15) {$c$};
\node at (0.75-0.2,-0.15) {$c$};
\node at (0.55,0.15) {$c$};
\node at (0.95,-0.15) {$q$};
\node at (0.75-0.2,-0.65) {$c$};
\draw[thick] (0.75,0) -- (0.75,-0.75);
\node at (0.9,0.65) {$c$};
\end{tikzpicture}
+
\begin{tikzpicture}[baseline={(0,-0.5ex)}]
\draw[thick] (0,0)--(1.5,0);
\draw[thick] (0.75,0) -- (0.75,0.75);
\node at (0.15,0.15) {$c$};
\node at (0.85+0.5,-0.15) {$c$};
\node at (0.9,0.15) {$q$};
\node at (0.75-0.2,-0.15) {$c$};
\node at (0.55,0.15) {$q$};
\node at (0.95,-0.15) {$q$};
\node at (0.75-0.2,-0.65) {$c$};
\draw[thick] (0.75,0) -- (0.75,-0.75);
\node at (0.9,0.65) {$c$};
\end{tikzpicture}
\nn\\
& =\frac{\lambda H_0^{4}}{8\,(p_1 p_2 p_3 p_4)^{3}}\!\left[
-\frac{p_1 p_2 p_3 p_4}{p_t}
+p_t\,\Bigl(\!\sum_{i<j} p_i p_j - \fft{4}{9}p_t^{\!2}\Bigr)
+\fft{1}{9}\Bigl(\sum_{i=1}^4 p_i^{3}\Bigr)\!\left(1 + 3\log\frac{e^{\gamma_E}p_t}{a H_0}\right)\right]\,.\label{eq: 4pt full}
\ee
In EFT, we take four momenta to be equally soft compared to any time scale, then the same four-point tree-level diagram gives
\be
\left[\begin{tikzpicture}[baseline={(0,-0.5ex)}]
\draw[thick] (0,0)--(1.5,0);
\draw[thick] (0.75,0) -- (0.75,0.75);
\node at (0.15,0.15) {$c$};
\node at (0.85+0.5,-0.15) {$c$};
\node at (0.9,0.15) {$c$};
\node at (0.75-0.2,-0.15) {$c$};
\node at (0.55,0.15) {$c$};
\node at (0.95,-0.15) {$q$};
\node at (0.75-0.2,-0.65) {$c$};
\draw[thick] (0.75,0) -- (0.75,-0.75);
\node at (0.9,0.65) {$c$};
\end{tikzpicture}
+
\begin{tikzpicture}[baseline={(0,-0.5ex)}]
\draw[thick] (0,0)--(1.5,0);
\draw[thick] (0.75,0) -- (0.75,0.75);
\node at (0.15,0.15) {$c$};
\node at (0.85+0.5,-0.15) {$c$};
\node at (0.9,0.15) {$q$};
\node at (0.75-0.2,-0.15) {$c$};
\node at (0.55,0.15) {$q$};
\node at (0.95,-0.15) {$q$};
\node at (0.75-0.2,-0.65) {$c$};
\draw[thick] (0.75,0) -- (0.75,-0.75);
\node at (0.9,0.65) {$c$};
\end{tikzpicture}\right]_{\rm EFT}=\fft{\lambda H_0^4}{24(p_1 p_2 p_3 p_4)^3} \sum_{i=1}^4 p_i^3\left(\log\fft{p_i}{\varepsilon a H_0}+\fft{1}{3}\right)\,.\label{eq: 4pt EFT}
\ee
It is then obvious to see \eqref{eq: Cs2 O5} adds to \eqref{eq: 4pt EFT} reproduces the full theory result in \eqref{eq: 4pt full}. This completes the non-Gaussian stochastic matching for $\dot{\phi}^q_s\dot{\phi}^q_s\dot{\phi}^q_s\dot{\phi}^q_s$.

It is natural to ask why $\mathcal{O}_3$, $\mathcal{O}_4$, and their gradient expansions can effectively play the same role as $\mathcal{O}_5$ by matching onto the Keldysh four-point correlator, but in a particular squeezed limit and order by order. In general, this should not be surprising, because in our hard cutoff scheme we did not use the equations of motion to remove operator redundancies. The second point we need to emphasize is that equal-time correlators are only a subset of all possible correlators in a full reduced density matrix description. Therefore, $\mathcal{O}_3$, $\mathcal{O}_4$, and $\mathcal{O}_5$ may serve their own purposes when matching other out-of-time-order correlators. We leave this investigation to future work.

Let us summarize the logic of this section. 
The number of $\phi_s^q$ insertions in the influence functional organizes the non-Gaussian diffusion terms. 
The quadratic terms reproduce the Gaussian diffusion discussed in section \ref{sec: EFT Gaussian}, while higher powers of $\phi_s^q$ give the non-Gaussian diffusion. 
Unlike the Gaussian case, the matching data are in general Wilson kernels rather than local Wilson coefficients. 
From the method-of-region perspective, these kernels compute the early-time region of full theory SK diagrams that is not reproduced by the dynamics of long wavelength modes. 
From the open EFT perspective, the same early-time contribution is encoded as terms in the influence functional. 

This makes the distinction between the stochastic channel and the ordinary Wilsonian channel manifest beyond Gaussian order. 
The former is associated with the time at which a mode crosses the time-dependent separation scale $k\sim \varepsilon aH_0$, while the latter comes from splitting loop momenta into hard and soft regions. 
The non-Gaussian diffusion terms discussed in this section belong to the stochastic channel, although they can still be further dressed by ordinary Wilsonian renormalization.

\section{Structure of EFT and master equations}
\label{sec:eft-structure}

\subsection{Structure of the resulting EFT and renormalization}

Let us now summarize the structure of the open EFT for stochastic inflation. The matching in the previous sections already determines the Wilson data in the SK/Liouville action. In this subsection we simply organize these data into the effective Liouville action and derive the corresponding master equations.

The results of previous calculation are organized as an open EFT for the reduced density matrix of long wavelength modes, rather than an ordinary in-out Wilsonian action. In Liouville space, the effective action takes the schematic form
\be
S^{\rm eff}_L=S_{L0}+ S_{{\rm IF}}^{\rm G} + S_{{\rm IF}}^{\rm NG}\,,
\ee
where $S_{L0}$ is the unitary sector
\be
S_{L0}=\int dt d^3x\left(a^3 \dot{\phi}^c_s \dot{\phi}^q_s-a \partial_i \phi^c_s \partial^i \phi^q_s-a^3 v[\phi^c_s,\phi^q_s]\right)\,.\label{eq: S free}
\ee
Moreover, $S_{\rm IF}^G$ denotes the Gaussian diffusion
\be
&S_{{\rm IF}}^{\rm G}=i \int dt d^3x_1 d^3x_2 a^6 \mathcal{G}(\varepsilon a H_0 |x_{12}|)\times\nn\\
& \Bigg[C_{s,2}^{11}[\phi^c_s(x_i),x_i]\dot{\phi}^q_s(t,x_1)\dot{\phi}^q_s(t,x_2)+C_{s,2}^{10}[\phi^c_s(x_i),x_i]\dot{\phi}^q_s(t,x_1)\phi^q_s(t,x_2)
+C_{s,2}^{01}[\phi^c_s(x_i),x_i]\dot{\phi}^q_s(t,x_2)\phi^q_s(t,x_1)\nn\\
&
+C_{s,2}^{00}[\phi^c_s(x_i),x_i]\phi^q_s(t,x_1)\phi^q_s(t,x_2)\Bigg]+\cdots\,,
\ee
where $\cdots$ denotes further gradient expansion in $p/(a H_0)$ expansion and $\dot{\phi}^c_s/H_0$ expansion. Let's be more precise about the Gaussian diffusive Wilson coefficients from previous sections \eqref{eq: SIF free}, \eqref{eq: second-diagram-IF} and \eqref{eq: third-fourth-IF} in $\varepsilon^0$ :
\be
 C_{s,2}^{11}[\phi^c_s(x_i),x_i]&=\fft{H_0^3}{8\pi^2}+\fft{H_0 \log_\varepsilon}{24\pi^2}\left(v''(\phi^c_s(t,x_1))+v''(\phi^c_s(t,x_2))\right) \nn\\
&+\fft{1}{144H_0 \pi^2}\left(\log_\varepsilon^2+\fft{\log_\varepsilon}{3}+\fft{\pi^2-8}{4}\right)
\left(v''(\phi^c_s(t,x_1))+v''(\phi^c_s(t,x_2))\right)^2\,,\nn\\
 C_{s,2}^{10}[\phi^c_s(x_i),x_i]&=C_{s,2}^{01}[\phi^c_s(x_i),x_i]=\fft{H_0^2}{48\pi^2}\left(v''(\phi^c_s(t,x_1))+v''(\phi^c_s(t,x_2))\right)\nn\\
& + \fft{1}{144\pi^2 }\left(\log_\varepsilon+\fft{1}{6}\right)\left(v''(\phi^c_s(t,x_1))+v''(\phi^c_s(t,x_2))\right)^2 \,,\nn\\
 C_{s,2}^{00}[\phi^c_s(x_i),x_i]&=\fft{H_0}{288\pi^2}\left(v''(\phi^c_s(t,x_1))+v''(\phi^c_s(t,x_2))\right)^2\,.\label{eq: Cs all}
\ee
This generalizes the result of \cite{Li:2025azq} beyond zero-modes.

Furthermore, $S_{{\rm IF}}^{\rm NG}$ denotes all possible non-Gaussian diffusion, schematically as
\be
& S_{\rm IF}^{\rm NG}=i\int dt d^3x_1 d^3x_2 a^6 \mathcal{G}(\varepsilon a H_0 |x_{12}|) \sum_{nm=0}^1 \tilde{C}_{s,2}^{nm}[\phi^q_s(x_i),\phi^c_s(x_i),x_i] \partial_t^{n} \phi^q_s(t,x_1)\partial_t^{m} \phi^q_s(t,x_2)\nn\\
&
+ \int \left(\prod_{i=1}^3 dt_i a_i^3\right)  \left(\prod_{i=1}^3 dx_i \right) \sum_{nmk=0}^1 C_{s,3}^{nmk}[\phi^q_s(x_i),\phi^c_s(x_i),x_i] \partial_{t_1}^{n} \phi^q_s(t_1,x_1)\partial_{t_2}^{m} \phi^q_s(t_2,x_2)\partial_{t_3}^{k} \phi^q_s(t_3,x_3)\nn\\
& 
+i \int \left(\prod_{i=1}^3 dt_i a_i^3\right)  \left(\prod_{i=1}^4 dx_i \right) \sum_{n_i=0}^1 C_{s,4}^{n_1 n_2 n_3 n_4}[\phi^q_s(x_i),\phi^c_s(x_i),x_i] \prod_{k=1}^4 \partial_{t_k}^{n_k}\phi^q_s(t_k,x_k)\nn\\
&+\cdots\,,
\ee
where $\cdots$ denotes gradient expansion in $p/(a H_0)$ expansion and $\dot{\phi}^c_s/H_0$ expansion. We absorb the function $\mathcal{G}$ into the definition of Wilson kernels $C_{s,k>2}^q$. Some explicit Wilson kernels for $\lambda \phi^4$ theory were computed in previous section, see \eqref{eq: IF O1 1}, \eqref{eq: IF O2}, \eqref{eq: IF O3 and O4} and \eqref{eq: IF O5} in momentum space for $\lambda \phi^4$ theory. Unlike the Gaussian sector, the non-Gaussian Wilson ``coefficients" are in general kernels that depend functionally on the history of $\phi_s^q$, reflecting the genuinely non-Markovian character of the open EFT beyond Gaussian order.

Nevertheless, we emphasize that what we schematically write here is the ``bare'' action in the sense of standard renormalization. Further renormalization in the Wilsonian channel should be performed, promoting to (see the organization in section \ref{sec: continuum EFT})
\be
[S_{L}^{\rm eff}]_R(\mu)=[S_{L0}]_R(\mu)+[S_{{\rm IF}}^{\rm G}]_R(\mu) +[S_{{\rm IF}}^{\rm NG}]_R(\mu)+S_L^{\rm ct}\,.
\ee
More specifically, beyond the usual renormalization of wavefunctions and vertices, the stochastic operators themselves also require renormalization. 
This generates logarithms of $\log(b/\varepsilon)$ and requires corresponding counterterms. 
In particular, the higher stochastic operators can mix with each other under the Wilsonian channel. 
We leave a systematic study of this stochastic operator renormalization to future work.

A useful way to represent the same influence functional is to perform the time integrals and localize the kernels near $\eta_\ast \sim -\varepsilon/p_i$. 
In this representation, the influence functional appears as a generally nonlocal surface term at early times, and it plays the role of a stochastic initial condition in SdSEFT \cite{Cohen:2020php,Cohen:2021fzf,Beneke:2026rtf} (see appendix \ref{app: relation to SdSEFT}). The Wilsonian channel then acts as the renormalization of this stochastic initial condition.

\subsection{The Liouville Hamiltonian and master equations}

Let us now follow the open quantum system discussion in section \ref{sec: path integral} to derive the Hamiltonian in Liouville space and use it to write down the master equations.  It is also important to note that the discussion in this subsection is formal: the Hamiltonian and master equations are written for the unrenormalized density matrix and therefore capture only the bare stochastic effects. They are also functional and in general nonlocal, which makes their practical use limited.
The procedure is standard, we first compute the canonical momentum
\be
\Pi^c_s(t,x)= \fft{\delta \mathcal{L}_{L}^{\rm eff}}{\delta \dot{\phi}^q_s(t,x)}\,,\quad \Pi^q_s(t,x)= \fft{\delta \mathcal{L}_{L}^{\rm eff}}{\delta \dot{\phi}^c_s(t,x)}\,,
\ee
and then the SK Hamiltonian density in the Liouville space
\be
\mathcal{H}_{L}^{\rm eff}(t,x)&=\Pi^c_s(t,x) \dot{\phi}^q_s(t,x)+\Pi^q_s(t,x)\dot{\phi}^c_s(t,x)- \mathcal{L}_{L}^{\rm eff}\nn\\
& = \mathcal{H}_{L0}+\mathcal{H}_{L}^{\rm G}+\mathcal{H}_{L}^{\rm NG}\,.
\ee
Let's first present the unitary and Gaussian sector and then discuss the non-Gaussian correction. The unitary sector is standard
\be
\mathcal{H}_{L0}(t,x)=\fft{\Pi^c_s(t,x)\Pi^q_s(t,x)}{a^3}+a^3 v[\phi^c_s(t,x),\phi^q_s(t,x)]\,.
\ee
The Gaussian sector gives the first nontrivial diffusion effects
\be
\mathcal{H}_{L}^{\rm G}(t,x_1)& =-i \int d^3x_2 \mathcal{G}(\varepsilon a H_0 |x_{12}|)\Bigg[C_{s,2}^{11}[\phi^c_s(x_i),x_i]\Pi^q_s(t,x_1)\Pi^q_s(t,x_2)\nn\\
& +2a^3 C_{s,2}^{10}[\phi^c_s(x_i),x_i]\Pi^q_s(t,x_1)\phi^q_s(t,x_2)+a^6 C_{s,2}^{00}[\phi^c_s(x_i),x_i]\phi^q_s(t,x_1)\phi^q_s(t,x_2)\Bigg]\,.
\ee

Let's now comment on the non-Gaussian diffusion, which is in general complicated because of the non-Markovianity. For two $\dot{\phi}^q_s$ with additional $\phi^q_s$ in the Wilson functional, we have similar structure as for Gaussian sector
\be
\mathcal{H}_{L,2}^{\rm NG}(t,x_1)& \supset  -i \int d^3x_2 \mathcal{G}(\varepsilon a H_0 |x_{12}|)\Bigg[\tilde{C}_{s,2}^{11}[\phi^q_s(x_i),\phi^c_s(x_i),x_i]\Pi^q_s(t,x_1)\Pi^q_s(t,x_2)\nn\\
& +2a^3 \tilde{C}_{s,2}^{10}[\phi^q_s(x_i),\phi^c_s(x_i),x_i]\Pi^q_s(t,x_1)\phi^q_s(t,x_2)+a^6 \tilde{C}_{s,2}^{00}[\phi^q_s(x_i),\phi^c_s(x_i),x_i]\phi^q_s(t,x_1)\phi^q_s(t,x_2)\Bigg]\,.
\ee
More nontrivial examples include three $\dot{\phi}^q_s$ and four $\dot{\phi}^q_s$. For example, we may schematically have (by cyclic symmetry of the Wilson kernels)
\be
 \mathcal{H}_{L,3}^{\rm NG}(t_1,x_1) & \supset -\int \prod_{i=2}^3 dt_i d^3x_i C_{s,3}^{111}[\phi^c_s(x_i),x_i]\prod_{i=1}^3 \Pi^q_s(t_i,x_i)\,,\nn\\
\mathcal{H}_{L,4}^{\rm NG}(t_1,x_1) & \supset -i \int \prod_{i=2}^4 dt_i d^3x_i C_{s,4}^{1111}[\phi^c_s(x_i),x_i]\prod_{i=1}^4 \Pi^q_s(t_i,x_i)\,,
\ee
where we only keep the terms without functional dependence on $\phi^q_s$ in the Wilsonian kernel. We also have similar terms such as $a_3^3 \Pi^q_{s}(t_1,x_1) \Pi^q_s(t_2,x_2) \phi^q_s(t_3,x_3)$ that describe the diffusion in phase space.

\subsubsection{Fokker-Planck equation for diagonal density matrix}

Having the Hamiltonian density, we can now follow \eqref{eq: FP general} to write down the FP functional equation. In general, the prescription around \eqref{eq: FP general} requires us to solve nonlinear equation of motion 
\be
& \partial_t(a^3 \dot{\phi}^c_s)-a \partial_i^2 \phi^c_s+a^3 \left(v'(\phi^c)+ \phi^q_s v''(\phi^c_s)+\cdots\right)+\mathcal{F}_1[\phi^q_s,\phi^c_s;\dot{\phi}^q_s,\dot{\phi}^c_s;\cdots]=0\,,\nn\\
& \partial_t(a^3 \dot{\phi}^q_s)-a \partial_i^2 \phi^q_s+a^3 \phi^q_s \left(v''(\phi^c)+\cdots\right)+\mathcal{F}_2[\phi^q_s,\phi^c_s;\dot{\phi}^q_s,\dot{\phi}^c_s;\cdots]=0\,,
\ee
with the boundary condition
\be
\phi^q_s(t_f,x)\equiv 0\,,\quad \phi^c_s(t_f,x)=\phi^c(y)\,,
\ee
where $\mathcal{F}_i$ denotes nonlinear functional terms coming from diffusion sector. This goes completely beyond the Langevin description. Luckily, we can estimate the solution around the final time slice where we evaluate the FP equation in slow roll limit. The effects of $\phi^q$ is pushed to be far away from leading few terms of slow-roll expansion and gradient expansion \cite{Li:2025azq}, and we find
\be
\phi^q_{\rm sad}(x)\simeq 0\,,\quad \Pi^c_{\rm sad}(x)\simeq -\fft{a^3}{3H_0}\left(v'\left(\phi^c_s(x)\right)+\fft{v'\left(\phi^c_s(x)\right)v''\left(\phi^c_s(x)\right)}{9H_0^2}+\cdots \right) \,.
\ee
Schematically, we have the non-Markovian FP functional equation (we now slip off the superscript $c$)
\be
& \fft{\partial}{\partial t}P\left[\phi(y),t\right]=-\int d^3 x \fft{\delta}{\delta \phi(x)} a^{-3}\Pi_{\rm sad}(x)P\left[\phi(y),t\right]\nn\\
& +\int d^3x_1 d^3x_2 \mathcal{G}(\varepsilon a H_0 |x_{12}|)\fft{\delta^2}{\delta \phi(x_1)\delta \phi(x_2)}C_{s,2}^{11}[\phi(x_i)]P\left[\phi(y),t\right]\nn\\
&- \int \left(\prod_{i=2}^3 dt_i\right)\left(\prod_{i=1}^3 d^3x_i\fft{\delta}{\delta \phi(x_i)}\right) C_{s,3}^{111}[\phi(x_i)]P\left[\phi(y),t\right]\nn\\
&- \int \left(\prod_{i=2}^4 dt_i\right)\left(\prod_{i=1}^4 d^3x_i\fft{\delta}{\delta \phi(x_i)}\right) C_{s,4}^{1111}[\phi(x_i)]P\left[\phi(y),t\right]\nn\\
& + \cdots\,.\label{eq: FP functional}
\ee
Note that when the functional derivatives act on the Wilson kernels, they may generate factors of $\delta^3(0)$. Equivalently, $(2\pi)^3\delta^3(0)$ is the comoving volume, which should be regulated as a finite volume $\Omega$. This factor can be understood as arising from the zero-modes contained in the Wilson kernels. Accordingly, both the identification $\phi^q\sim -i\delta/\delta\phi^c$ and the density matrix functional $P$ should be normalized consistently with this volume regularization. This point is also related to the fact that the density matrix has not yet been properly renormalized in the Wilsonian channel.

This equation displays the EFT organization directly. 
The unitary sector gives the drift term, the Gaussian stochastic sector gives the second functional derivatives, and the higher stochastic sector gives higher functional derivatives. 
Thus the FP equation is obtained as a projection of the open EFT master equation, rather than imposed as an independent stochastic equation.

\subsubsection{Klein-Kramers equation for the Wigner functional and Lindbladian}

We can also formally obtain the Klein-Kramers functional equation for Wigner functional using \eqref{eq: Wigner general}. Let's just explicitly write down the Gaussian sector, as it is generally lengthy to enumerate the non-Gaussian corrections
\be
& \fft{\partial}{\partial t} W[\phi(y),\Pi(y),t]=-\int d^3x a^{-3}  \Pi(x) \fft{\delta}{\delta \phi(x)}W[\phi(y),\Pi(y),t]\nn\\
&+\int d^3x a^3\Big\{\Big\{ v\left[\phi(x),i\fft{\delta}{\delta \Pi(x)}\right],W[\phi(y),\Pi(y),t]\Big\}\Big\}\nn\\
&+\int d^3x_1 d^3x_2 \mathcal{G}(\varepsilon a H_0 |x_{12}|)\Big[\fft{\delta^2}{\delta \phi(x_1)\delta\phi(x_2)}C_{s,2}^{11}[\phi(x_i)]\nn\\
& -2a^3 \fft{\delta^2}{\delta \phi(x_1)\delta\Pi(x_2)}C_{s,2}^{10}[\phi(x_i)]+a^6 \fft{\delta^2}{\delta \Pi(x_1)\delta\Pi(x_2)}C_{s,2}^{00}[\phi(x_i)]\Big] W[\phi(y),\Pi(y),t]\nn\\
& + \text{non-Gaussian}\,.\label{eq: Wigner eq}
\ee

Following \eqref{eq: Lindbladian} and the Brownian-particle discussion  in appendix \ref{app: Brownian}, we can interpret the Gaussian sector of Klein-Kramers equation \eqref{eq: Wigner eq} as a nonlocal ``Lindblad'' functional equation. We emphasize that this Lindbladian form is a structural consequence of the open EFT construction, rather than being assumed. Specifically, we find that we have the functional ``dissipator''
\be
\fft{\partial \rho}{\partial t}=-i H_{L0}\rho- \fft{\gamma}{2} L_{\rm dis} \, \rho\,,
\ee
where
\be
L_{\rm dis}\, \rho=\int d^3x_1 d^3x_2 \mathcal{G}(\varepsilon a H_0 |x_{12}|)L^q(x_1,x_2;x_1)L^q(x_1,x_2;x_2)\circ \rho\,.\label{eq: Lindblad functional}
\ee
We have defined the jump density $L^q(x)$
\be
L(x_1,x_2;x)=\left[1+\alpha(x_1,x_2)\right]\Pi(x_1)+\fft{a^3}{6H_0}\left[v''(\phi^c(x_1))+v''(\phi^c(x_2))\right]\left[1+\beta(x_1,x_2)\right]\phi(x_1)\,,
\ee
where
\be
& \alpha(x_1,x_2)=\fft{v''(\phi^c(x_1))+v''(\phi^c(x_2))}{6H_0^2}\log_\varepsilon+
\fft{\left[v''(\phi^c(x_1))+v''(\phi^c(x_2))\right]^2}{432 H_0^4}\left(6\log_\varepsilon^2+4\log_\varepsilon+3\pi^2-24\right)\,,\nn\\
&\beta(x_1,x_2)=\fft{v''(\phi^c(x_1))+v''(\phi^c(x_2))}{18H_0^2}\left(1+3\log_\varepsilon\right)\,,\quad \gamma=\fft{H_0^3}{4\pi^2}\,.
\ee

\subsection{Simplifications for ``zero-modes''}
\label{subsec: zero-modes}

As discussed above, the SK Hamiltonian and the master functional equations are, in general, difficult to use in practice because of their nonlocal, non-Markovian, and functional nature. Nevertheless, significant simplifications occur for observables in the extreme IR limit, where the dynamics collapses onto nearly zero-modes. The resulting reduced open quantum mechanics of zero-modes is useful for performing the secular resummation of certain late-time ``smeared observables'', but it inherently fails to capture other observables that are sensitive to the separation of inflaton.

In position space, the EFT has a spatial resolution scale
$\ell_{\rm EFT}\sim 1/(\varepsilon aH_0)$.
The extreme IR limit corresponds to taking $x_{ij}\ll \ell_{\rm EFT}$
for all separations between inflaton. In this regime, the EFT cannot resolve their relative positions, and the inflaton evolve collectively through the spatially averaged mode $\int d^3x\,\phi(x)$. This effectively defines the one-point probability distribution of the long wavelength inflaton \cite{Starobinsky:1994bd,Gorbenko:2019rza} (see also appendix \ref{app: one and two-pt probability}), and is equivalent to projecting onto the zero-modes sector. This is also the regime described by the separate universe approximation \cite{Starobinsky:1982ee,Starobinsky:1985ibc,sasaki1996general,Wands:2000dp,Rigopoulos:2003ak}. We leave the simplification for the two-point  probability distribution to appendix \ref{app: one and two-pt probability}. 

In momentum space, the same limit is implemented by taking $p_i=p_0\rightarrow 0$. 
More explicitly, we denote the zero-modes by $\phi_0$, corresponding to a constant inflaton profile $\phi(t,x)\sim \phi_0(t)$. Equivalently, in momentum space we define
\be
\phi(t,p_i\sim 0)=\phi_{0}(t)(2\pi)^3 \delta^3(p_i)\,.\label{eq: zero-modes}
\ee
In this limit, the effective action naively diverges, because a constant mode is smeared over the whole universe. We should therefore regulate the comoving volume by replacing $(2\pi)^3 \delta^3(0)$ with a finite volume $\Omega$. Physically, this reflects the fact that, at sufficiently long distances, the inflaton effectively probes a compact global de Sitter slice rather than an infinite flat slice. To keep the kinetic term normalization, we simply rescale $\phi^q\rightarrow \phi^q/\Omega$. The prescription for reducing to zero-modes is then to first integrate out $p_i$ and send $\varepsilon\rightarrow 0$. We note that this prescription takes the order of integration and limits opposite to that used in the matching prescription. This reversal is precisely what enables secular resummation, rather than a perturbative reproduction of IR and secular divergent observables.\footnote{See \cite{Christie:2025knc} for a different ``zero-modes" projection.}

Let's first discuss the Gaussian sector, where we simply reduce to the findings in \cite{Li:2025azq}
\be
S_{\rm IF}^{\rm G}=i\int dt a^6\left(C_{s,2}^{11}[\phi^c_0](\dot{\phi}^q_0)^2+2 C_{s,2}^{10}[\phi^c_0]\dot{\phi}^q_0 \phi^q_0+C_{s,2}^{00}[\phi^c_0](\phi^q_0)^2\right)\,,\label{eq: zero-modes IF}
\ee
where $C_{s,2}$ can be found in \eqref{eq: Cs all} by setting all $\phi^c_s(x_i)=\phi^c_0$. Correspondingly, we have \cite{Li:2025azq}
\be
& \fft{\partial}{\partial t} P[\phi_0,t]=-a^{-3}\fft{\partial}{\partial \phi_0}\left(\Pi_{{\rm sad}0} P[\phi_0,t]\right)+ \fft{\partial^2}{\partial \phi_0^2}\left(C_{s,2}^{11}[\phi_0]P[\phi_0,t]\right)\,,\nn\\
& \fft{\partial}{\partial t} W[\phi_0,\Pi_0,t]=- a^{-3}\Pi_0 \fft{\partial}{\partial \phi_0}W[\phi_0,\Pi_0,t]+ a^3 \left\{\left\{v[\phi_0,i\fft{\partial}{\partial \Pi_0}],W[\phi_0,\Pi_0,t]\right\}\right\}\nn\\
& + \left[\fft{\partial^2}{\partial \phi_0^2}C_{s,2}^{11}[\phi_0]-2a^3 \fft{\partial^2}{\partial \phi_0 \partial \Pi_0}C_{s,2}^{10}[\phi_0]+a^6 \fft{\partial^2}{\partial\Pi_0^2}C_{s,2}^{00}[\phi_0]\right]W[\phi_0,\Pi_0,t]\,.\label{eq: FP and KK zero-modes}
\ee
It was verified in \cite{Li:2025azq,Green:2025hmo}, that, in appropriate limits, the FP equation in the first line can be consistently derived from the Klein-Kramers equation in the second line.

An important conceptual point is that, in our framework, the zero-modes projection does not remove non-Markovianity for non-Gaussian diffusion. To obtain a Markovian description, one must additionally  take temporal multipole expansion. This expansion is not justified for $\phi^q$, since it probes the history of the density matrix. However, it is a good approximation for $\dot{\phi}^q$: $G^{cq}(t_1,t_2)$ remembers $(t_1,t_2)$ equally, while $\partial_{t_1} G^{cq}(t_1,t_2)$ loses track of $t_2$ in the long wavelength limit. Therefore, we generically obtain Markovian non-Gaussian diffusion in configuration space, but non-Markovian non-Gaussian diffusion in phase space.

Let's now consider the diffusion term in configuration space, for example, for $S_{\rm IF}^{\mathcal{O}_3}$ in \eqref{eq: IF O3 and O4}. Following our prescription, we first integrate $p_i$, send $\varepsilon \rightarrow 0$, and then use the zero-modes definition \eqref{eq: zero-modes}. In the end, we apply the temporal multipole expansion $\dot{\phi}^q(t_i)\simeq\dot{\phi}^q(t_1),\dot{\phi}^c(t_i)\simeq\dot{\phi}^c(t_1)$ and then integrate out $t_2,t_3$. Our estimate gives the action and its FP term as follows
\be
S_{\rm IF}^{\mathcal{O}_3}= \int dt a^9 C_{s,3}^{111}[\phi^c_0](\dot{\phi}^q_0)^3\rightarrow -\fft{\partial^3}{\partial \phi_0^3}\left(C_{s,3}^{111}[\phi_0]P[\phi_0]\right)
\ee
where
\be
C_{s,3}^{111}[\phi^c_0]=\fft{\lambda H_0^3}{53760\pi^4}\left(943-252\gamma_E+576\log 2-810 \log 3-252 \log\varepsilon\right) \phi^c_0\,.\label{eq: C cubic}
\ee

This coefficient differs significantly from the corresponding prediction of SdSEFT~\cite{Cohen:2021fzf}. We expect this discrepancy to be resolved by a consistent renormalization of the non-Gaussian stochastic operators in the Wilsonian channel, which would amount to computing the operator mixing as in~\cite{Cohen:2021fzf,Premkumar:2021mlz,Cohen:2021jbo,Beneke:2026rtf,Beneke:2026ksj}, which we leave to future work. Three structural differences are likely relevant: first, our framework  obtains a master equation for the density matrix directly, whereas~\cite{Cohen:2021fzf,Beneke:2026ksj} reinterprets the master equation as a dynamical RG and studies the corresponding RG equation of operator mixing. Second, Wilsonian channel renormalization may introduce fixed time counterterms that enter the master equation. Third, we do not take care of the redundancies of effective operators. It will therefore be important to incorporate a clean independent operator basis and the renormalization in a consistent scheme and resolve this discrepancy. We thus now do not interpret this coefficient as a renormalized or scheme-independent stochastic diffusion coefficient.

\section{Toward the continuum EFT: stochastic renormalization}
\label{sec: continuum EFT}

We have studied the open EFT of stochastic inflation in the hard cutoff scheme, as traditionally used in the literature. In particular, we have focused only on the stochastic channel. It is well known that the hard cutoff scheme is not convenient for renormalization, since it generates many scheme dependent artifacts that are difficult to organize. More importantly, regulator dependent power corrections can spoil the perturbative expansion. This prevents us from systematically studying the Wilsonian channel of the open EFT of stochastic inflation.

The continuum EFT, formulated with dimensional regularization or other analytic regulators, is better for studying renormalization and matching, and provides a more systematically well-defined EFT framework. It is therefore important to at least define the Wilsonian channel within a continuum EFT framework, using either dimensional regularization or a de Sitter analytic regulator, together with an additional IR regulator \cite{Cohen:2020php,Cohen:2021fzf,Beneke:2023wmt}. Since the stochastic channel is, in general, independent of the Wilsonian channel, it is natural to consider a mixed scheme, in which the hard cutoff scheme is used for the stochastic channel while an analytic regulator is used for the Wilsonian channel. This is analogous to SdSEFT, where the stochastic initial conditions are obtained as matching data in the hard cutoff scheme after performing the time integral \cite{Cohen:2020php,Cohen:2021fzf,Beneke:2026rtf,Beneke:2026ksj}, and one then studies their renormalization.

The mixed scheme above is a natural intermediate step, but it does not yet place the open EFT of stochastic inflation entirely within a continuum framework. In this section, we therefore develop a continuum description of the stochastic channel itself. We first discuss the organization of operators and Wilson coefficients appropriate to this channel in the continuum, then introduce a stochastic analytic regulator adapted to the mode-crossing region, and explain how renormalization and matching are performed within this scheme. We illustrate the construction using a massive scalar as a toy example, which already exhibits the essential features of the procedure. A complete continuum open EFT for stochastic inflation requires combining this construction with dimensional regularization in the Wilsonian channel, which we leave to future work.

\subsection{The organization of the continuum EFT}

We organize the continuum open EFT as before
\be
S_{L}^{\rm eff}=S_{L0}+S_{\rm IF}^{\rm G}+S_{\rm IF}^{\rm NG}+S_{L0}^{\rm ct}+S_{\rm IF}^{\rm ct}\,,
\ee
where $S_{L0}$ is the standard unitary sector \eqref{eq: S free}, and $S_L^{\rm ct}$ denotes the corresponding counterterms in the unitary sector. The term $S_{\rm IF}^{\rm ct}$ denotes the counterterms for the influence functional, whose detailed structure is left for future study. Specifically, we now replace $S_{\rm IF}$ by
\be
& S_{\rm IF}^{\rm G}=i\int \fft{d^{d-1} p}{(2\pi)^{d-1}} \sum_i C_{2,i}(\nu,\mu)\mathcal{O}_{2,i}(\nu,\mu;p)\,,\nn\\
& S_{\rm IF}^{\rm NG}=\sum_{n>2} i^{n-1} \int  \prod_{k=1}^{n}\fft{d^{d-1}p_k}{(2\pi)^{d-1}}(2\pi)^{d-1}\delta^{d-1}\left(\sum_{j=1}^n p_j\right)\sum_i C_{n,i}(\nu,\mu)\mathcal{O}_{n,i}(\nu,\mu;p_k)\,,
\ee
where we introduce a ``stochastic scale'' $\nu$ that characterizes the stochastic channel and the crossing region. In the hard cutoff scheme, this scale is roughly $\nu\sim \varepsilon H_0$, and the corresponding crossing time is set by $t^\ast\sim \log(p/\nu)$ (see appendix \ref{app: relation to SdSEFT}).

We note that in this organization, the Wilson coefficients are always local numbers, nevertheless the effective stochastic operators $\mathcal{O}_{n,i}$ are intrinsically nonlocal (with nontrivial momentum dependent kernels) and non-Markov, as we also absorb the time integral into the definition of those operators. $\mathcal{O}_{n,i}$ here denotes that there are $n$ $\phi^q$ regardless of the number of derivatives. $i$ counts the number of operators, as we include the systematic gradient expansion in $p/(a H_0)$. We emphasize that this organization by $(n,i)$ does not yet provide a complete power counting for the general nonlocal and non-Markovian EFT. In particular, the number $n$ of response fields is an organization of the Gaussian and non-Gaussian sectors, rather than an EFT suppression parameter. A general ordering of arbitrary non-Markovian kernel structures remains an important problem for future investigation.

\subsection{Stochastic analytic regulator, renormalization and operator mixing}

Let's now propose the continuum analytic regulator for the stochastic renormalization.

We start with the free theory in BD vacuum, because we know that in hard cutoff scheme, we must have from \eqref{eq: SIF free}
\be
S_{\rm IF}^{\rm free}\sim iH_0^3 \pi\int \fft{d^3p}{(2\pi)^3} \int dt a^6 \fft{\delta(p-\varepsilon a H_0)}{4\pi p^2} \dot{\phi}^q_s(t,p)\dot{\phi}^q_s(t,-p)+\mathcal{O}\left(\fft{p^2}{a^2 H_0^2}\right)\,.\label{eq: SIF free momentum}
\ee
According to our organization, it is reasonable to define the nonlocal operator
\be
\mathcal{O}_{2,1}(\varepsilon;p)= \int dt a^6 \fft{\delta(p-\varepsilon a H_0)}{4\pi p^2} \dot{\phi}^q(t,p)\dot{\phi}^q(t,-p)\,.
\ee
To introduce an analytic regulator, we note
\be
\lim_{\varepsilon\rightarrow 0}\delta(p-\varepsilon a H_0)=-\lim_{\delta\rightarrow 0}\fft{\delta}{p^{1+\delta}}(\nu a)^\delta\,.\label{eq: distributional replacement}
\ee
Our proposal is then to replace all delta functions that describe the crossing by this analytic function. We thus have
\be
\mathcal{O}_{2,1}(\nu;p)=- \int dt  a^{2(d-1)+\delta} \fft{\delta \nu^\delta}{4\pi p^{d-1+\delta}}\dot{\phi}^q_s(t,p)\dot{\phi}^q_s(t,-p)\,,
\ee
which reads in position space\footnote{Since this operator describes how short wavelength modes cross into the long wavelength sector that can be measured, it may be regarded as a natural generalization of detector operators \cite{Caron-Huot:2022eqs,Henriksson:2023cnh,Chang:2025zib,Li:2025knf} in dS. See \cite{De:2025yls} for other possible candidates for detector operators in dS.}
\be
\mathcal{O}_{2,1}(\nu;x_1,x_2)=-\fft{\Gamma\left(-\fft{\delta}{2}\right) \delta \nu^\delta \mu^{-d}}{2^{\delta+d+1}\pi^{\fft{d+1}{2}}\Gamma\left(\fft{\delta+d-1}{2}\right)}\int dt a^{2(d-1)+\delta} |x_{12}|^{\delta}\dot{\phi}^q_s(t,x_1)\dot{\phi}^q_s(t,x_2)\,.
\ee
Similarly, one can introduce analytic regulators for the non-Gaussian operators. This analytic regulator is closely analogous to the analytic regulator used in SdSEFT \cite{Cohen:2020php,Cohen:2021fzf}, since it also deforms the exponent of the Keldysh correlator. We emphasize, however, that there is an important difference: in our case the operator is nonlocal, obtained by integrating along time, whereas the operators in SdSEFT are defined at a fixed early-time $t^\ast$.

We thus now need to study the RG equation of stochastic operators along both $\nu$ and $\mu$. In general, we expect the following structure
\be
\nu \fft{d}{d\nu} \mathcal{O}_{n,i}(\nu,\mu;p_k)=\gamma_{n,ij}^\nu \mathcal{O}_{n,j}(\nu,\mu;p_k)\,,\quad \mu \fft{d}{d\mu} \mathcal{O}_{n,i}(\nu,\mu;p_k)=\gamma_{n,ij}^\mu \mathcal{O}_{n,j}(\nu,\mu;p_k)\,.
\ee
Correspondingly, we have
\be
\nu \fft{d}{d\nu} C_{n,i}(\nu,\mu)=-\left(\gamma_{n,ij}^\nu\right)^T C_{n,j}(\nu,\mu)\,,\quad \mu \fft{d}{d\mu} C_{n,i}(\nu,\mu)=-\left(\gamma_{n,ij}^\mu \right)^T  C_{n,j}(\nu,\mu)\,.
\ee
The workflow is therefore as follows. We first study the RG equations for the operators, which determine the RG equations for the Wilson coefficients. We then perform the matching computation onto the UV theory at a scale $\nu\sim\mu\sim H_0$, followed by RG evolution down to the IR scale at which the observables are measured.

\subsection{Example: $m^2 \phi^2$}

In this subsection, we study a massive scalar as a toy example to demonstrate stochastic renormalization in our proposed continuum EFT. The simplification is that there is no Wilsonian channel to worry about and we work in $d=4$. We leave a systematic study of $\lambda \phi^4$ theory to future work.

In the free theory, we know that $\mathcal{O}_{2,1}$ must be included, together with other operators that contain fewer time derivatives but more spatial derivatives, as shown in \eqref{eq: SIF free momentum}. 
These operators are not suppressed by the gradient expansion. We have:
\be
& \mathcal{O}_{2,1}(\nu;p)=\int dt a^{6+\delta} \dot{\phi}^q(t,p)\dot{\phi}^q(t,-p)  a^6 \fft{\delta \nu^\delta}{p^{3+\delta}}\,,\quad \mathcal{O}_{2,2}(\nu;p)=-2 \fft{m^2}{H_0} \int dt  a^{6+\delta} \dot{\phi}^q(t,p)\phi^q(t,-p)  \fft{\delta \nu^\delta}{p^{3+\delta}}\,,\nn\\
&  \mathcal{O}_{2,3}(\nu;p)= -\fft{m^4}{H_0^2} \int dt a^{6+\delta}\phi^q(t,p)\phi^q(t,-p)\fft{\delta \nu^\delta}{p^{3+\delta}} \,.
\ee
Matching to the free theory yields $C_{2,1}=H_0^3/(8\pi^2)$ as expected. 

Let us now study the renormalization. Since we have not removed the redundant operators using the equations of motion and field redefinitions, the full operator-mixing matrix is not uniquely determined by the projected matrix elements considered here. For this example, we therefore consider a diagonal renormalization prescription. We find
\be
\langle  (\mathcal{O}_{2,1})_b\rangle\sim  -\fft{2m^2}{3H_0^2\delta}\left(1+\fft{m^2}{9 H_0^2}\right)\langle  (\mathcal{O}_{2,1})_b\rangle\,,\quad  \langle (\mathcal{O}_{2,2})_b\rangle\sim -\fft{2m^2}{3H_0^2\delta} \langle (\mathcal{O}_{2,2})_b\rangle\,,\quad \langle (\mathcal{O}_{2,3})_b\rangle\sim -\fft{m^4}{18 H_0^5 p^3 \pi}\,,
\ee
where $\langle\cdot\rangle:=\langle \phi^c(t)|\cdot|\phi^c(0)\rangle$.
To remove the ``secular divergences'', we perform the renormalization 
\be
(\mathcal{O}_{2,i})_R=Z^{-1}_{ij} (\mathcal{O}_{2,j})_b\,,\quad Z^{-1}=\left(
\begin{array}{ccc}
 1+\fft{2m^2}{3H_0^2 \delta}+\fft{2m^4}{27 H_0^4 \delta} & 0 & 0 \\
 0 & 1+\fft{2m^2}{3H_0^2 \delta} & 0 \\
 0 & 0 & 1 \\
\end{array}
\right)\,.
\ee
It is important to note that our ``bare" operator depends on the scale explicitly, where we find
\be
\nu \fft{d}{d\nu} \langle(\mathcal{O}_{2,i})_b\rangle=\Gamma_{ij}\langle(\mathcal{O}_{2,j})_b\rangle\,.
\ee
The $\Gamma$ matrix is
\be
\Gamma=\left(
\begin{array}{ccc}
 -\fft{2m^2}{3H_0^2}-\fft{2m^4}{27 H_0^4} & 0 & 0 \\
 0 & -\fft{2m^2}{3H_0^2} & 0 \\
 0 & 0 & 0 \\
\end{array}
\right)\,.
\ee
Since the subtraction matrix $Z$ is independent of $\nu$ in this prescription, we obtain the anomalous dimensions $\gamma=Z^{-1}\Gamma Z$, which gives
\be
\gamma_{2,11}=-\fft{2m^2}{3H_0^2}-\fft{2m^4}{27 H_0^4}\,, \quad \gamma_{2,22}=-\fft{2m^2}{3H_0^2}\,,
\ee
and others are zero. Correspondingly, we have the stochastic RG equation for Wilson coefficients
\be
\nu \fft{d}{d\nu} C_{2,1}=-\gamma_{2,11}C_{2,1}\,,\quad \nu \fft{d}{d\nu}C_{2,2}=-\gamma_{2,22} C_{2,2}\,.
\ee
Solving these equations yields
\be
C_{2,1}(\nu)=C_{2,1}(\nu_0)\left(\fft{\nu}{\nu_0}\right)^{-\gamma_{2,11}}\,,\quad C_{2,2}(\nu)=C_{2,2}(\nu_0)\left(\fft{\nu}{\nu_0}\right)^{-\gamma_{2,22}}\,.\label{eq: RG sol}
\ee
We thus see the logarithmic structures are enforced if we expand in $m\rightarrow 0$. We can now match onto the Keldysh correlator in massive theory up to the same order by following the prescription in section \ref{sec: EFT Gaussian}, where the matching scale is $\nu_0=H_0/2\, e^{2-\gamma_E}$. The matching conditions admit more than one solution, reflecting the redundancy of the operator basis under equations of motion and field redefinitions. For definiteness, we choose the following solution
\be
C_{2,1}(\nu_0)= \fft{H_0^3}{8\pi^2}+ m^4\fft{\pi^2-8}{144 H_0 \pi^2}\,,\quad C_{2,2}(\nu_0)= \fft{H_0^3}{12\pi^2}+  \fft{m^2 H_0}{108 \pi^2}\,,\quad C_{2,3}= \fft{ H_0^3}{72\pi^2}\,.
\ee
We then use \eqref{eq: RG sol} to flow the Wilson coefficients down to the scale $\nu=\varepsilon H_0$, which reproduces the Wilson coefficients obtained in hard cutoff scheme \eqref{eq: Cs all}. As a consistency check, to the order contributing through $m^4$ to the influence functional, this solution also satisfies
\be
\nu \fft{d}{d\nu}C_{2,1}=\fft{m^2}{H_0^2 }C_{2,2}\,,\quad \nu \fft{d}{d\nu} C_{2,2}=4 \fft{m^2}{H_0^2 }C_{2,3}\,.
\ee

\section{Summary}
\label{sec: summary}

In this paper we have developed the open EFT of Starobinsky's
stochastic inflation. Building on the proposal of \cite{Li:2025azq} that stochastic
inflation should be viewed as an open quantum system, we have constructed the
EFT of the reduced density matrix of long wavelength modes
systematically beyond both the zero-modes limit and Gaussian order, working
throughout in the SK/Liouville-space language.

The first structural lesson is that, although our coarse-graining
follows the Wilsonian step, the time-dependent separation scale
$\Lambda(t)=\varepsilon a H_{0}$ continuously transfers modes from the bath into the
system around the crossing surface $\varepsilon a H_{0}$. The resulting EFT is therefore intrinsically
open, and its Wilson coefficients run under \emph{two} RG flows: a standard Wilsonian flow in $\log(b/\varepsilon)$ that
renormalizes the unitary sector and dresses diffusion, and a new stochastic flow in $\log\varepsilon$, generated by ``horizon-crossing'', that has no counterpart in conventional Wilsonian EFT and acts directly on the diffusion sector.

At the technical level, we have constructed the influence functional. At Gaussian order, we determine the diffusion Wilson coefficients for a generic potential $v(\phi)$. Beyond Gaussian order, the matching coefficients are no longer local Wilson coefficients but \emph{Wilson kernels}: because $\phi^q_s$ does not admit a
temporal multipole expansion, the kernels depend on multiple crossing times
and on the history of $\phi^q_s$, making the non-Gaussian sector
intrinsically nonlocal in space and non-Markovian in time. The relevant
kernels in $\lambda\phi^{4}$ theory were verified against the equal-time
form factor of $(\phi^c)^{2}$, the squeezed limit of the equal-time
four-point function, and the full equal-time four-point function. In each
case, matching admits a clean \emph{method-of-region in time}
interpretation: the full theory SK diagram splits around the crossing
time into a late-time region reproduced by the same diagram in EFT and an early-time
region encoded in the influence functional. Therefore, the stochastic Wilson
kernels are not phenomenological inputs but are uniquely fixed by matching to
the perturbative full theory.

Collecting the influence functional into the effective Liouville action, we then
derived the corresponding master equations for the reduced cosmological
density matrix, including a functional FP equation for the diagonal density matrix, and a functional Kramers-Klein equation for the Wigner functional. We also identified a nonlocal
Lindblad form for the Gaussian sector. In the extreme-IR limit these reduce
to the familiar zero-modes Starobinsky FP equation \cite{Starobinsky:1986fx,Starobinsky:1994bd} and its phase space generalization \cite{Bellini:1996uh,Calzetta:1989vk,Gorbenko:2019rza,Li:2025azq,Green:2025hmo},
with nevertheless a sharpened identification of the diffusion coefficients in terms of
matched Wilson data. Beyond zero-modes and beyond Gaussian order, however,
the master equations retain non-Markovianity and acquire higher
functional derivatives that encode the non-Gaussian Wilson kernels.

Finally, we took a first step toward placing this open EFT within a
continuum framework, by trading the hard cutoff for an analytic regulator
$\nu$ adapted to the horizon-crossing. In this scheme the open EFT is
organized in terms of nonlocal stochastic operators with Wilson
coefficients that satisfy RG equations along both the stochastic
scale and the conventional renormalization scale. The warm-up example
$v=\tfrac{1}{2}m^{2}\phi^{2}$ illustrates our key procedure: a small set of
stochastic operators flow under the stochastic RG equation with computable anomalous dimensions, and matching at a UV scale followed by running down to
$\nu\sim\varepsilon H_{0}$ reproduces the hard cutoff Wilson coefficients computed
previously. This shows that stochastic renormalization is well-defined in a
continuum scheme.

Several directions are interesting to pursue in the future. The most important step is to construct a fully continuum open EFT in which our stochastic analytic regulator is
combined with dimensional regularization (or another analytic regulator) in
the Wilsonian channel, placing the two RG channels on equal footing and
removing the regulator-dependent power corrections that make the hard cutoff
scheme useless. A second direction is the systematic renormalization of the
non-Gaussian stochastic operators, including their operator mixing, which we
expect to resolve the discrepancy observed at non-Gaussian order between our
zero-modes diffusion and the corresponding prediction of SdSEFT \cite{Cohen:2021fzf}.

More broadly, it will be important to clarify the precise relation of the
open EFT developed here to two alternative frameworks for long-distance
physics in dS: SdSET \cite{Cohen:2020php} and the exact RG approach to stochastic
inflation \cite{Green:2025hmo,Cespedes:2026fdp}. The primary connection to SdSET established in
appendix~\ref{app: relation to SdSEFT}
suggests that the two frameworks are best understood as
representations of the same data, related by the prescription for
integrating out the early-time and hard-momentum region. It is also important to clarify the relation to the exact RG approach \cite{Green:2025hmo,Cespedes:2026fdp}, where a hard cutoff is replaced by a smooth
window function. Understanding these connections precisely
would unify the three frameworks and clarify the scheme dependence of the
stochastic channel.

It is also interesting to generalize to other choices of the vacuum and other ``topology" in dS. We have focused on the BD vacuum in flat slices, but the same open EFT construction can in principle be
performed for other vacua such as $\alpha$-states \cite{Allen:1985ux,Einhorn:2002nu,Goldstein:2003ut,Einhorn:2003xb}. Similarly, an analogous open EFT can be developed for other slicings of dS, such as global dS \cite{Li:2025azq}, the static patch \cite{Mirbabayi:2019qtx}, and the hyperbolic slice that is relevant to the open inflation \cite{Bucher:1994gb}. These
generalizations would test the robustness of the two-channel RG structure
and would extend the framework to settings of direct phenomenological
relevance.

Beyond these technical directions, the open EFT framework opens several
phenomenological avenues. The reduced density matrix and its master
equations provide a natural starting point for studying decoherence in the
early universe \cite{Hu:1992ig,morikawa1987origin,morikawa1990dissipation,Lombardo:1995fg,Calzetta:1995ys,Burgess:2006jn,Burgess:2014eoa,Sano:2025ird,Liu:2016aaf,Lombardo:2004fr,Lombardo:2005iz,DaddiHammou:2022itk,deKruijf:2024ufs,Lopez:2025arw,LopezNacir:2011kk,Salcedo:2024smn,Colas:2024lse,Salcedo:2025ezu,Christie:2025knc}, the regime of eternal inflation \cite{Linde:1986fd,Goncharov:1987ir,Guth:2007ng,Creminelli:2008es,mijic1990random,Linde:1993nz,Linde:1993xx,Tolley:2008na,Cohen:2021jbo}, and the formation of
primordial black holes from large quantum fluctuations \cite{Pattison:2017mbe,Kuhnel:2019xes,Ezquiaga:2019ftu,Vennin:2020kng,Ballesteros:2020sre,Figueroa:2021zah,Saha:2026cay,Ye:2026saa}, all of which might be
sensitive to the non-Gaussian and non-Markovian structure that we have made
explicit. Generalizations to multi-field inflation \cite{Pinol:2018euk,Pinol:2020cdp}, to richer environment
sectors including heavy fields \cite{LopezNacir:2011kk,Burgess:2014eoa,Salcedo:2024smn,Colas:2024lse,Salcedo:2025ezu,Colas:2025app,Colas:2025ind,Salcedo:2026cqb,Salcedo:2026sdn,Cespedes:2025ple} and gravitational degrees of freedom \cite{Calzetta:1999zr,Hu:2008rga,Ye:2026saa,Saha:2026cay} as gauged open EFT \cite{Kaplanek:2025moq,Kaplanek:2026kpp} are natural next steps. We hope that the open EFT developed in this paper could provide a unified language in which these directions can be pursued systematically.

\textbf{Note added:} During the preparation of this paper, a preprint appeared \cite{Cespedes:2026fdp} that studies stochastic inflation using non-equilibrium renormalization. Although the techniques are different, there are conceptual overlaps, including free diffusion, the clarification of unitary and stochastic sectors, and non-locality in non-Gaussian diffusion.

\section*{Acknowledgements}	

We are grateful to Nicola Bartolo, Martin Beneke, Simon Caron-Huot, Sebastián Céspedes, Thomas Colas, Lennard Dufner, Daniel Green, Victor Gorbenko, Patrick Hagerb, Bei Lok Hu, Jajie Mei, Enrico Pajer, Rajat Kumar Panda, Shinsei Ryu, Andrea Sanfilippo, Cheng Shang, David Simmons-Duffin, Raman Sundrum, Abbas Tinwala, Junsei Tokuda, Shao-jiang Wang, Xiao-Quan Ye, Wucheng Zhang for useful discussions. We are especially grateful to Daniel Green for helpful clarifications on SdSEFT and valuable comments on the draft. We are also grateful to the JHEP referee for pointing out a mistake in the previous version concerning the stochastic RG in the continuum scheme. This work is supported by Simons Foundation grant No. 917464 (Simons Collaboration on Confinement and QCD Strings).

\appendix 

\section{Warm-up: a Brownian particle}
\label{app: Brownian}

In this appendix, we present a toy example of the SK formalism for density matrix: a Brownian particle arising from a particle coupling to an external bath \cite{schwinger1961brownian,massignan2015quantum,maniscalco2004lindblad,lampo2016lindblad,kamenev2023field}. We will derive the corresponding master equation by integrating out the bath, which yields a Fokker–Planck equation for the diagonal density matrix and a Klein–Kramers equation for the Wigner function. These equations are essentially the Caldeira–Leggett equation in different representations.

We consider the full density matrix
\be
\rho\left(r[x,p;Q],t\right)=N^2 \int^{r[x,p;Q]}_{x^\pm(\mp i\infty)=0,\Omega_\beta(Q)} [Dx^c Dx^q Dp^c Dp^q DQ^c DQ^q] e^{i \left(S_{L,x}+S_{L,Q}+S_{L,{\rm int}}\right)}\,,
\ee
where $\Omega_\beta(\Omega)$ refers to the thermal equilibrium state of $Q$. The action is
\be
S_x=\int dt\left(\fft{1}{2}\dot{x}^2-V(x)\right)\,,\quad S_{\rm int}=\int dt\, Q x\,,
\ee
such that the Liouville action is
\be
S_{L,x}=\int dt\left(\dot{x}^c \dot{x}^q
-V(x^+)+V(x^-)\right)\,,\quad S_{L,{\rm int}}=\int dt\left(Q^q x^c+Q^c x^q\right)\,.
\ee
Here, $Q$ refers to the collective degrees of freedom of the environment. To emphasize the philosophy of effective theory, we will not specify the details of $Q$ in this discussion, thus we do not need $S_{L,Q}$. This form of effective coupling is not new; it is essentially analogous to the familiar dipolar electromagnetic interaction in optical physics, where $Q$ plays the role of the dipole moment and $x$ is replaced by the electromagnetic field. A similar structure has also been recently adopted in the study of gravitational wave physics, where $Q$ refers to the multipole moments of compact objects, and $x$ is replaced by operators constructed from gravitational waves \cite{Goldberger:2009qd,Saketh:2023bul,Ivanov:2024sds,Glazer:2024eyi,Caron-Huot:2025tlq,Ivanov:2026icp}.

We now integrate out $Q$, i.e., trace over the environmental degrees of freedom. To proceed, we assume weak coupling between the environment and the system, which is equivalent to assuming that linear response theory for $Q$ is sufficient. We therefore perturbatively expand $S_{L,{\rm int}}$ and evaluate its expectation value under $e^{i S_{L,Q}}$. The resulting influence functional encodes all correlation functions of $Q$, where we assume for simplicity that the thermal vacuum satisfies $\langle Q \rangle_\beta = 0$. At leading order, we find
\be
S_{{\rm IF}}\simeq \fft{i}{2}\int dt dt'\left(\langle Q^c(t)Q^q(t')\rangle
x^q(t)x^c(t')+\langle Q^q(t)Q^c(t')\rangle
x^c(t)x^q(t')+\langle Q^c(t)Q^c(t')\rangle
x^q(t)x^q(t')\right)\,.
\ee
It is known that in the Keldysh basis, $-i\langle Q^c Q^q \rangle = G_R$ corresponds to the retarded Green function (i.e., the response function), and $-i\langle Q^c Q^c \rangle = G_K$ is referred to as the Keldysh Green function. The logic of effective theory is that, although we do not know the microscopic details of $Q$ and its correlators, we can perform a low-frequency expansion of the correlation functions. Since the analyticity and causality properties of the retarded Green function are more transparent, we focus on its Taylor expansion
\be
G_R(\omega)=\sum_{n=0} c_n (i\omega)^n\,.\label{eq: GR expansion}
\ee
This, when inverse Fourier transformed back to the time domain, yields local-in-time terms of the form $\partial_t^n \delta(t - t')$. Therefore, such an ``EFT'' expansion precisely corresponds to the Markov approximation. An example is the Lorentz–Drude spectral density, which encodes the microscopic details but still admits the expansion \eqref{eq: GR expansion} at low frequency. This type of EFT has also been considered recently in gravitational wave physics, where the coefficients $c_n$ are referred to as Love numbers and characterize how compact objects respond to external gravitational waves. 

Since we assume the environment is in the thermal equilibrium vacuum, the Keldysh Green function is not independent, but is instead determined by the fluctuation–dissipation theorem
\be
G_K(\omega)=i \coth\fft{\beta\omega}{2}\, {\rm Im}\, G_R(\omega)
\ee
which is non-local in time and encodes all memory effects. Nevertheless, we can still perform a Taylor expansion around $\omega \to 0$ as a good approximation. This corresponds to a secular approximation around zero frequency and ensures the Markov property for the full bath excitations.

We consider the simplest example, focusing on the leading dissipative effect in \eqref{eq: GR expansion}\footnote{Conservative responses are neglected, as they can be absorbed into a redefinition of the potential involving time derivatives and are not essential to our discussion.}
\be
G_R(\omega)=-i\gamma \omega+\cdots \,.\label{eq: GR Brownian}
\ee
This is sometimes known as the Ohmic bath. The corresponding Keldysh Green function is
\be
G_K(\omega)=-i\gamma \omega \coth\fft{\beta\omega}{2}\simeq -2i\fft{\gamma}{\beta}+\cdots\,.
\ee
For simplicity, we perform an integration by parts to rewrite $x^c \dot{x}^q$ as $-x^q \dot{x}^c$. This alters the Liouville Hamiltonian if boundary terms are neglected; nevertheless, the modification is harmless and corresponds to the procedure of ``Hamiltonian renormalization'' in the Lindblad model of a Brownian particle discussed in \cite{maniscalco2004lindblad,lampo2016lindblad}.
We thus find
\be
S_{{\rm IF}}=\int dt\left(-\gamma x^q \dot{x}^c+i \fft{\gamma}{\beta}\left(x^q\right)^2+\cdots\right)\,.\label{eq: IF Brownian}
\ee
It is then manifest that, in deriving such a local influence functional, we have made the Born approximation (weak coupling), the Markov approximation (Taylor expansion of the response function), and the secular or rotating wave approximation (Taylor expansion of the Keldysh correlator).

Before turning to the discussion of the evolution of the density matrix, we make a side comment on the stochastic perspective of the influence functional \eqref{eq: IF Brownian}. In the stochastic approach, one typically rewrites $(x^q)^2$ in terms of $x^q \xi$ by introducing a white noise field $\xi$
\be
e^{iS_{L,x}+i S_{\rm IF}}=\left(\fft{\beta}{4\pi \gamma}\right)^{\fft{1}{2}}\int [D\xi] e^{i S_{x}+i\int dt \left(- \gamma x^q \dot{x}^c+x^q \xi\right)}\, e^{-\fft{\beta}{4\gamma}\int dt\,  \xi(t)^2}\,.
\ee
This then clearly explains why the model describes a Brownian particle. Classically, the particle satisfies an equation of motion obtained by varying $x^q$
\be
\ddot{x}^c+\gamma \dot{x}^c+V'(x^c)-\xi=0\,,\label{eq: Lagenvin Brownian}
\ee
subject to a Gaussian random variable $\xi$ with $\langle \xi(t) \xi(t') \rangle = \frac{2\gamma}{\beta} \delta(t - t')$. This is precisely the Langevin equation.

Let's now return to deriving the master equation. Having obtained the full effective action, we can readily extract the Liouville Hamiltonian and project it onto any representation of interest
\be
& H_L^{\rm eff}=p^c p^q  -V\left(x^c-\fft{x^q}{2}\right)+V\left(x^c+\fft{x^q}{2}\right)+H_{L,{\rm IF}}\,,\quad H_{L,{\rm IF}}=\gamma p^c x^q-i \fft{\gamma}{\beta}\left(x^q\right)^2\,,\nn\\
& p^c=\dot{x}^c\,,\quad p^q=\dot{x}^q-\gamma x^q\,.
\ee
We should now notice that this effective Hamiltonian is indeed the one that generates the Caldeira–Leggett equation for a quantum Brownian particle \cite{caldeira1985influence}
\be
\fft{\partial}{\partial t}\rho=-i[H_x,\rho]-i\fft{\gamma}{2}[x,\{p,\rho\}]-\fft{\gamma}{\beta}[x,[x,\rho]]\,,\label{eq: CL equation}
\ee
as we note, the SK formalism gives $[x,\{p,\rho\}] \rightarrow x^q p^c \rho$ and $[x,[x,\rho]] \rightarrow (x^q)^2 \rho$. In the position representation, we simply replace $p^{c/q} = -i\partial/\partial x^{c/q}$ in the evolution equation $\partial \rho/\partial t = -i H_L \rho$, which, however, is not straightforward to analyze. More interestingly, let us focus on the diagonal density matrix in position space and the Wigner function.

\begin{itemize}
\item[I.] Diagonal density matrix

We have
\be
\partial_t P(x,t)=-i \int^{x^c(t)=x,x^q(t)=0}_{x^\pm(\mp i\infty)=0} [Dx^c Dx^q] H_L^{\rm eff}e^{i S_{L}^{\rm eff}}\,.
\ee
To obtain a closed-form expression, we consider the semiclassical limit $x^q \to 0$ as well as the overdamped limit $ \ddot{x} \ll \dot{x} \sim V'(x) \ll \gamma$. The overdamped limit is essentially the rotating wave approximation applied to the Hamiltonian, reducing the quadratic kinetic term to a linear one. This is also analogous to the slow-roll limit in the context of inflation. We then need to solve the following equation of motion, subject to the boundary conditions and the approximations stated above
\be
-\ddot{x}^c-\gamma \dot{x}^c-V'(x^c) +2i\fft{\gamma}{\beta}x^q=0\,,\quad -\ddot{x}^q+\gamma \dot{x}^q-V''(x^c)x^q=0\,.
\ee
We find that, in our limit, $x^q$ can be eliminated at leading order (thus the boundary condition $x^q(t)=0$ is not used), yielding
\be
\dot{x}^c=-\fft{V'(x^c)}{\gamma}-2i\fft{p^q}{\gamma \beta}\,,\quad x^q=\fft{p^q \gamma}{V''(x^c)-\gamma^2}\simeq -\fft{p^q}{\gamma}+\cdots\,.
\ee
Thus, at leading order we have
\be
H_L^{\rm eff}\simeq -\fft{V'(x^c)}{\gamma}p^q -\fft{i}{\beta\gamma}\left(p^q\right)^2\,.
\ee
Replacing $p^q = -i\partial/\partial x^c$ yields the Fokker–Planck equation for a Brownian particle associated with the Langevin equation \eqref{eq: Lagenvin Brownian} in the overdamped limit
\be
\partial_t P(x,t)=\left(\fft{1}{\gamma}V'(x) \partial_x+ \fft{1}{\beta\gamma}\partial_x^2\right)P(x,t)\,.\label{eq: FP Brownian}
\ee

\item[II.] The Wigner function

It is actually more straightforward to obtain the evolution equation for the Wigner function, and we still work in the $x^q \to 0$ limit. We simply replace $(p^q = -i\partial/\partial x,\ x^q = i\partial/\partial p)$ in $H_L^{\rm eff}$
\be
\partial_t W(x,p)=- p \partial_x W(x,p)+\partial_p\left(p W(x,p)\right)+\partial_p\left(V'(x)W(x,p)\right)+\fft{\gamma}{\beta}\partial_p^2 W(x,p)\,.
\ee
This is known as the Klein–Kramers equation, which governs the phase-space evolution of a Brownian particle.
\end{itemize}

We emphasize that our strategy here is different from the Martin–Siggia–Rose (MSR) formalism for stochastic processes, where the Langevin equation is imposed as a functional identity to generate a path integral \cite{kamenev2023field}. Our derivation follows an ``EFT'' perspective, providing a consistent and automatic construction from coarse-graining over unknown degrees of freedom.

A final comment is that, in the regime we consider in \eqref{eq: GR Brownian}, the resulting open system is not Lindbladian. Indeed, the Caldeira–Leggett model is known not to be Lindbladian, as it does not preserve the positivity of the density matrix. Nevertheless, it represents the low-frequency limit of a more complete Lindbladian system. We also note that if we restrict to the diagonal components of the density matrix, those elements satisfy a Lindblad equation, as shown in \eqref{eq: FP Brownian}, as it is known that the Fokker–Planck equation is the semiclassical limit of a Lindblad equation, with the jump operator simply given by $p^q$ \cite{sieberer2016keldysh}. See \cite{hu1992quantum} for the non-Markov generalization of \eqref{eq: CL equation} from a generic bath, which is known as Hu-Paz-Zhang master equation.

\section{Full influence functional and FDR of a massive scalar}
\label{app: FDR massive}

In this appendix, we present the full influence functional and the corresponding FDR for a massive scalar in dS, generalizing the result of free field theory in \eqref{eq: SIF free} and \eqref{eq: FDR dS}.

We can first write down the general influence functional
\be
S_{{\rm IF},m}&=i \fft{H_0^3}{8\pi^2}\int dt d^3x_1 d^3x_2 a^6 \mathcal{G}(\varepsilon a H_0 |x_{12}|)\times \nn\\
& \left(C_{2,1,m}\partial_t \phi^q_s(x_1)\partial_t \phi^q_s(x_2)+C_{2,2,m}\left(\partial_t \phi^q_s(x_1) \phi^q_s(x_2)+\partial_t \phi^q_s(x_2) \phi^q_s(x_1)\right)+C_{2,3,m}  \phi^q_s(x_1)\phi^q_s(x_2)\right)\,.
\ee
Following the matching procedure in section \ref{sec: EFT Gaussian}, we match to the massive correlators in dS and find
\be
& C_{2,1,m}=\fft{\pi}{2}\varepsilon^3 \left(J_{\nu}(\varepsilon)^2+Y_{\nu}(\varepsilon)^2\right)\nn\\
& C_{2,2,m}=\frac{H_{0}\pi}{4}\,\varepsilon^{3}  \left[ 2\varepsilon \left( J_{\nu-1}(\varepsilon)\,J_{\nu}(\varepsilon) + Y_{\nu-1}(\varepsilon)\,Y_{\nu}(\varepsilon) \right) - (2\nu - 3)\left( J_{\nu}(\varepsilon)^{2} + Y_{\nu}(\varepsilon)^{2} \right) \right]\,,\nn\\
& C_{2,3,m}=\frac{H_{0}^{2}\pi}{8}\,\varepsilon^{3}  \Bigl[ -4\varepsilon(\nu - 3)\left( J_{\nu-1}(\varepsilon)\,J_{\nu}(\varepsilon) + Y_{\nu-1}(\varepsilon)\,Y_{\nu}(\varepsilon) \right) - 3(4\nu - 3)\left( J_{\nu}(\varepsilon)^{2} + Y_{\nu}(\varepsilon)^{2} \right) \nn\\
& + \varepsilon^{2}\left( 3 J_{\nu-1}(\varepsilon)^{2} + J_{\nu+1}(\varepsilon)^{2} + 3 Y_{\nu-1}(\varepsilon)^{2} + Y_{\nu+1}(\varepsilon)^{2} \right) \Bigr]\,.
\ee
Correspondingly, the FDR reads
\begin{align}
& G^{cc}_m(p,t_1,t_2)=-\fft{H_0^3}{2} \int dt' a'^6 \fft{\delta(p-\varepsilon a' H_0)}{p^2}\Bigg(-C_{2,1,m}\partial_{t'}G^{cq}(p,t_1,t')\partial_{t'}G^{cq}(p,t_2,t')\nn\\
&\quad +C_{2,2,m}\left(\partial_{t'}G^{cq}(p,t_1,t')G^{cq}(p,t_2,t')-\left(1\leftrightarrow 2\right)\right) +C_{2,3,m} G^{cq}(p,t_1,t')G^{cq}(p,t_2,t')\Bigg) \,.
\end{align}

This full expression allows us to have the gradient expansion to arbitrary orders. For example, to the order $m^4$ and $(p/(a H_0))^3$, we find
\be
& C_{2,1,m}=1 + \varepsilon^{2} + m^{2}\left[ \frac{\varepsilon^{4}}{3 H_{0}^{2}} + \frac{2\log_{\varepsilon}}{3 H_{0}^{2}} + \frac{2\varepsilon^{2}\left(1 + \log_{\varepsilon}\right)}{3 H_{0}^{2}} - \frac{2\varepsilon^{3}\pi}{9 H_{0}^{2}} \right] \nn\\
& + m^{4}\left[ \frac{\varepsilon^{4}\left(7 + 6\log_{\varepsilon}\right)}{27 H_{0}^{4}} - \frac{2\varepsilon^{3}\pi}{27 H_{0}^{4}} + \frac{\varepsilon^{2}\left(4 + 28\log_{\varepsilon} + 12\log_{\varepsilon}^{2} + 3\pi^{2}\right)}{54 H_{0}^{4}} + \frac{4\log_{\varepsilon} + 12\log_{\varepsilon}^{2} + 3\left(\pi^{2} - 8\right)}{54 H_{0}^{4}} \right]\,,\nn\\
& C_{2,2,m}=\varepsilon^{2} H_{0} + m^{2}\left[ \frac{1}{3 H_{0}} + \frac{2\varepsilon^{4}}{3 H_{0}} + \frac{\varepsilon^{2}\left(3 + 2\log_{\varepsilon}\right)}{3 H_{0}} - \frac{\varepsilon^{3}\pi}{3 H_{0}} \right] \nn\\
& + m^{4}\left[ \frac{1 + 6\log_{\varepsilon}}{27 H_{0}^{3}} + \frac{\varepsilon^{4}\left(17 + 12\log_{\varepsilon}\right)}{27 H_{0}^{3}} - \frac{\varepsilon^{3}\pi}{9 H_{0}^{3}} + \frac{\varepsilon^{2}\left(18 + 40\log_{\varepsilon} + 12\log_{\varepsilon}^{2} + 3\pi^{2}\right)}{54 H_{0}^{3}} \right]\,,\\
& C_{2,3,m}=\varepsilon^{4} H_{0}^{2} + \left[ \frac{2\varepsilon^{2}}{3} + \frac{2}{3}\varepsilon^{4}\left(2 + \log_{\varepsilon}\right) \right] m^{2} + \nn\\
& m^{4}\left[ \frac{1}{9 H_{0}^{2}} + \frac{\varepsilon^{2}\left(17 + 12\log_{\varepsilon}\right)}{27 H_{0}^{2}} - \frac{2\varepsilon^{3}\pi}{9 H_{0}^{2}} + \frac{\varepsilon^{4}\left(80 + 52\log_{\varepsilon} + 12\log_{\varepsilon}^{2} + 3\pi^{2}\right)}{54 H_{0}^{2}} \right]\,,
\ee
which is consistent with \eqref{eq: SIF free} in the limit of $m\rightarrow 0$, and also consistent with \eqref{eq: Cs all} in the limit of $\varepsilon\rightarrow 0$.

\section{Influence functional as stochastic initial condition}
\label{app: relation to SdSEFT}

In this appendix, we comment on the relation between the influence functional and the stochastic initial condition in the language of SdSEFT \cite{Cohen:2020php}.

Let's start with the free Gaussian term \eqref{eq: SIF free} at the leading order of the gradient expansion $\varepsilon^0$ to illustrate the connection. In momentum space we have \eqref{eq: SIF free}. Now we integrate $t$, which sets $t=t^\ast=H_0^{-1}\log(p/(\varepsilon H_0))$, we thus have
\be
S_{\rm IF}^{\rm free}=i \fft{H_0^3}{4} \int \fft{d^3p}{(2\pi)^3} \fft{1}{p^3} a_\ast^6\, \dot{\phi^q}(t^\ast,p)\dot{\phi^q}(t^\ast,-p)\,.\label{eq: IF free initial}
\ee
The key difference from SdSEFT is that we are describing the density matrix, while SdSEFT is only capturing the equal-time correlators. If we persist to the equal-time correlators, we know that $\dot{\phi}^q$ can only contract with $\phi^c$ and gives $a_\ast^{-3}$, therefore we can define $\varphi^\pm$ such that $\phi\sim  \varphi^+ + a^{-3}\varphi^{-}$, \eqref{eq: IF free initial} can thus be equivalently written as
\be
\mathcal{P}_{t}\circ S_{\rm IF}^{\rm free}\sim F^{\rm free}[\varphi^\pm]=i \fft{H_0^3}{2} \int \fft{d^3p}{(2\pi)^3} \fft{1}{p^3} \varphi^{+}(t^\ast,p)\varphi^{-}(t^\ast,-p)\,,
\ee
where $\mathcal{P}_{t}$ denotes the equal-time projection.

Similar discussion applies to non-Gaussian influence functional, which becomes effectively the non-Gaussian stochastic initial condition. Let's take $(\dot{\phi^q})^4)$ in \eqref{eq: IF O5} as an example. In general, we have, for example
\be
S_{\rm IF}^{\mathcal{O}_5}\sim i \int \left(\prod_{i=1}^4 \fft{d^3p_i}{(2\pi)^3} (a_i^\ast)^3\right) C_{s,4}(p_i) (2\pi)^{d-1}\delta^{d-1}(p_1+p_2+p_3+p_4) \left(\prod_{i=1}^4 \dot{\phi}^q_s(t_i^\ast,p_i)\right) \,,
\ee
This is not a standard initial condition, because different modes require different effective ``initial times'' $t_i^\ast$. Nevertheless, we should perform the equal-time projection, which gives
\be
& \mathcal{P}_{t}\circ S_{\rm IF}^{\mathcal{O}_5}\sim i \int \prod_{i=1}^4 \fft{d^3p_i}{(2\pi)^3} C_{s,4}(p_i) (2\pi)^{d-1}\delta^{d-1}(p_1+p_2+p_3+p_4) \left(\prod_{i=1}^3\varphi^+(t^\ast,p_i)\right) \varphi^-(t^\ast,p_4)\,,
\ee
up to permutation and symmetry factors, where $t^\ast$ is determined by the minimal momentum scale.

We leave a deeper and more systematic study of the relation between open EFT and SdSEFT for future work.

\section{One and two-point probability distribution}
\label{app: one and two-pt probability}

In this appendix, we briefly clarify the relation between the one-point probability distribution \cite{Starobinsky:1994bd} and the zero-modes simplification used in section \ref{sec:eft-structure}. We also explain how the two-point probability distribution \cite{Starobinsky:1994bd} is encoded in the full reduced density matrix. For simplicity, we only focus on the Gaussian sector.

The traditional one-point probability distribution is defined by
\be
P_1(\phi_0)=\int d\phi \, \delta(\phi(x)-\phi_0)P[\phi(y),t]\,.
\ee
We note that $P[\phi(y),t]$ is the diagonal density matrix functional, defined as
\be
P[\phi(y),t]\sim \langle \phi|\rho|\phi\rangle\,,\label{eq: one-point}
\ee
as a functional over the full field configuration space. The one-point probability distribution amounts to selecting the subset of diagonal matrix elements for homogeneous configurations $\phi_0$, namely $\langle \phi_0|\rho|\phi_0\rangle$. This projection is implemented by the delta function. Therefore, it can be directly understood as setting $\phi(x)$ to its zero-modes $\phi_0$, as in section \ref{sec:eft-structure}. An equivalent way to see this is to consider $\partial_t P_1(\phi_0,t)$. Using the definition \eqref{eq: one-point} together with the functional FP equation \eqref{eq: FP functional}, and then integrating by parts with respect to the field variable $\phi(y)$
\be
\int d\phi\, \delta(\phi(x)-\phi_0) \fft{\delta}{\delta \phi(x)}=-\int d\phi \left(\fft{\delta}{\delta \phi(x)}\right) \delta(\phi(x)-\phi_0) =\fft{\partial}{\partial \phi_0}\int d\phi\, \delta(\phi(y)-\phi_0)\,.
\ee
This indeed reproduces the first line of \eqref{eq: FP and KK zero-modes}.

We now discuss the two-point probability distribution, which is defined as
\be
P_2(\phi_1,\phi_2,t)=\int d\phi\, \delta(\phi(x_1,t)-\phi_1)\delta(\phi(x_2,t)-\phi_2) P[\phi(y),t]\,.
\ee
Following the same trick of integration-by-parts, we find \cite{Starobinsky:1994bd}
\be
& \partial_t P_2(\phi_1,\phi_2,t)= -i \mathcal{L}_{\phi_1}P_2(\phi_1,\phi_2,t) -i \mathcal{L}_{\phi_2}P_2(\phi_1,\phi_2,t)\nn\\
&+\fft{\partial^2}{\partial \phi_1\partial \phi_2}\left[\mathcal{G}(\epsilon a H_0 |x_{12}|)\left(C_{2,1}(\phi_1)+C_{2,1}(\phi_2)\right)P_2(\phi_1,\phi_2,t)\right]\,,
\ee
where $\mathcal{L}_\phi$ is the Liouville operator for one-point distribution. This is the equal-time two-point probability, which in general fails to compute out-of-time correlators. We can also define a more general two-point probability using the EFT
\be
P_2(\phi_1,t_1;\phi_2,t_2)=\int D\phi^{\pm}\, \delta(\phi(x_1,t_1)-\phi_1)\delta(\phi(x_2,t_2)-\phi_2) e^{i S_{L}^{\rm eff}}\,.
\ee
Generally, we should understand the two-point distribution as the two-point function of the projection operator
\be
P_2={\rm Tr}[|\phi_1\rangle\langle \phi_1|\rho |\phi_2\rangle\langle \phi_2| ]\,.
\ee

\bibliographystyle{JHEP}
\bibliography{ref}

\providecommand{\href}[2]{#2}\begingroup\raggedright\begin{thebibliography}{100}

\bibitem{Bauer:2000ew}
C.~W. Bauer, S.~Fleming and M.~E. Luke, \emph{{Summing Sudakov logarithms in $B
  \to X_s \gamma $in effective field theory.}},
  \href{http://dx.doi.org/10.1103/PhysRevD.63.014006}{\emph{Phys. Rev. D} {\bf
  63} (2000) 014006}, [\href{https://arxiv.org/abs/hep-ph/0005275}{{\tt
  hep-ph/0005275}}].

\bibitem{Bauer:2000yr}
C.~W. Bauer, S.~Fleming, D.~Pirjol and I.~W. Stewart, \emph{{An Effective field
  theory for collinear and soft gluons: Heavy to light decays}},
  \href{http://dx.doi.org/10.1103/PhysRevD.63.114020}{\emph{Phys. Rev. D} {\bf
  63} (2001) 114020}, [\href{https://arxiv.org/abs/hep-ph/0011336}{{\tt
  hep-ph/0011336}}].

\bibitem{becher2015introduction}
T.~Becher, A.~Broggio, A.~Ferroglia et~al., \emph{Introduction to
  soft-collinear effective theory}, vol.~10.
\newblock Springer, 2015.

\bibitem{Rothstein:2016bsq}
I.~Z. Rothstein and I.~W. Stewart, \emph{{An Effective Field Theory for Forward
  Scattering and Factorization Violation}},
  \href{http://dx.doi.org/10.1007/JHEP08(2016)025}{\emph{JHEP} {\bf 08} (2016)
  025}, [\href{https://arxiv.org/abs/1601.04695}{{\tt 1601.04695}}].

\bibitem{feynman1963theory}
R.~P. Feynman and F.~L. Vernon~Jr, \emph{The theory of a general quantum system
  interacting with a linear dissipative system}, {\emph{Annals of physics} {\bf
  24} (1963) 118--173}.

\bibitem{schwinger1961brownian}
J.~Schwinger, \emph{Brownian motion of a quantum oscillator}, {\emph{Journal of
  Mathematical Physics} {\bf 2} (1961) 407--432}.

\bibitem{keldysh2024diagram}
L.~V. Keldysh, \emph{Diagram technique for nonequilibrium processes},  in
  \emph{Selected Papers of Leonid V Keldysh}, pp.~47--55.
\newblock World Scientific, 2024.

\bibitem{Crossley:2015evo}
M.~Crossley, P.~Glorioso and H.~Liu, \emph{{Effective field theory of
  dissipative fluids}},
  \href{http://dx.doi.org/10.1007/JHEP09(2017)095}{\emph{JHEP} {\bf 09} (2017)
  095}, [\href{https://arxiv.org/abs/1511.03646}{{\tt 1511.03646}}].

\bibitem{Liu:2018kfw}
H.~Liu and P.~Glorioso, \emph{{Lectures on non-equilibrium effective field
  theories and fluctuating hydrodynamics}},
  \href{http://dx.doi.org/10.22323/1.305.0008}{\emph{PoS} {\bf TASI2017} (2018)
  008}, [\href{https://arxiv.org/abs/1805.09331}{{\tt 1805.09331}}].

\bibitem{Goldberger:2009qd}
W.~D. Goldberger and A.~Ross, \emph{{Gravitational radiative corrections from
  effective field theory}},
  \href{http://dx.doi.org/10.1103/PhysRevD.81.124015}{\emph{Phys. Rev. D} {\bf
  81} (2010) 124015}, [\href{https://arxiv.org/abs/0912.4254}{{\tt
  0912.4254}}].

\bibitem{Saketh:2023bul}
M.~V.~S. Saketh, Z.~Zhou and M.~M. Ivanov, \emph{{Dynamical tidal response of
  Kerr black holes from scattering amplitudes}},
  \href{http://dx.doi.org/10.1103/PhysRevD.109.064058}{\emph{Phys. Rev. D} {\bf
  109} (2024) 064058}, [\href{https://arxiv.org/abs/2307.10391}{{\tt
  2307.10391}}].

\bibitem{Ivanov:2024sds}
M.~M. Ivanov, Y.-Z. Li, J.~Parra-Martinez and Z.~Zhou, \emph{{Gravitational
  Raman Scattering in Effective Field Theory: A Scalar Tidal Matching at
  O(G3)}}, \href{http://dx.doi.org/10.1103/PhysRevLett.132.131401}{\emph{Phys.
  Rev. Lett.} {\bf 132} (2024) 131401},
  [\href{https://arxiv.org/abs/2401.08752}{{\tt 2401.08752}}].

\bibitem{Glazer:2024eyi}
D.~Glazer, A.~Joyce, M.~J. Rodriguez, L.~Santoni, A.~R. Solomon and L.~F.
  Temoche, \emph{{Higher-Dimensional Black Holes and Effective Field Theory}},
  \href{https://arxiv.org/abs/2412.21090}{{\tt 2412.21090}}.

\bibitem{Caron-Huot:2025tlq}
S.~Caron-Huot, M.~Correia, G.~Isabella and M.~Solon, \emph{{Gravitational Wave
  Scattering via the Born Series: Scalar Tidal Matching to $\mathcal{O}(G^7)$
  and Beyond}},  \href{https://arxiv.org/abs/2503.13593}{{\tt 2503.13593}}.

\bibitem{Ivanov:2026icp}
M.~M. Ivanov, Y.-Z. Li, J.~Parra-Martinez and Z.~Zhou, \emph{{Gravitational
  Raman Scattering: a Systematic Toolkit for Tidal Effects in General
  Relativity}},  \href{https://arxiv.org/abs/2602.06951}{{\tt 2602.06951}}.

\bibitem{LopezNacir:2011kk}
D.~Lopez~Nacir, R.~A. Porto, L.~Senatore and M.~Zaldarriaga, \emph{{Dissipative
  effects in the Effective Field Theory of Inflation}},
  \href{http://dx.doi.org/10.1007/JHEP01(2012)075}{\emph{JHEP} {\bf 01} (2012)
  075}, [\href{https://arxiv.org/abs/1109.4192}{{\tt 1109.4192}}].

\bibitem{Burgess:2014eoa}
C.~P. Burgess, R.~Holman, G.~Tasinato and M.~Williams, \emph{{EFT Beyond the
  Horizon: Stochastic Inflation and How Primordial Quantum Fluctuations Go
  Classical}}, \href{http://dx.doi.org/10.1007/JHEP03(2015)090}{\emph{JHEP}
  {\bf 03} (2015) 090}, [\href{https://arxiv.org/abs/1408.5002}{{\tt
  1408.5002}}].

\bibitem{Salcedo:2024smn}
S.~A. Salcedo, T.~Colas and E.~Pajer, \emph{{The open effective field theory of
  inflation}}, \href{http://dx.doi.org/10.1007/JHEP10(2024)248}{\emph{JHEP}
  {\bf 10} (2024) 248}, [\href{https://arxiv.org/abs/2404.15416}{{\tt
  2404.15416}}].

\bibitem{Colas:2024lse}
T.~Colas, \emph{{Open Effective Field Theories for cosmology}},  in \emph{{58th
  Rencontres de Moriond on Cosmology}}, 5, 2024.
\newblock \href{https://arxiv.org/abs/2405.09639}{{\tt 2405.09639}}.

\bibitem{Salcedo:2025ezu}
S.~A. Salcedo, T.~Colas, L.~Dufner and E.~Pajer, \emph{{An Open System Approach
  to Gravity}},  \href{https://arxiv.org/abs/2507.03103}{{\tt 2507.03103}}.

\bibitem{Colas:2025app}
T.~Colas, \emph{{Lectures on Open Effective Field Theories}},  9, 2025.
\newblock \href{https://arxiv.org/abs/2510.00140}{{\tt 2510.00140}}.

\bibitem{Colas:2025ind}
T.~Colas, Z.~Qin and X.~Tong, \emph{{Open Effective Field Theory and the
  Physics of Cosmological Collider Signals}},
  \href{https://arxiv.org/abs/2512.07941}{{\tt 2512.07941}}.

\bibitem{Salcedo:2026cqb}
S.~A. Salcedo, T.~Colas, L.~Dufner and E.~Pajer, \emph{{Phenomenology of an
  Open Effective Field Theory of Dark Energy}},
  \href{https://arxiv.org/abs/2603.12321}{{\tt 2603.12321}}.

\bibitem{Salcedo:2026sdn}
S.~A. Salcedo, T.~Colas, P.~Suman, B.~Zhang, J.~Fergusson and E.~P.~S.
  Shellard, \emph{{Primordial non-Gaussianity constraints on dissipative
  inflation}},  \href{https://arxiv.org/abs/2603.13473}{{\tt 2603.13473}}.

\bibitem{Cespedes:2025ple}
S.~Cespedes, Z.~Qin and D.-G. Wang,
  \emph{{{\ensuremath{\lambda}}{\ensuremath{\phi}}$^{4}$ as an effective theory
  in de Sitter}}, \href{http://dx.doi.org/10.1007/JHEP05(2026)143}{\emph{JHEP}
  {\bf 05} (2026) 143}, [\href{https://arxiv.org/abs/2510.25826}{{\tt
  2510.25826}}].

\bibitem{Starobinsky:1986fx}
A.~A. Starobinsky, \emph{{STOCHASTIC DE SITTER (INFLATIONARY) STAGE IN THE
  EARLY UNIVERSE}},
  \href{http://dx.doi.org/10.1007/3-540-16452-9_6}{\emph{Lect. Notes Phys.}
  {\bf 246} (1986) 107--126}.

\bibitem{Mukhanov:1990me}
V.~F. Mukhanov, H.~A. Feldman and R.~H. Brandenberger, \emph{{Theory of
  cosmological perturbations. Part 1. Classical perturbations. Part 2. Quantum
  theory of perturbations. Part 3. Extensions}},
  \href{http://dx.doi.org/10.1016/0370-1573(92)90044-Z}{\emph{Phys. Rept.} {\bf
  215} (1992) 203--333}.

\bibitem{Maldacena:2002vr}
J.~M. Maldacena, \emph{{Non-Gaussian features of primordial fluctuations in
  single field inflationary models}},
  \href{http://dx.doi.org/10.1088/1126-6708/2003/05/013}{\emph{JHEP} {\bf 05}
  (2003) 013}, [\href{https://arxiv.org/abs/astro-ph/0210603}{{\tt
  astro-ph/0210603}}].

\bibitem{Acquaviva:2002ud}
V.~Acquaviva, N.~Bartolo, S.~Matarrese and A.~Riotto, \emph{{Second order
  cosmological perturbations from inflation}},
  \href{http://dx.doi.org/10.1016/S0550-3213(03)00550-9}{\emph{Nucl. Phys. B}
  {\bf 667} (2003) 119--148},
  [\href{https://arxiv.org/abs/astro-ph/0209156}{{\tt astro-ph/0209156}}].

\bibitem{Durrer:2004fx}
R.~Durrer, \emph{{Cosmological perturbation theory}},
  \href{http://dx.doi.org/10.1007/978-3-540-31535-3_2}{\emph{Lect. Notes Phys.}
  {\bf 653} (2004) 31--70}, [\href{https://arxiv.org/abs/astro-ph/0402129}{{\tt
  astro-ph/0402129}}].

\bibitem{Baumann:2009ds}
D.~Baumann, \emph{{Inflation}},  in \emph{{Theoretical Advanced Study Institute
  in Elementary Particle Physics}: {Physics of the Large and the Small}},
  pp.~523--686, 2011.
\newblock \href{https://arxiv.org/abs/0907.5424}{{\tt 0907.5424}}.
\newblock \href{http://dx.doi.org/10.1142/9789814327183_0010}{DOI}.

\bibitem{wang2014inflation}
Y.~Wang, \emph{Inflation, cosmic perturbations and non-gaussianities},
  {\emph{Communications in Theoretical Physics} {\bf 62} (2014) 109}.

\bibitem{Linde:1986fd}
A.~D. Linde, \emph{{Eternally Existing Selfreproducing Chaotic Inflationary
  Universe}}, \href{http://dx.doi.org/10.1016/0370-2693(86)90611-8}{\emph{Phys.
  Lett. B} {\bf 175} (1986) 395--400}.

\bibitem{Goncharov:1987ir}
A.~S. Goncharov, A.~D. Linde and V.~F. Mukhanov, \emph{{The Global Structure of
  the Inflationary Universe}},
  \href{http://dx.doi.org/10.1142/S0217751X87000211}{\emph{Int. J. Mod. Phys.
  A} {\bf 2} (1987) 561--591}.

\bibitem{Guth:2007ng}
A.~H. Guth, \emph{{Eternal inflation and its implications}},
  \href{http://dx.doi.org/10.1088/1751-8113/40/25/S25}{\emph{J. Phys. A} {\bf
  40} (2007) 6811--6826}, [\href{https://arxiv.org/abs/hep-th/0702178}{{\tt
  hep-th/0702178}}].

\bibitem{Creminelli:2008es}
P.~Creminelli, S.~Dubovsky, A.~Nicolis, L.~Senatore and M.~Zaldarriaga,
  \emph{{The Phase Transition to Slow-roll Eternal Inflation}},
  \href{http://dx.doi.org/10.1088/1126-6708/2008/09/036}{\emph{JHEP} {\bf 09}
  (2008) 036}, [\href{https://arxiv.org/abs/0802.1067}{{\tt 0802.1067}}].

\bibitem{Carr:1974nx}
B.~J. Carr and S.~W. Hawking, \emph{{Black holes in the early Universe}},
  \href{http://dx.doi.org/10.1093/mnras/168.2.399}{\emph{Mon. Not. Roy. Astron.
  Soc.} {\bf 168} (1974) 399--415}.

\bibitem{Carr:1975qj}
B.~J. Carr, \emph{{The Primordial black hole mass spectrum}},
  \href{http://dx.doi.org/10.1086/153853}{\emph{Astrophys. J.} {\bf 201} (1975)
  1--19}.

\bibitem{Sasaki:2018dmp}
M.~Sasaki, T.~Suyama, T.~Tanaka and S.~Yokoyama, \emph{{Primordial black
  holes{\textemdash}perspectives in gravitational wave astronomy}},
  \href{http://dx.doi.org/10.1088/1361-6382/aaa7b4}{\emph{Class. Quant. Grav.}
  {\bf 35} (2018) 063001}, [\href{https://arxiv.org/abs/1801.05235}{{\tt
  1801.05235}}].

\bibitem{Starobinsky:1994bd}
A.~A. Starobinsky and J.~Yokoyama, \emph{{Equilibrium state of a
  selfinteracting scalar field in the De Sitter background}},
  \href{http://dx.doi.org/10.1103/PhysRevD.50.6357}{\emph{Phys. Rev. D} {\bf
  50} (1994) 6357--6368}, [\href{https://arxiv.org/abs/astro-ph/9407016}{{\tt
  astro-ph/9407016}}].

\bibitem{mijic1990random}
M.~Miji{\'c}, \emph{Random walk after the big bang}, {\emph{Physical Review D}
  {\bf 42} (1990) 2469}.

\bibitem{Linde:1993nz}
A.~D. Linde and A.~Mezhlumian, \emph{{Stationary universe}},
  \href{http://dx.doi.org/10.1016/0370-2693(93)90187-M}{\emph{Phys. Lett. B}
  {\bf 307} (1993) 25--33}, [\href{https://arxiv.org/abs/gr-qc/9304015}{{\tt
  gr-qc/9304015}}].

\bibitem{Linde:1993xx}
A.~D. Linde, D.~A. Linde and A.~Mezhlumian, \emph{{From the Big Bang theory to
  the theory of a stationary universe}},
  \href{http://dx.doi.org/10.1103/PhysRevD.49.1783}{\emph{Phys. Rev. D} {\bf
  49} (1994) 1783--1826}, [\href{https://arxiv.org/abs/gr-qc/9306035}{{\tt
  gr-qc/9306035}}].

\bibitem{Tolley:2008na}
A.~J. Tolley and M.~Wyman, \emph{{Stochastic Inflation Revisited: Non-Slow Roll
  Statistics and DBI Inflation}},
  \href{http://dx.doi.org/10.1088/1475-7516/2008/04/028}{\emph{JCAP} {\bf 04}
  (2008) 028}, [\href{https://arxiv.org/abs/0801.1854}{{\tt 0801.1854}}].

\bibitem{Cohen:2021jbo}
T.~Cohen, D.~Green and A.~Premkumar, \emph{{A tail of eternal inflation}},
  \href{http://dx.doi.org/10.21468/SciPostPhys.14.5.109}{\emph{SciPost Phys.}
  {\bf 14} (2023) 109}, [\href{https://arxiv.org/abs/2111.09332}{{\tt
  2111.09332}}].

\bibitem{tsamis2005stochastic}
N.~C. Tsamis and R.~P. Woodard, \emph{Stochastic quantum gravitational
  inflation}, {\emph{Nuclear Physics B} {\bf 724} (2005) 295--328}.

\bibitem{Gorbenko:2019rza}
V.~Gorbenko and L.~Senatore, \emph{{$\lambda \phi^4$ in dS}},
  \href{https://arxiv.org/abs/1911.00022}{{\tt 1911.00022}}.

\bibitem{Woodard:2025cez}
R.~P. Woodard, \emph{{Recent Developments in Stochastic Inflation}},
  \href{https://arxiv.org/abs/2501.15843}{{\tt 2501.15843}}.

\bibitem{rey1987dynamics}
S.-J. Rey, \emph{Dynamics of inflationary phase transition}, {\emph{Nuclear
  Physics B} {\bf 284} (1987) 706--728}.

\bibitem{nambu1988stochastic}
Y.~Nambu and M.~Sasaki, \emph{Stochastic stage of an inflationary universe
  model}, {\emph{Physics Letters B} {\bf 205} (1988) 441--446}.

\bibitem{nambu1989stochastic}
Y.~Nambu and M.~Sasaki, \emph{Stochastic approach to chaotic inflation and the
  distribution of universes}, {\emph{Physics Letters B} {\bf 219} (1989)
  240--246}.

\bibitem{kandrup1989stochastic}
H.~E. Kandrup, \emph{Stochastic inflation as a time-dependent random walk},
  {\emph{Physical Review D} {\bf 39} (1989) 2245}.

\bibitem{habib1992stochastic}
S.~Habib, \emph{Stochastic inflation: Quantum phase-space approach},
  {\emph{Physical Review D} {\bf 46} (1992) 2408}.

\bibitem{Enqvist:2008kt}
K.~Enqvist, S.~Nurmi, D.~Podolsky and G.~I. Rigopoulos, \emph{{On the
  divergences of inflationary superhorizon perturbations}},
  \href{http://dx.doi.org/10.1088/1475-7516/2008/04/025}{\emph{JCAP} {\bf 04}
  (2008) 025}, [\href{https://arxiv.org/abs/0802.0395}{{\tt 0802.0395}}].

\bibitem{Hu:2018nxy}
B.-L. Hu, \emph{{Infrared Behavior of Quantum Fields in Inflationary Cosmology
  -- Issues and Approaches: an overview}},
  \href{https://arxiv.org/abs/1812.11851}{{\tt 1812.11851}}.

\bibitem{Baumgart:2019clc}
M.~Baumgart and R.~Sundrum, \emph{{De Sitter Diagrammar and the Resummation of
  Time}}, \href{http://dx.doi.org/10.1007/JHEP07(2020)119}{\emph{JHEP} {\bf 07}
  (2020) 119}, [\href{https://arxiv.org/abs/1912.09502}{{\tt 1912.09502}}].

\bibitem{Pattison:2017mbe}
C.~Pattison, V.~Vennin, H.~Assadullahi and D.~Wands, \emph{{Quantum diffusion
  during inflation and primordial black holes}},
  \href{http://dx.doi.org/10.1088/1475-7516/2017/10/046}{\emph{JCAP} {\bf 10}
  (2017) 046}, [\href{https://arxiv.org/abs/1707.00537}{{\tt 1707.00537}}].

\bibitem{Kuhnel:2019xes}
F.~Kuhnel and K.~Freese, \emph{{On Stochastic Effects and Primordial Black-Hole
  Formation}},
  \href{http://dx.doi.org/10.1140/epjc/s10052-019-7466-0}{\emph{Eur. Phys. J.
  C} {\bf 79} (2019) 954}, [\href{https://arxiv.org/abs/1906.02744}{{\tt
  1906.02744}}].

\bibitem{Ezquiaga:2019ftu}
J.~M. Ezquiaga, J.~Garc{\'\i}a-Bellido and V.~Vennin, \emph{{The exponential
  tail of inflationary fluctuations: consequences for primordial black holes}},
  \href{http://dx.doi.org/10.1088/1475-7516/2020/03/029}{\emph{JCAP} {\bf 03}
  (2020) 029}, [\href{https://arxiv.org/abs/1912.05399}{{\tt 1912.05399}}].

\bibitem{Vennin:2020kng}
V.~Vennin, \emph{{Stochastic inflation and primordial black holes}}.
\newblock PhD thesis, AstroParticule et Cosmologie, France, U. Paris-Saclay, 6,
  2020.
\newblock \href{https://arxiv.org/abs/2009.08715}{{\tt 2009.08715}}.

\bibitem{Ballesteros:2020sre}
G.~Ballesteros, J.~Rey, M.~Taoso and A.~Urbano, \emph{{Stochastic inflationary
  dynamics beyond slow-roll and consequences for primordial black hole
  formation}},
  \href{http://dx.doi.org/10.1088/1475-7516/2020/08/043}{\emph{JCAP} {\bf 08}
  (2020) 043}, [\href{https://arxiv.org/abs/2006.14597}{{\tt 2006.14597}}].

\bibitem{Figueroa:2021zah}
D.~G. Figueroa, S.~Raatikainen, S.~Rasanen and E.~Tomberg, \emph{{Implications
  of stochastic effects for primordial black hole production in ultra-slow-roll
  inflation}},
  \href{http://dx.doi.org/10.1088/1475-7516/2022/05/027}{\emph{JCAP} {\bf 05}
  (2022) 027}, [\href{https://arxiv.org/abs/2111.07437}{{\tt 2111.07437}}].

\bibitem{Saha:2026cay}
P.~Saha, Y.~Tada and Y.~Urakawa, \emph{{Nonlinear Lattice Framework for
  Inflation: Bridging stochastic inflation and the $\delta N$ formalism}},
  \href{https://arxiv.org/abs/2604.00978}{{\tt 2604.00978}}.

\bibitem{Ye:2026saa}
X.-Q. Ye and S.-J. Wang, \emph{{Nonperturbative stochastic inflation in
  perturbative dynamical background}},
  \href{https://arxiv.org/abs/2604.15219}{{\tt 2604.15219}}.

\bibitem{Calzetta:1993qe}
E.~Calzetta and B.~L. Hu, \emph{{Noise and fluctuations in semiclassical
  gravity}}, \href{http://dx.doi.org/10.1103/PhysRevD.49.6636}{\emph{Phys. Rev.
  D} {\bf 49} (1994) 6636--6655},
  [\href{https://arxiv.org/abs/gr-qc/9312036}{{\tt gr-qc/9312036}}].

\bibitem{PerreaultLevasseur:2013kfq}
L.~Perreault~Levasseur, \emph{{Lagrangian formulation of stochastic inflation:
  Langevin equations, one-loop corrections and a proposed recursive approach}},
  \href{http://dx.doi.org/10.1103/PhysRevD.88.083537}{\emph{Phys. Rev. D} {\bf
  88} (2013) 083537}, [\href{https://arxiv.org/abs/1304.6408}{{\tt
  1304.6408}}].

\bibitem{Fujita:2013cna}
T.~Fujita, M.~Kawasaki, Y.~Tada and T.~Takesako, \emph{{A new algorithm for
  calculating the curvature perturbations in stochastic inflation}},
  \href{http://dx.doi.org/10.1088/1475-7516/2013/12/036}{\emph{JCAP} {\bf 12}
  (2013) 036}, [\href{https://arxiv.org/abs/1308.4754}{{\tt 1308.4754}}].

\bibitem{Fujita:2014tja}
T.~Fujita, M.~Kawasaki and Y.~Tada, \emph{{Non-perturbative approach for
  curvature perturbations in stochastic $\delta N$ formalism}},
  \href{http://dx.doi.org/10.1088/1475-7516/2014/10/030}{\emph{JCAP} {\bf 10}
  (2014) 030}, [\href{https://arxiv.org/abs/1405.2187}{{\tt 1405.2187}}].

\bibitem{PerreaultLevasseur:2014ziv}
L.~Perreault~Levasseur and E.~McDonough, \emph{{Backreaction and Stochastic
  Effects in Single Field Inflation}},
  \href{http://dx.doi.org/10.1103/PhysRevD.91.063513}{\emph{Phys. Rev. D} {\bf
  91} (2015) 063513}, [\href{https://arxiv.org/abs/1409.7399}{{\tt
  1409.7399}}].

\bibitem{Vennin:2015hra}
V.~Vennin and A.~A. Starobinsky, \emph{{Correlation Functions in Stochastic
  Inflation}},
  \href{http://dx.doi.org/10.1140/epjc/s10052-015-3643-y}{\emph{Eur. Phys. J.
  C} {\bf 75} (2015) 413}, [\href{https://arxiv.org/abs/1506.04732}{{\tt
  1506.04732}}].

\bibitem{Grain:2017dqa}
J.~Grain and V.~Vennin, \emph{{Stochastic inflation in phase space: is slow
  roll a stochastic attractor?}},
  \href{http://dx.doi.org/10.1088/1475-7516/2017/05/045}{\emph{JCAP} {\bf 05}
  (2017) 045}, [\href{https://arxiv.org/abs/1703.00447}{{\tt 1703.00447}}].

\bibitem{Firouzjahi:2018vet}
H.~Firouzjahi, A.~Nassiri-Rad and M.~Noorbala, \emph{{Stochastic Ultra Slow
  Roll Inflation}},
  \href{http://dx.doi.org/10.1088/1475-7516/2019/01/040}{\emph{JCAP} {\bf 01}
  (2019) 040}, [\href{https://arxiv.org/abs/1811.02175}{{\tt 1811.02175}}].

\bibitem{Pinol:2018euk}
L.~Pinol, S.~Renaux-Petel and Y.~Tada, \emph{{Inflationary stochastic
  anomalies}}, \href{http://dx.doi.org/10.1088/1361-6382/ab097f}{\emph{Class.
  Quant. Grav.} {\bf 36} (2019) 07LT01},
  [\href{https://arxiv.org/abs/1806.10126}{{\tt 1806.10126}}].

\bibitem{Firouzjahi:2020jrj}
H.~Firouzjahi, A.~Nassiri-Rad and M.~Noorbala, \emph{{Stochastic nonattractor
  inflation}}, \href{http://dx.doi.org/10.1103/PhysRevD.102.123504}{\emph{Phys.
  Rev. D} {\bf 102} (2020) 123504},
  [\href{https://arxiv.org/abs/2009.04680}{{\tt 2009.04680}}].

\bibitem{Pattison:2021oen}
C.~Pattison, V.~Vennin, D.~Wands and H.~Assadullahi, \emph{{Ultra-slow-roll
  inflation with quantum diffusion}},
  \href{http://dx.doi.org/10.1088/1475-7516/2021/04/080}{\emph{JCAP} {\bf 04}
  (2021) 080}, [\href{https://arxiv.org/abs/2101.05741}{{\tt 2101.05741}}].

\bibitem{Cruces:2021iwq}
D.~Cruces and C.~Germani, \emph{{Stochastic inflation at all order in slow-roll
  parameters: Foundations}},
  \href{http://dx.doi.org/10.1103/PhysRevD.105.023533}{\emph{Phys. Rev. D} {\bf
  105} (2022) 023533}, [\href{https://arxiv.org/abs/2107.12735}{{\tt
  2107.12735}}].

\bibitem{Cohen:2021fzf}
T.~Cohen, D.~Green, A.~Premkumar and A.~Ridgway, \emph{{Stochastic Inflation at
  NNLO}}, \href{http://dx.doi.org/10.1007/JHEP09(2021)159}{\emph{JHEP} {\bf 09}
  (2021) 159}, [\href{https://arxiv.org/abs/2106.09728}{{\tt 2106.09728}}].

\bibitem{Aldabergenov:2025ulq}
Y.~Aldabergenov, D.~Ding, W.~Lin and Y.~Wan, \emph{{Towards Stochastic
  Inflation in Higher-Curvature Gravity}},
  \href{https://arxiv.org/abs/2506.21423}{{\tt 2506.21423}}.

\bibitem{Panda:2025tpu}
R.~K. Panda, S.~Panda and A.~Tinwala, \emph{{Revisiting Lagrangian Formulation
  of Stochastic inflation}},  \href{https://arxiv.org/abs/2510.03171}{{\tt
  2510.03171}}.

\bibitem{Kawasaki:2026hnx}
M.~Kawasaki and T.~Kuroda, \emph{{Numerical simulation of the stochastic
  formalism including non-Markovianity}},
  \href{https://arxiv.org/abs/2602.11652}{{\tt 2602.11652}}.

\bibitem{morikawa1990dissipation}
M.~Morikawa, \emph{Dissipation and fluctuation of quantum fields in expanding
  universes}, {\emph{Physical Review D} {\bf 42} (1990) 1027}.

\bibitem{hosoya1989stochastic}
A.~Hosoya, M.~Morikawa and K.~Nakayama, \emph{Stochastic dynamics of scalar
  field in the inflationary universe}, {\emph{International Journal of Modern
  Physics A} {\bf 4} (1989) 2613--2625}.

\bibitem{Hu:1992ig}
B.~L. Hu, J.~P. Paz and Y.~Zhang, \emph{{Quantum origin of noise and
  fluctuations in cosmology}},  in \emph{{The Origin of Structure in the
  Universe}}, 1992.
\newblock \href{https://arxiv.org/abs/gr-qc/9512049}{{\tt gr-qc/9512049}}.

\bibitem{Hu:1994iw}
B.~L. Hu, \emph{{Quantum statistical field theory in gravitation and
  cosmology}},  in \emph{{3rd Workshop on Thermal Field Theories and their
  Applications}}, 1, 1994.
\newblock \href{https://arxiv.org/abs/gr-qc/9403061}{{\tt gr-qc/9403061}}.

\bibitem{Garbrecht:2013coa}
B.~Garbrecht, G.~Rigopoulos and Y.~Zhu, \emph{{Infrared correlations in de
  Sitter space: Field theoretic versus stochastic approach}},
  \href{http://dx.doi.org/10.1103/PhysRevD.89.063506}{\emph{Phys. Rev. D} {\bf
  89} (2014) 063506}, [\href{https://arxiv.org/abs/1310.0367}{{\tt
  1310.0367}}].

\bibitem{Garbrecht:2014dca}
B.~Garbrecht, F.~Gautier, G.~Rigopoulos and Y.~Zhu, \emph{{Feynman Diagrams for
  Stochastic Inflation and Quantum Field Theory in de Sitter Space}},
  \href{http://dx.doi.org/10.1103/PhysRevD.91.063520}{\emph{Phys. Rev. D} {\bf
  91} (2015) 063520}, [\href{https://arxiv.org/abs/1412.4893}{{\tt
  1412.4893}}].

\bibitem{Moss:2016uix}
I.~Moss and G.~Rigopoulos, \emph{{Effective long wavelength scalar dynamics in
  de Sitter}},
  \href{http://dx.doi.org/10.1088/1475-7516/2017/05/009}{\emph{JCAP} {\bf 05}
  (2017) 009}, [\href{https://arxiv.org/abs/1611.07589}{{\tt 1611.07589}}].

\bibitem{Prokopec:2017vxx}
T.~Prokopec and G.~Rigopoulos, \emph{{Functional renormalization group for
  stochastic inflation}},
  \href{http://dx.doi.org/10.1088/1475-7516/2018/08/013}{\emph{JCAP} {\bf 08}
  (2018) 013}, [\href{https://arxiv.org/abs/1710.07333}{{\tt 1710.07333}}].

\bibitem{Collins:2017haz}
H.~Collins, R.~Holman and T.~Vardanyan, \emph{{The quantum Fokker-Planck
  equation of stochastic inflation}},
  \href{http://dx.doi.org/10.1007/JHEP11(2017)065}{\emph{JHEP} {\bf 11} (2017)
  065}, [\href{https://arxiv.org/abs/1706.07805}{{\tt 1706.07805}}].

\bibitem{Andersen:2021lii}
J.~O. Andersen, M.~Eriksson and A.~Tranberg, \emph{{Stochastic inflation from
  quantum field theory and the parametric dependence of the effective noise
  amplitude}}, \href{http://dx.doi.org/10.1007/JHEP02(2022)121}{\emph{JHEP}
  {\bf 02} (2022) 121}, [\href{https://arxiv.org/abs/2111.14503}{{\tt
  2111.14503}}].

\bibitem{Tokuda:2017fdh}
J.~Tokuda and T.~Tanaka, \emph{{Statistical nature of infrared dynamics on de
  Sitter background}},
  \href{http://dx.doi.org/10.1088/1475-7516/2018/02/014}{\emph{JCAP} {\bf 02}
  (2018) 014}, [\href{https://arxiv.org/abs/1708.01734}{{\tt 1708.01734}}].

\bibitem{Tokuda:2018eqs}
J.~Tokuda and T.~Tanaka, \emph{{Can all the infrared secular growth really be
  understood as increase of classical statistical variance?}},
  \href{http://dx.doi.org/10.1088/1475-7516/2018/11/022}{\emph{JCAP} {\bf 11}
  (2018) 022}, [\href{https://arxiv.org/abs/1806.03262}{{\tt 1806.03262}}].

\bibitem{Cruces:2022imf}
D.~Cruces, \emph{{Review on Stochastic Approach to Inflation}},
  \href{http://dx.doi.org/10.3390/universe8060334}{\emph{Universe} {\bf 8}
  (2022) 334}, [\href{https://arxiv.org/abs/2203.13852}{{\tt 2203.13852}}].

\bibitem{Miyachi:2023fss}
T.~Miyachi, J.~Soda and J.~Tokuda, \emph{{Stochastic Tunneling in de Sitter
  Spacetime}},
  \href{http://dx.doi.org/10.3390/universe10070292}{\emph{Universe} {\bf 10}
  (2024) 292}, [\href{https://arxiv.org/abs/2309.07440}{{\tt 2309.07440}}].

\bibitem{Kamenshchik:2024ybm}
A.~Kamenshchik and P.~Petriakova, \emph{{IR finite correlation functions in de
  Sitter space, a smooth massless limit, and an autonomous equation}},
  \href{http://dx.doi.org/10.1007/JHEP04(2025)127}{\emph{JHEP} {\bf 04} (2025)
  127}, [\href{https://arxiv.org/abs/2410.16226}{{\tt 2410.16226}}].

\bibitem{Pinol:2020cdp}
L.~Pinol, S.~Renaux-Petel and Y.~Tada, \emph{{A manifestly covariant theory of
  multifield stochastic inflation in phase space: solving the discretisation
  ambiguity in stochastic inflation}},
  \href{http://dx.doi.org/10.1088/1475-7516/2021/04/048}{\emph{JCAP} {\bf 04}
  (2021) 048}, [\href{https://arxiv.org/abs/2008.07497}{{\tt 2008.07497}}].

\bibitem{Kamenshchik:2025ses}
A.~Kamenshchik and P.~Petriakova, \emph{{From the Fokker-Planck equation to
  perturbative QFT{\textquoteright}s results in de Sitter space}},
  \href{http://dx.doi.org/10.1007/JHEP08(2025)063}{\emph{JHEP} {\bf 08} (2025)
  063}, [\href{https://arxiv.org/abs/2504.20646}{{\tt 2504.20646}}].

\bibitem{Cruces:2026yvs}
D.~Cruces and T.~Kuroda, \emph{{A consistent formulation of stochastic
  inflation I: Non-Markovian effects and issues beyond linear perturbations}},
  \href{https://arxiv.org/abs/2605.00476}{{\tt 2605.00476}}.

\bibitem{Palma:2023idj}
G.~A. Palma and S.~Sypsas, \emph{{Non-Gaussian statistics of de Sitter
  spectators: a perturbative derivation of stochastic dynamics}},
  \href{http://dx.doi.org/10.1007/JHEP12(2025)170}{\emph{JHEP} {\bf 12} (2025)
  170}, [\href{https://arxiv.org/abs/2309.16474}{{\tt 2309.16474}}].

\bibitem{Palma:2025oux}
G.~A. Palma, S.~Sypsas and D.~Tapia, \emph{{Confronting infrared divergences in
  de Sitter: loops, logarithms and the stochastic formalism}},
  \href{https://arxiv.org/abs/2507.21310}{{\tt 2507.21310}}.

\bibitem{Li:2025azq}
Y.-Z. Li, \emph{{Stochastic Inflation as an Open Quantum System}},
  \href{http://dx.doi.org/10.1103/hlfz-8lhg}{\emph{Phys. Rev. Lett.} {\bf 136}
  (2026) 071501}, [\href{https://arxiv.org/abs/2507.02070}{{\tt 2507.02070}}].

\bibitem{Green:2025hmo}
D.~Green and K.~Gupta, \emph{{Quantum Walks and Exact RG in de Sitter Space}},
  \href{https://arxiv.org/abs/2512.13842}{{\tt 2512.13842}}.

\bibitem{Cespedes:2023aal}
S.~C{\'e}spedes, A.-C. Davis and D.-G. Wang, \emph{{On the IR divergences in de
  Sitter space: loops, resummation and the semi-classical wavefunction}},
  \href{http://dx.doi.org/10.1007/JHEP04(2024)004}{\emph{JHEP} {\bf 04} (2024)
  004}, [\href{https://arxiv.org/abs/2311.17990}{{\tt 2311.17990}}].

\bibitem{Cespedes:2026fdp}
S.~C{\'e}spedes and T.~Colas, \emph{{Stochastic inflation from a
  non-equilibrium renormalization group}},
  \href{https://arxiv.org/abs/2605.11096}{{\tt 2605.11096}}.

\bibitem{Beneke:2026ksj}
M.~Beneke, P.~Hager and A.~F. Sanfilippo, \emph{{Quantum correction to the
  diffusion term in stochastic inflation from composite-operator matching in
  Soft de Sitter Effective Theory}},
  \href{https://arxiv.org/abs/2604.14283}{{\tt 2604.14283}}.

\bibitem{hartle1983wave}
J.~B. Hartle and S.~W. Hawking, \emph{Wave function of the universe},
  {\emph{Physical Review D} {\bf 28} (1983) 2960}.

\bibitem{kamenev2023field}
A.~Kamenev, \emph{Field theory of non-equilibrium systems}.
\newblock Cambridge University Press, 2023.

\bibitem{manzano2020short}
D.~Manzano, \emph{A short introduction to the lindblad master equation},
  {\emph{Aip advances} {\bf 10} (2020) }.

\bibitem{Jana:2020vyx}
C.~Jana, R.~Loganayagam and M.~Rangamani, \emph{{Open quantum systems and
  Schwinger-Keldysh holograms}},
  \href{http://dx.doi.org/10.1007/JHEP07(2020)242}{\emph{JHEP} {\bf 07} (2020)
  242}, [\href{https://arxiv.org/abs/2004.02888}{{\tt 2004.02888}}].

\bibitem{Ivo:2024ill}
V.~Ivo, Y.-Z. Li and J.~Maldacena, \emph{{The no boundary density matrix}},
  \href{http://dx.doi.org/10.1007/JHEP02(2025)124}{\emph{JHEP} {\bf 02} (2025)
  124}, [\href{https://arxiv.org/abs/2409.14218}{{\tt 2409.14218}}].

\bibitem{Witten:2021nzp}
E.~Witten, \emph{{A Note On Complex Spacetime Metrics}},
  \href{https://arxiv.org/abs/2111.06514}{{\tt 2111.06514}}.

\bibitem{Winitzki:1999ve}
S.~Winitzki and A.~Vilenkin, \emph{{Effective noise in stochastic description
  of inflation}},
  \href{http://dx.doi.org/10.1103/PhysRevD.61.084008}{\emph{Phys. Rev. D} {\bf
  61} (2000) 084008}, [\href{https://arxiv.org/abs/gr-qc/9911029}{{\tt
  gr-qc/9911029}}].

\bibitem{polchinski1984renormalization}
J.~Polchinski, \emph{Renormalization and effective lagrangians}, {\emph{Nuclear
  Physics B} {\bf 231} (1984) 269--295}.

\bibitem{bunch1978quantum}
T.~S. Bunch and P.~C. Davies, \emph{Quantum field theory in de sitter space:
  renormalization by point-splitting}, {\emph{Proceedings of the Royal Society
  of London. A. Mathematical and Physical Sciences} {\bf 360} (1978) 117--134}.

\bibitem{breuer2002theory}
H.-P. Breuer and F.~Petruccione, \emph{The theory of open quantum systems}.
\newblock OUP Oxford, 2002.

\bibitem{Green:2024cmx}
D.~Green and G.~Sun, \emph{{Effective field theory and in-in correlators}},
  \href{http://dx.doi.org/10.1007/JHEP04(2025)166}{\emph{JHEP} {\bf 04} (2025)
  166}, [\href{https://arxiv.org/abs/2412.02739}{{\tt 2412.02739}}].

\bibitem{Cohen:2020php}
T.~Cohen and D.~Green, \emph{{Soft de Sitter Effective Theory}},
  \href{http://dx.doi.org/10.1007/JHEP12(2020)041}{\emph{JHEP} {\bf 12} (2020)
  041}, [\href{https://arxiv.org/abs/2007.03693}{{\tt 2007.03693}}].

\bibitem{Beneke:2023wmt}
M.~Beneke, P.~Hager and A.~F. Sanfilippo, \emph{{Cosmological correlators in
  massless {\ensuremath{\phi}}$^{4}$-theory and the method of regions}},
  \href{http://dx.doi.org/10.1007/JHEP04(2024)006}{\emph{JHEP} {\bf 04} (2024)
  006}, [\href{https://arxiv.org/abs/2312.06766}{{\tt 2312.06766}}].

\bibitem{Beneke:2026rtf}
M.~Beneke, P.~Hager and A.~F. Sanfilippo, \emph{{Renormalisation and matching
  of massless scalar correlation functions in Soft de Sitter Effective
  Theory}},  \href{https://arxiv.org/abs/2603.09438}{{\tt 2603.09438}}.

\bibitem{Starobinsky:1982ee}
A.~A. Starobinsky, \emph{{Dynamics of Phase Transition in the New Inflationary
  Universe Scenario and Generation of Perturbations}},
  \href{http://dx.doi.org/10.1016/0370-2693(82)90541-X}{\emph{Phys. Lett. B}
  {\bf 117} (1982) 175--178}.

\bibitem{Starobinsky:1985ibc}
A.~A. Starobinsky, \emph{{Multicomponent de Sitter (Inflationary) Stages and
  the Generation of Perturbations}}, {\emph{JETP Lett.} {\bf 42} (1985)
  152--155}.

\bibitem{sasaki1996general}
M.~Sasaki and E.~D. Stewart, \emph{A general analytic formula for the spectral
  index of the density perturbations produced during inflation},
  {\emph{Progress of Theoretical Physics} {\bf 95} (1996) 71--78}.

\bibitem{Wands:2000dp}
D.~Wands, K.~A. Malik, D.~H. Lyth and A.~R. Liddle, \emph{{A New approach to
  the evolution of cosmological perturbations on large scales}},
  \href{http://dx.doi.org/10.1103/PhysRevD.62.043527}{\emph{Phys. Rev. D} {\bf
  62} (2000) 043527}, [\href{https://arxiv.org/abs/astro-ph/0003278}{{\tt
  astro-ph/0003278}}].

\bibitem{Rigopoulos:2003ak}
G.~I. Rigopoulos and E.~P.~S. Shellard, \emph{{The separate universe approach
  and the evolution of nonlinear superhorizon cosmological perturbations}},
  \href{http://dx.doi.org/10.1103/PhysRevD.68.123518}{\emph{Phys. Rev. D} {\bf
  68} (2003) 123518}, [\href{https://arxiv.org/abs/astro-ph/0306620}{{\tt
  astro-ph/0306620}}].

\bibitem{Christie:2025knc}
R.~Christie, J.~Joo, G.~Kaplanek, V.~Vennin and D.~Wands, \emph{{Cosmic
  Lockdown: When Decoherence Saves the Universe from Tunneling}},
  \href{https://arxiv.org/abs/2512.14204}{{\tt 2512.14204}}.

\bibitem{Premkumar:2021mlz}
A.~Premkumar, \emph{{Regulating loops in de Sitter spacetime}},
  \href{http://dx.doi.org/10.1103/PhysRevD.109.045003}{\emph{Phys. Rev. D} {\bf
  109} (2024) 045003}, [\href{https://arxiv.org/abs/2110.12504}{{\tt
  2110.12504}}].

\bibitem{Caron-Huot:2022eqs}
S.~Caron-Huot, M.~Kologlu, P.~Kravchuk, D.~Meltzer and D.~Simmons-Duffin,
  \emph{{Detectors in weakly-coupled field theories}},
  \href{http://dx.doi.org/10.1007/JHEP04(2023)014}{\emph{JHEP} {\bf 04} (2023)
  014}, [\href{https://arxiv.org/abs/2209.00008}{{\tt 2209.00008}}].

\bibitem{Henriksson:2023cnh}
J.~Henriksson, P.~Kravchuk and B.~Oertel, \emph{{Missing local operators,
  zeros, and twist-4 trajectories}},
  \href{http://dx.doi.org/10.1007/JHEP07(2024)248}{\emph{JHEP} {\bf 07} (2024)
  248}, [\href{https://arxiv.org/abs/2312.09283}{{\tt 2312.09283}}].

\bibitem{Chang:2025zib}
C.-H. Chang, H.~Chen, D.~Simmons-Duffin and H.~X. Zhu, \emph{{Seeing through
  the confinement screen: DGLAP/BFKL mixing and light-ray matching in QCD}},
  \href{http://dx.doi.org/10.1007/JHEP02(2026)251}{\emph{JHEP} {\bf 02} (2026)
  251}, [\href{https://arxiv.org/abs/2506.06431}{{\tt 2506.06431}}].

\bibitem{Li:2025knf}
Y.-Z. Li and D.~Simmons-Duffin, \emph{{Regge trajectories, detectors, and
  distributions in the critical O(N) model}},
  \href{http://dx.doi.org/10.1007/JHEP02(2026)149}{\emph{JHEP} {\bf 02} (2026)
  149}, [\href{https://arxiv.org/abs/2506.06419}{{\tt 2506.06419}}].

\bibitem{De:2025yls}
S.~De and Y.~Landau, \emph{{De Sitter light-ray operators}},
  \href{http://dx.doi.org/10.1007/JHEP04(2026)025}{\emph{JHEP} {\bf 04} (2026)
  025}, [\href{https://arxiv.org/abs/2512.10846}{{\tt 2512.10846}}].

\bibitem{Bellini:1996uh}
M.~Bellini, H.~Casini, R.~Montemayor and P.~Sisterna, \emph{{Stochastic
  approach to inflation: Classicality conditions}},
  \href{http://dx.doi.org/10.1103/PhysRevD.54.7172}{\emph{Phys. Rev. D} {\bf
  54} (1996) 7172--7180}.

\bibitem{Calzetta:1989vk}
E.~Calzetta and B.~L. Hu, \emph{{Wigner Distribution Function and Phase Space
  Formulation of Quantum Cosmology}},
  \href{http://dx.doi.org/10.1103/PhysRevD.40.380}{\emph{Phys. Rev. D} {\bf 40}
  (1989) 380--389}.

\bibitem{Allen:1985ux}
B.~Allen, \emph{{Vacuum States in de Sitter Space}},
  \href{http://dx.doi.org/10.1103/PhysRevD.32.3136}{\emph{Phys. Rev. D} {\bf
  32} (1985) 3136}.

\bibitem{Einhorn:2002nu}
M.~B. Einhorn and F.~Larsen, \emph{{Interacting quantum field theory in de
  Sitter vacua}},
  \href{http://dx.doi.org/10.1103/PhysRevD.67.024001}{\emph{Phys. Rev. D} {\bf
  67} (2003) 024001}, [\href{https://arxiv.org/abs/hep-th/0209159}{{\tt
  hep-th/0209159}}].

\bibitem{Goldstein:2003ut}
K.~Goldstein and D.~A. Lowe, \emph{{A Note on alpha vacua and interacting field
  theory in de Sitter space}},
  \href{http://dx.doi.org/10.1016/j.nuclphysb.2003.07.014}{\emph{Nucl. Phys. B}
  {\bf 669} (2003) 325--340}, [\href{https://arxiv.org/abs/hep-th/0302050}{{\tt
  hep-th/0302050}}].

\bibitem{Einhorn:2003xb}
M.~B. Einhorn and F.~Larsen, \emph{{Squeezed states in the de Sitter vacuum}},
  \href{http://dx.doi.org/10.1103/PhysRevD.68.064002}{\emph{Phys. Rev. D} {\bf
  68} (2003) 064002}, [\href{https://arxiv.org/abs/hep-th/0305056}{{\tt
  hep-th/0305056}}].

\bibitem{Mirbabayi:2019qtx}
M.~Mirbabayi, \emph{{Infrared dynamics of a light scalar field in de Sitter}},
  \href{http://dx.doi.org/10.1088/1475-7516/2020/12/006}{\emph{JCAP} {\bf 12}
  (2020) 006}, [\href{https://arxiv.org/abs/1911.00564}{{\tt 1911.00564}}].

\bibitem{Bucher:1994gb}
M.~Bucher, A.~S. Goldhaber and N.~Turok, \emph{{An open universe from
  inflation}}, \href{http://dx.doi.org/10.1103/PhysRevD.52.3314}{\emph{Phys.
  Rev. D} {\bf 52} (1995) 3314--3337},
  [\href{https://arxiv.org/abs/hep-ph/9411206}{{\tt hep-ph/9411206}}].

\bibitem{morikawa1987origin}
M.~Morikawa, \emph{The origin of the density fluctuations in de sitter space},
  {\emph{Progress of theoretical physics} {\bf 77} (1987) 1163--1177}.

\bibitem{Lombardo:1995fg}
F.~Lombardo and F.~D. Mazzitelli, \emph{{Coarse graining and decoherence in
  quantum field theory}},
  \href{http://dx.doi.org/10.1103/PhysRevD.53.2001}{\emph{Phys. Rev. D} {\bf
  53} (1996) 2001--2011}, [\href{https://arxiv.org/abs/hep-th/9508052}{{\tt
  hep-th/9508052}}].

\bibitem{Calzetta:1995ys}
E.~Calzetta and B.~L. Hu, \emph{{Quantum fluctuations, decoherence of the mean
  field, and structure formation in the early universe}},
  \href{http://dx.doi.org/10.1103/PhysRevD.52.6770}{\emph{Phys. Rev. D} {\bf
  52} (1995) 6770--6788}, [\href{https://arxiv.org/abs/gr-qc/9505046}{{\tt
  gr-qc/9505046}}].

\bibitem{Burgess:2006jn}
C.~P. Burgess, R.~Holman and D.~Hoover, \emph{{Decoherence of inflationary
  primordial fluctuations}},
  \href{http://dx.doi.org/10.1103/PhysRevD.77.063534}{\emph{Phys. Rev. D} {\bf
  77} (2008) 063534}, [\href{https://arxiv.org/abs/astro-ph/0601646}{{\tt
  astro-ph/0601646}}].

\bibitem{Sano:2025ird}
F.~Sano and J.~Tokuda, \emph{{False and genuine decoherence in the early
  universe: a local observer and time-averaged observables}},
  \href{https://arxiv.org/abs/2504.10472}{{\tt 2504.10472}}.

\bibitem{Liu:2016aaf}
J.~Liu, C.-M. Sou and Y.~Wang, \emph{{Cosmic Decoherence: Massive Fields}},
  \href{http://dx.doi.org/10.1007/JHEP10(2016)072}{\emph{JHEP} {\bf 10} (2016)
  072}, [\href{https://arxiv.org/abs/1608.07909}{{\tt 1608.07909}}].

\bibitem{Lombardo:2004fr}
F.~C. Lombardo, \emph{{Influence functional approach to decoherence during
  inflation}},
  \href{http://dx.doi.org/10.1590/S0103-97332005000300005}{\emph{Braz. J.
  Phys.} {\bf 35} (2005) 391--396},
  [\href{https://arxiv.org/abs/gr-qc/0412069}{{\tt gr-qc/0412069}}].

\bibitem{Lombardo:2005iz}
F.~C. Lombardo and D.~Lopez~Nacir, \emph{{Decoherence during inflation: The
  Generation of classical inhomogeneities}},
  \href{http://dx.doi.org/10.1103/PhysRevD.72.063506}{\emph{Phys. Rev. D} {\bf
  72} (2005) 063506}, [\href{https://arxiv.org/abs/gr-qc/0506051}{{\tt
  gr-qc/0506051}}].

\bibitem{DaddiHammou:2022itk}
A.~Daddi~Hammou and N.~Bartolo, \emph{{Cosmic decoherence: primordial power
  spectra and non-Gaussianities}},
  \href{http://dx.doi.org/10.1088/1475-7516/2023/04/055}{\emph{JCAP} {\bf 04}
  (2023) 055}, [\href{https://arxiv.org/abs/2211.07598}{{\tt 2211.07598}}].

\bibitem{deKruijf:2024ufs}
J.~de~Kruijf and N.~Bartolo, \emph{{The effect of quantum decoherence on
  inflationary gravitational waves}},
  \href{http://dx.doi.org/10.1088/1475-7516/2024/11/041}{\emph{JCAP} {\bf 11}
  (2024) 041}, [\href{https://arxiv.org/abs/2408.02563}{{\tt 2408.02563}}].

\bibitem{Lopez:2025arw}
F.~Lopez and N.~Bartolo, \emph{{Quantum signatures and decoherence during
  inflation from deep subhorizon perturbations}},
  \href{https://arxiv.org/abs/2503.23150}{{\tt 2503.23150}}.

\bibitem{Calzetta:1999zr}
E.~A. Calzetta, B.~L. Hu and F.~D. Mazzitelli, \emph{{Coarse grained effective
  action and renormalization group theory in semiclassical gravity and
  cosmology}},
  \href{http://dx.doi.org/10.1016/S0370-1573(01)00043-6}{\emph{Phys. Rept.}
  {\bf 352} (2001) 459--520}, [\href{https://arxiv.org/abs/hep-th/0102199}{{\tt
  hep-th/0102199}}].

\bibitem{Hu:2008rga}
B.~L. Hu and E.~Verdaguer, \emph{{Stochastic Gravity: Theory and
  Applications}}, \href{http://dx.doi.org/10.12942/lrr-2008-3}{\emph{Living
  Rev. Rel.} {\bf 11} (2008) 3}, [\href{https://arxiv.org/abs/0802.0658}{{\tt
  0802.0658}}].

\bibitem{Kaplanek:2025moq}
G.~Kaplanek, M.~Mylova and A.~J. Tolley, \emph{{Gauging Open EFTs from the top
  down}},  \href{https://arxiv.org/abs/2512.17089}{{\tt 2512.17089}}.

\bibitem{Kaplanek:2026kpp}
G.~Kaplanek, M.~Mylova and A.~J. Tolley, \emph{{Schwinger-Keldysh Path Integral
  for Gauge theories}},  \href{https://arxiv.org/abs/2604.26941}{{\tt
  2604.26941}}.

\bibitem{massignan2015quantum}
P.~Massignan, A.~Lampo, J.~Wehr and M.~Lewenstein, \emph{Quantum brownian
  motion with inhomogeneous damping and diffusion}, {\emph{Physical Review A}
  {\bf 91} (2015) 033627}.

\bibitem{maniscalco2004lindblad}
S.~Maniscalco, J.~Piilo, F.~Intravaia, F.~Petruccione and A.~Messina,
  \emph{Lindblad-and non-lindblad-type dynamics of a quantum brownian
  particle}, {\emph{Physical Review A—Atomic, Molecular, and Optical Physics}
  {\bf 70} (2004) 032113}.

\bibitem{lampo2016lindblad}
A.~Lampo, S.~H. Lim, J.~Wehr, P.~Massignan and M.~Lewenstein, \emph{Lindblad
  model of quantum brownian motion}, {\emph{Physical Review A} {\bf 94} (2016)
  042123}.

\bibitem{caldeira1985influence}
A.~O. Caldeira and A.~J. Leggett, \emph{Influence of damping on quantum
  interference: An exactly soluble model}, {\emph{Physical Review A} {\bf 31}
  (1985) 1059}.

\bibitem{sieberer2016keldysh}
L.~M. Sieberer, M.~Buchhold and S.~Diehl, \emph{Keldysh field theory for driven
  open quantum systems}, {\emph{Reports on Progress in Physics} {\bf 79} (2016)
  096001}.

\bibitem{hu1992quantum}
B.~L. Hu, J.~P. Paz and Y.~Zhang, \emph{Quantum brownian motion in a general
  environment: Exact master equation with nonlocal dissipation and colored
  noise}, {\emph{Physical Review D} {\bf 45} (1992) 2843}.

\end{thebibliography}\endgroup

\end{document}